\documentclass[preprint,3p]{elsarticle}
\makeatletter
\def\ps@pprintTitle{%
	\let\@oddhead\@empty
	\let\@evenhead\@empty
	\def\@oddfoot{}%
	\let\@evenfoot\@oddfoot}
\makeatother

\usepackage[utf8]{inputenc}
\usepackage[english]{babel}
\usepackage{amsfonts,amsmath,amsthm,amssymb}
\usepackage[algo2e,linesnumbered,ruled]{algorithm2e}
\usepackage{mathtools}
\usepackage{subcaption}
\usepackage[export]{adjustbox}
\usepackage[labelfont=bf,font=sf]{caption}
\usepackage{hyperref}
\usepackage{cleveref}
\usepackage{autonum}

\newcommand{\x}{\vct{x}}
\newcommand{\y}{\vct{y}}

\renewcommand{\k}{\vct{k}}
\renewcommand{\r}{\vct{r}}
\newcommand{\Id}{\operatorname*{Id}}

\newcommand{\eps}{\varepsilon}

\newcommand{\norm}[1]{\lVert #1 \rVert}
\newcommand{\abs}[1]{| #1 |}

\newcommand{\tp}{^T}

\renewcommand{\d}{\operatorname{d}\!}

\renewcommand{\O}{\mathcal{O}}
\newcommand{\conv}{*}
\newcommand{\atanh}{\operatorname{atanh}}

\renewcommand{\vec}[1]{\pmb{#1}}
\newcommand{\vct}[1]{\vec{#1}}
\newcommand{\ten}[1]{\pmb{#1}}
\newcommand{\tns}[1]{\ten{#1}}

\DeclarePairedDelimiter\ceil{\lceil}{\rceil}

\newcommand{\grad}{\nabla}

\newcommand{\R}{\mathbb{R}} 
\newcommand{\N}{\mathbb{N}} 

\renewcommand{\L}[1]{\mathrm{L}^{#1}} 

\newcommand{\E}{\operatorname{\mathbb{E}}}

\newcommand{\Gaussian}{\mathcal{N}}

\newcommand{\VaR}{\operatorname{VaR}}
\newcommand{\CVaR}{\operatorname{CVaR}}
\newcommand{\Var}{\operatorname{Var}}
\newcommand{\SD}{d} 
\newcommand{\SDdata}{\SD^{\ast}} 
\newcommand{\dist}{\rho} 

\newcommand{\BesselK}[1]{\mathcal{K}_{#1}}

\newcounter{rownumber}

\newcommand{\Domain}{\Omega} 
\newcommand{\dom}{\Domain} 	
\newcommand{\Phase}{\varphi} 	
\newcommand{\Int}{u} 		

\newcommand{\cov}{\mathcal{C}}

\newcommand{\vfm}{\bar{\Phase}} 
\newcommand{\corrlen}{\ell}

\newcommand{\Matern}[1]{\mathcal{M}_{#1}} 

\newcommand{\TargMat}{\Phase_{\ast}}

\newcommand{\thickness}{\tau}

\newcommand{\loss}{J}

\newcommand{\strain}{\tns{\varepsilon}}
\newcommand{\stress}{\tns{\sigma}}

\newdefinition{remark}{Remark}

\newdefinition{example}{Example}

\graphicspath{{./figures/}}

\begin{document}
	
	\title{Statistically equivalent surrogate material models and the impact of random imperfections on elasto-plastic response }

	\author[1]{Ustim Khristenko\corref{corr1}}\ead{khristen@ma.tum.de}
	\cortext[corr1]{Corresponding author}
	
	\author[2]{Andrei Constantinescu}\ead{andrei.constantinescu@polytechnique.edu}
	
	\author[2]{Patrick Le~Tallec}\ead{patrick.letallec@polytechnique.edu}
	
	\author[1]{Barbara Wohlmuth}\ead{wohlmuth@ma.tum.de}
	
	\address[1]{Department of Mathematics, Technical University of Munich, Garching, Germany}
	\address[2]{Laboratoire de M{\'e}canique des Solides, CNRS -  {\'E}cole Polytechnique - Institut Polytechnique de Paris, Palaiseau, France}

	\begin{abstract}
		Manufactured materials usually contain random imperfections due to the fabrication process, e.g., the 3D-printing, casting, etc.
		These imperfections affect significantly the effective material properties and result in uncertainties in the mechanical response.
		Numerical analysis of the effects of the imperfections and the uncertainty quantification (UQ) can be often done by use of digital stochastic surrogate material models.
		In this work, we present a new flexible class of surrogate models depending on a small number of parameters with special focus on two-phase materials.
		The surrogate models are constructed as the level-set of a linear combination of an intensity field representing the topological shape and a Gaussian perturbation representing the imperfections.
		The mathematical design parameters of the model are related to physical ones and thus easy to interpret.
		The calibration of the model parameters is performed using progressive batching sub-sampled quasi-Newton minimization, using a designed distance measure between the synthetic samples and the data.
		Then, employing a fast sampling algorithm, an arbitrary number of synthetic samples can be generated to use in Monte Carlo type methods.		
		In particular, we illustrate the method in application to UQ of the elasto-plastic response of an imperfect octet-truss lattice which plays an important role in additive manufacturing.
		To this end, we study the effective material properties of the lattice unit cell under elasto-plastic deformations and investigate the sensitivity of the effective Young's modulus to the imperfections.
	\end{abstract}


	\maketitle
	
	{\bf Dedication:}
		The authors would like to express their deepest gratitude to Dr. J. Tinsley Oden.  We admire him as a great scientist with unique and exceptional contributions in many areas,  as a kind and gentle person, as a dedicated teacher who has influenced many of us in our career and scientific life, and as a person to whom we are deeply attached and
		thankful.
		Without him, the paper would never have been started and would never have been completed.
		Throughout the project, he was a reliable source of inspiration, advice and encouragement. We want to dedicate this paper to him. 
	

	\section{Introduction}\label{sec:Intro}

In structural design, the shape and architecture of manufactured materials are usually optimized assuming perfect, defect-free geometries.
However, the as-manufactured material may differ from the as-designed one, in particular, owing to defects and imperfections induced by the manufacturing process.
Typical examples are the structures manufactured using 3D-printing technologies.
Moreover, the process-induced imperfections may significantly affect the effective properties of the manufactured material with respect to the as-designed structure~\cite{liu2017elastic,pasini2019imperfect,snow2020review,gavazzoni2022cyclic,korshunova2021image}.
In this framework, more robust material design requires to take the imperfections into account during topology optimization.
A general pipeline for such imperfections-aware topology optimization was proposed in~\cite{moussa2021topology}.
Since the imperfections are random in nature, the characterization of the properties of manufactured materials is performed using statistical methods.
Therefore, given that the number of real as-manufactured samples is limited, a common strategy is to develop a so-called \textit{digital twin} -- a mathematical model of a surrogate material given by a random field reproducing the topological shape and imitating statistical properties of the target real-world material.
The philosophy is to introduce a flexible class of surrogate materials characterized by several design parameters that need to be tuned.
Then, a numerical model can be employed using methods of \textit{uncertainty quantification} (UQ), wherein the statistics of physical models with uncertainties is accounted for, and in which random fields typically enter as simulation inputs.

The use of modern machine learning (ML) methods has become a popular and powerful approach for the construction of surrogate models; see, e.g., \cite{bessa2017framework,gayon2020pores,garland2020deep,debroy2021metallurgy}.
However, ML is often used as a closed black box tool, in which the created model may depend on a large number of non-interpretable parameters, and requires a large amount of training data.
In order to simplify such approaches, we focus on cases in which the topological shape of the structure is known a priori, and the uncertainties include only the process-induced imperfections.
In this situation, only a few geometrically interpretable parameters are sufficient to describe the model.

In this work, we represent the material as a random phase field.
Considering a class of surrogate material models characterized by several \textit{design} parameters, we look for a projection of the target material onto this class by minimizing the distance between vectors of statistical descriptors associated with surrogate samples and the data, respectively.

We discuss a unified form of surrogate models for two-phase heterogeneous materials which present deterministic topological shapes but are subject to uncertain imperfections.
We formulate the surrogate phase field via the level-set method; see, e.g, \cite{allaire2002level,allaire2004structural,wang2004color,allaire2014multi,nika2019design,korshunova2021uncertainty}.
The introduced level-set function (called the intensity) is a random field given by a combination of the known topological structure and a random perturbation field representing the imperfections.
Several given examples reproducing particular two-phase structure types such as pores, beams, cracks, lattices, etc., illustrate the flexibility of the proposed surrogate model class.

We then apply the proposed model for a 3D-printed octet-truss lattice, which is a popular architecture in additive manufacturing~\cite{tancogne2018elastically,qi2019mechanical,balit2021crushing}.
First, we calibrate the model such that it reproduces statistical properties of several manufactured material samples obtained from X-ray computed tomography (CT) measurements.
To this end, we formulate and solve a stochastic optimization problem to identify the model design parameters minimizing a specific misfit measure between the synthetic samples and the data.
In order to avoid oversampling, a progressive batching strategy is employed, when an appropriate number of samples is estimated at each iteration and is adaptively updated using a specific test; see~\cite{byrd2012sample,bollapragada2019exact,roosta2019sub,xie2020constrained,beiser2020adaptive}.
Such efficient stochastic programming methods are important for large-scale decision-making problems, especially in engineering design, where oversampling is significantly costly.
The implementation also benefits from algorithmic differentiation for learning the model parameters.
Once the surrogate model is calibrated, we use it to generate synthetic samples and to quantify the uncertainties in the elasto-plastic response of the lattice material.
In particular, we estimate the expected values and the standard deviation of the effective tangent modulus for the lattice unit cell in different loading directions.
Besides, perturbing the model parameters, we also study the effects of the imperfections on the effective Young's modulus.


The structure of the paper is as follows.
In~\Cref{sec:Preliminaries}, we formulate the general problem and introduce notation.
In~\Cref{sec:Surrogate}, we introduce the surrogate material model and discuss several examples of the structure.
In~\Cref{sec:Application}, we focus on a particular example, the octet-truss lattice, to demonstrate the process of calibration of the surrogate model and its application.
First, we optimize the model parameters by fitting the CT data;
the technical details of the optimization algorithm are given in~\ref{sec:BFGS}.
Then, we use the calibrated model to generate synthetic samples and to quantify the uncertainties in the elasto-plastic response.
In addition, we briefly discuss possible generalizations of the model in~\Cref{sec:Extensions}.
Conclusion is given in~\Cref{sec:Conclusion}.

	\section{Preliminaries}\label{sec:Preliminaries}

Let $\dom=[0,1]^3$ be the $3$-dimensional unit cube with periodic boundaries associated with a \textit{representative volume element} (RVE) of the architected material.
We consider a two-phase (binary) material, defined by the phase field
\begin{equation}\label{eq:two-phase}
\Phase(\x) = 
\left\{
\begin{array}{ll}
0, &\quad\text{if } \x\in\Domain_1, \\
1, &\quad\text{if } \x\in\Domain_2,
\end{array}
\right.
\end{equation}
where the domain $\Domain_1$ corresponds to the phase one and $\Domain_2 = \Domain\setminus\Domain_1$ to the phase two.
Owing to uncertainties in the manufacturing process, the phase field~$\Phase$ is a random field.
That is, for each $\x\in\Omega$, the value $\Phase(\x)$ is a random value correlated with all other points in~$\Omega$.
Thus, a digital 3D image representation of the phase field~$\Phase$ is given by a multivariate random vector.

We shall consider a series of particular realizations of the material distribution expressed by the phase field $\Phase(\x; \omega)$, where $\omega$ is a sample indicator.
For simplicity, we will also write $\Phase(\omega):=\Phase(\cdot; \omega)$.
In particular, in our numerical experiments, we associate~$\omega$ with the seed of a pseudo-random number generator.  
We also denote the mean value for any given functional~$f$ by $\E_{\omega}[f(\Phase(\omega))]$.

Let us further denote by~$\TargMat$ the phase field characterizing the real manufactured material, which we will call the \textit{target material}.
In practice, this distribution is unknown, and only a small number of samples $X_i=\TargMat(\omega_i)$, $i=1\ldots N_{data}$, is accessible (e.g., as CT scans).
Our purpose is to construct a surrogate mathematical model (a so-called \textit{digital twin}) which allows the generation of synthetic samples with small computational cost and providing statistical properties similar to the target material~$\TargMat$.

\begin{figure}[ht!]
	\centering\noindent
	\includegraphics[width=\textwidth]{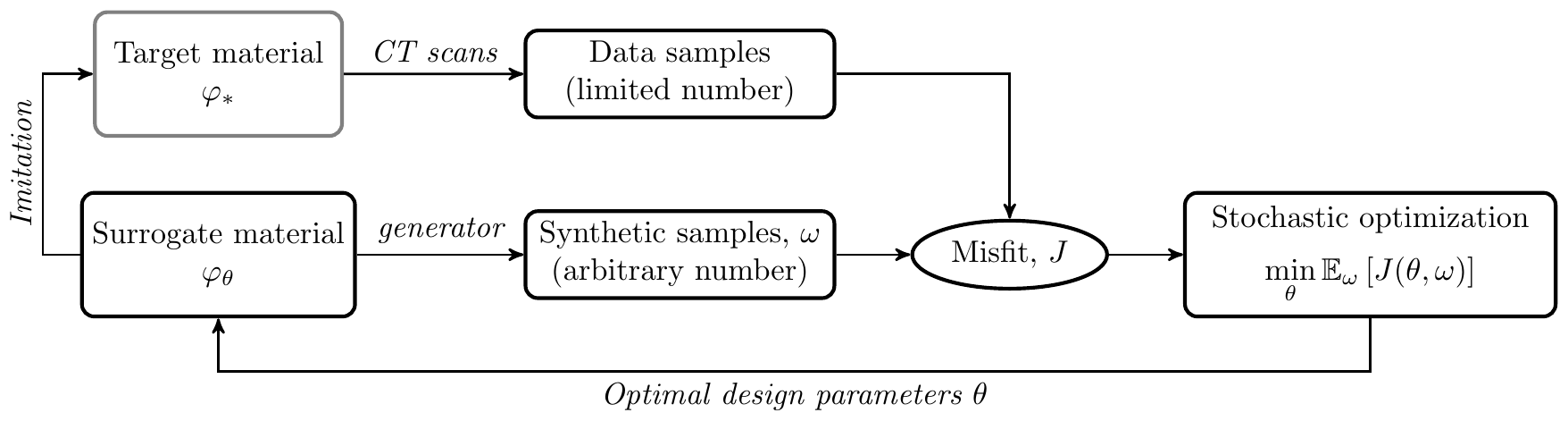}
	\caption{
		Visual scheme of the model calibration process.
		The target material distribution $\TargMat$ is unknown, however, a limited number of its samples (data) is given, e.g., as CT scans. In order to construct a surrogate which imitates the statistical properties of the target, we calibrate the design parameters $\theta$ of the surrogate model~$\Phase_{\theta}$ minimizing the expectation of the misfit $\loss$ between the synthetic samples (indexed by $\omega$) and the data.
	}
	\label{fig:VisualScheme}
\end{figure}

Let us define a \textit{surrogate material model} $\mathfrak{M}$ as a parametrized family of random phase fields $\Phase_{\theta} =\Phase(\theta)$, where $\theta$ is the vector of \textit{design parameters}.
Then, our digital twin is understood as a projection of $\TargMat$ onto the set~$\mathfrak{M}$.
That is, we want to find the \textit{optimal} design parameters minimizing some distance measure between the distributions $\Phase_\theta$ and $\TargMat$, formulated in terms of specific statistical descriptors.
However, given only a set of samples of $\TargMat$, we can minimize only the distance between the random field $\Phase(\theta)$ and the data set $\vec{X}=\{X_1, \ldots\}$ of available realizations.
In particular, we consider an objective function $\bar{\loss}(\theta) := \E_{\omega}[\loss(\Phase(\theta,\omega), \vec{X}) ]$, where $\loss(\Phase(\theta,\omega), \vec{X})$ denotes the distance from a particular surrogate sample~$\omega$ to the data set~$\vec{X}$.
A choice of such a distance measure will be further discussed in detail in~\Cref{sec:Calibration}, where we formulate the corresponding stochastic optimization problem.
A visual diagram of the calibration process of the surrogate material model is given in~\Cref{fig:VisualScheme}.

In this way, extracting statistical information from the morphology of the given samples of the architected material, we construct a random field, which captures the principal geometric features and imitates the statistical properties of the target material samples.
Then, employing a fast sampling procedure, we have access to an arbitrary number of synthetic samples that can be used for computational purposes, in particular, in Monte Carlo methods, for uncertainty quantification of the material response, quantities of interest, etc..

	\section{Surrogate material model}\label{sec:Model}

\label{sec:Surrogate}

A common strategy to define a synthetic phase field is the \textit{level-set method} (gray-scale thresholding), where the phases are defined as level-sets of some continuous \textit{intensity field} (gray-scale); see~\cite{allaire2002level,allaire2004structural,wang2004color,allaire2014multi,nika2019design,korshunova2021uncertainty}.
That is, we define the phase field as
\begin{equation}\label{eq:levelcut}
\Phase(\x; \theta, \omega) = 
\left\{
\begin{array}{ll}
0, &\quad\text{if } \Int(\x; \theta, \omega) < 0, \\
1, &\quad\text{otherwise},
\end{array}
\right.
\end{equation}
$\x\in\dom$, where the random intensity field~$\Int$ is parametrized with $\theta$, and $\omega$ is a sample indicator.

Our purpose is to construct the field which combines both a deterministic  topological shape and uncertain imperfections.
An example of such combination can be found, for example, in~\cite{korshunova2021uncertainty}, where the intensity is considered as a Gaussian random field with the mean non-constant in~$\Omega$.
In the current work, we consider a generalized \textit{hybrid} model, where the intensity field is given by a linear combination of two intensities:
\begin{equation}\label{eq:HybridIntensity}
\Int(\x;\theta,\omega) = (1-\alpha)\cdot\Int_1(\x;\theta,\omega) + \alpha\cdot\Int_2(\x;\theta,\omega),
\qquad\alpha\in[0,1].
\end{equation}
Here, $\Int_1$ defines the topological support, and $\Int_2$ is a random perturbation representing the imperfections.
Note that the topological support $\Int_1$ can also exhibit uncertainties and that $\Int_1$ and $\Int_2$ are statistically independent.
The parameter~$\alpha\in[0;\,1]$ defines the perturbation level, controlling the contribution of each term.
In particular, $\alpha = 0$ corresponds to a structure without imperfections, and $\alpha = 1$ to an unstructured statistically homogeneous random media.



\subsection{Perturbation field. Gaussian model}
\label{sec:GaussianModel}

We define the intensity perturbation $\Int_2$ as a centered Gaussian random field with a given covariance function $\cov(\x,\y):=\E[\Int_2(\x)\,\Int_2(\y)]$. 
As $\Int_2$ is already scaled with $\alpha$ in~\eqref{eq:HybridIntensity}, we can assume, without loss of generality, that it has unit marginal variance $\cov(\x,\x)=1$ for all $\x$.
Moreover, we assume that the random field $\Int_2$ is \textit{statistically homogeneous} (stationary); that is, the covariance function is of the form ~$\cov(\x,\y)=\cov(\x-\y)$.

A stationary Gaussian random field $\Int_2$ can be formally written as a convolution; see~\cite{oliver1995moving}:
\begin{equation}\label{eq:GaussianModel}
\Int_2(\x) = \cov_{\frac{1}{2}}\conv\eta(\x),
\end{equation}
$\x\in\dom$, where $\conv$ denotes the convolution product in $\R^3$, and $\eta$ denotes the white Gaussian noise in $\R^3$ \cite{hida2013white,kuo2018white}.
The convolution kernel $\cov_{\frac{1}{2}}$ is the \textit{square root} of the the covariance operator in the sense that $\cov_{\frac{1}{2}}\conv\cov_{\frac{1}{2}} = \cov$.

The covariance function defines how the values~$\Int_2(\x)$ are correlated at distinct points~$\x$.
We will consider herein a stationary covariance function of the form
\begin{equation}\label{eq:MaternCovariance}
\cov(\vct{x}-\vct{y}) =  \Matern{\nu}\left(\frac{\sqrt{2\nu}}{\corrlen}\,\norm{\x-\y}\right),
\end{equation}
where $\norm{\cdot}$ stands for the Euclidean norm, and $\Matern{\nu}(x)$ is the normalized Mat\'ern kernel~\cite{matern1986spatial,stein2012interpolation}:
\begin{equation}\label{eq:MaternFunction}
\Matern{\nu}(x) = \frac{x^\nu \,\BesselK{\nu}(x)}{2^{\nu-1}\Gamma(\nu)}.
\end{equation}
Here, $\Gamma(x)$ and $\BesselK{\nu}(x)$ denote the Euler Gamma function and the modified Bessel function of the second kind~\cite{abramowitz1965handbook,watson1995treatise}, respectively.
The covariance function~\eqref{eq:MaternCovariance} is parametrized using two parameters: regularity $\nu>0$ and correlation length $\corrlen>0$, which thus are part of the design parameters vector of the surrogate model.
The Mat\'ern covariance is widely used in statistics~\cite{stein2012interpolation,gneiting2012studies}, geostatistics~\cite{minasny2005matern} and machine learning~\cite{williams2006gaussian}.
It represents a large class of covariance kernels, varying from the exponential ($\nu=0.5$) to the squared-exponential ($\nu\to\infty$).
In particular, the parameter~$\nu$ controls the regularity of the function at zero: from weak singularity to infinite smoothness.
The regularity of the kernel at the origin is directly related to the smoothness of the level-set of the corresponding Gaussian field.
However, because of the voxelized nature of the digital samples, the accuracy of the parameter~$\nu$ is limited by the resolution of the image.
Moreover, as shown in~\cite{korshunova2021uncertainty}, when fitting the data to a Mat\'ern model, there arises a manifold of \textit{near-optimal} pairs $(\nu,\corrlen)$; see also~\cite{de2000bayesian}.
This makes the identification of a unique regularity $\nu$ and correlation length~$\corrlen$ problematic.

The Gaussian model admits an efficient sampling procedure by use of the \textit{spectral simulation method} employing the \textit{Fast Fourier transform} (FFT) algorithm on a regular grid; see~\cite{le2000fft,abrahamsen2018simulation} and the references therein.
Indeed, by the convolution theorem, the convolution product in~\eqref{eq:GaussianModel} reduces under Fourier transformation to a point-wise product of two fields:
\begin{equation}\label{eq:GaussianModel_Fourier}
\hat{\Int}(\k) = \hat{\cov}_{\frac{1}{2}}(\k)\,\hat{\eta}(\k),
\end{equation}
where $\k$ corresponds to the wavevector, $\hat{\Int}$ and $\hat{\cov}$ denote Fourier transforms of $u$ and $\cov$, respectively, and the Fourier transform $\hat{\eta}$ of the Gaussian noise is a complex Gaussian noise.
The use of the FFT algorithm makes it possible to reduce the complexity of the generation of one sample from $\O(N^2)$ for the direct convolution approximation to $\O(N\log N)$, where $N$ corresponds to the number of voxels.
The Fourier transform in $\R^d$ of the Mat\'ern kernel~\eqref{eq:MaternCovariance} is given explicitly by
\begin{equation}\label{eq:MaternFourier}
\hat{\cov}(\k) = \abs{\hat{\cov}_{\frac{1}{2}}(\k)}^2
=\left(\frac{2\pi\corrlen^2}{\nu}\right)^{\frac{d}{2}} \;\frac{\Gamma(\nu+\frac{d}{2})}{\Gamma(\nu)}
\; \left(1 + \frac{\corrlen^2}{2\nu}\,\norm{\k}^2\right)^{-(\nu + \frac{d}{2})},
\end{equation}
see \cite[Vol.II, section~$8.13$, formula~$(3)$]{bateman1954tables}.

\begin{remark}
	A remarkable feature of a Mat\'ern kernel is that it is associated with the Green's function for the fractional differential equation of the form (up to a factor)
	\begin{equation}\label{key}
		 \left(1 - \frac{\corrlen^2}{2\nu}\,\Delta\right)^{(\nu + \frac{d}{2})/2}\Int_2 = \eta,
	\end{equation}
	where $\Delta$ is the Laplace operator; see~\cite{whittle1954stationary,whittle1963stochastic}.
	Thus, imposing boundary conditions and using variable coefficients, $u_2$ can be extended to a more general non-stationary random field, see, e.g., \cite{roininen2014whittle,keith2021fractional,lindgren2021spde}.
	Though it is appropriate to mention these extension options, such models are beyond the scope of the current paper.
\end{remark}

\begin{figure}[!p]
	\centering
	\newcommand{\size}{0.3\textwidth}
	
	\begin{subfigure}{\size}
		\centering
		\includegraphics[width=\textwidth]{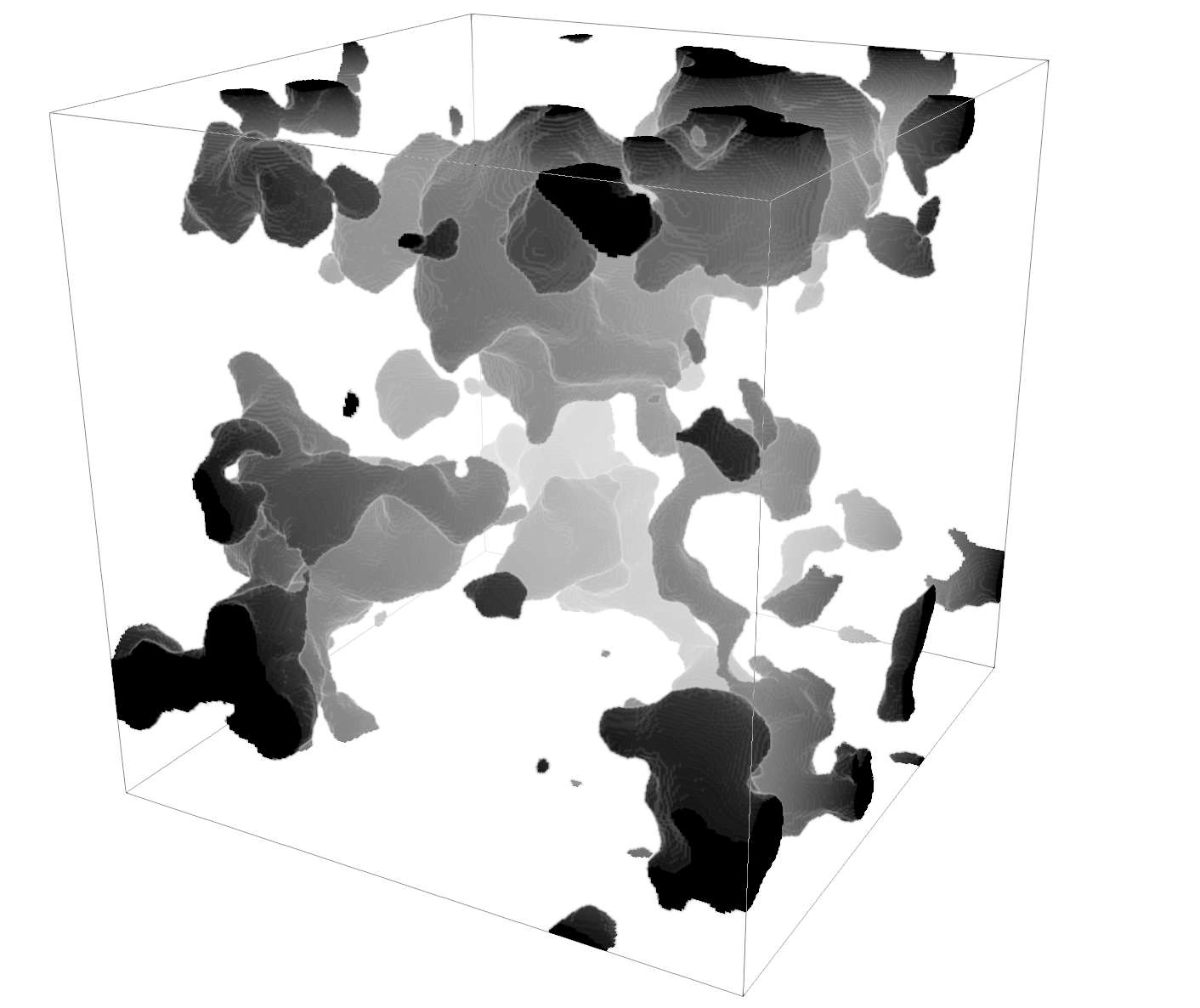}
		\caption{\scriptsize $\vfm = 0.1$, $\nu = 2$, $\corrlen=0.1$}\label{sbfig:ex1}
	\end{subfigure}
	\hfill
	\begin{subfigure}{\size}
		\centering
		\includegraphics[width=\textwidth]{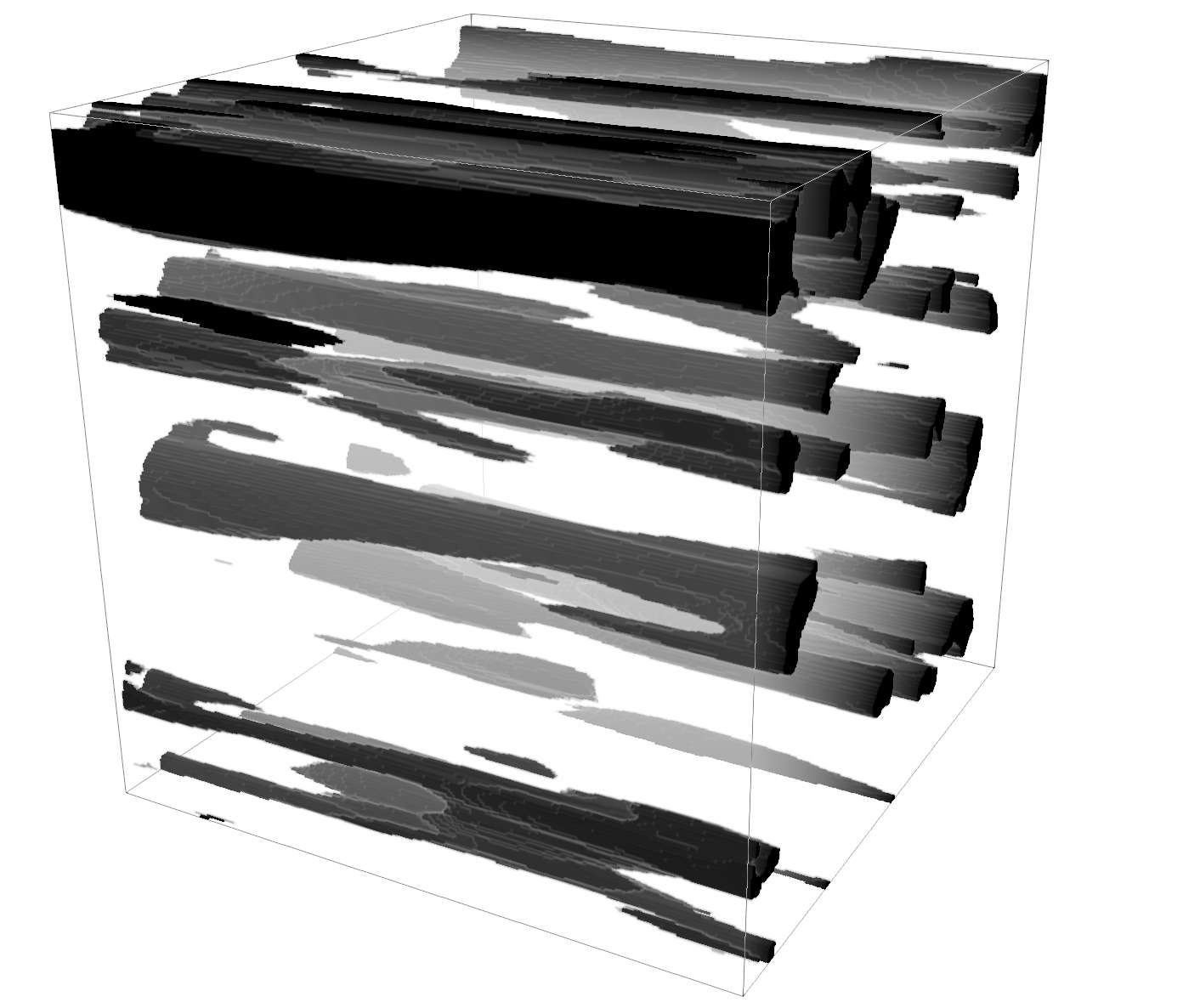}
		\caption{\scriptsize $\vfm = 0.1$, $\nu = 2$,  $\corrlen=0.05$, \\ lengthscales $[10, 1, 1]$}\label{sbfig:ex2}
	\end{subfigure}
	\hfill
	\begin{subfigure}{\size}
		\centering
		\includegraphics[width=\textwidth]{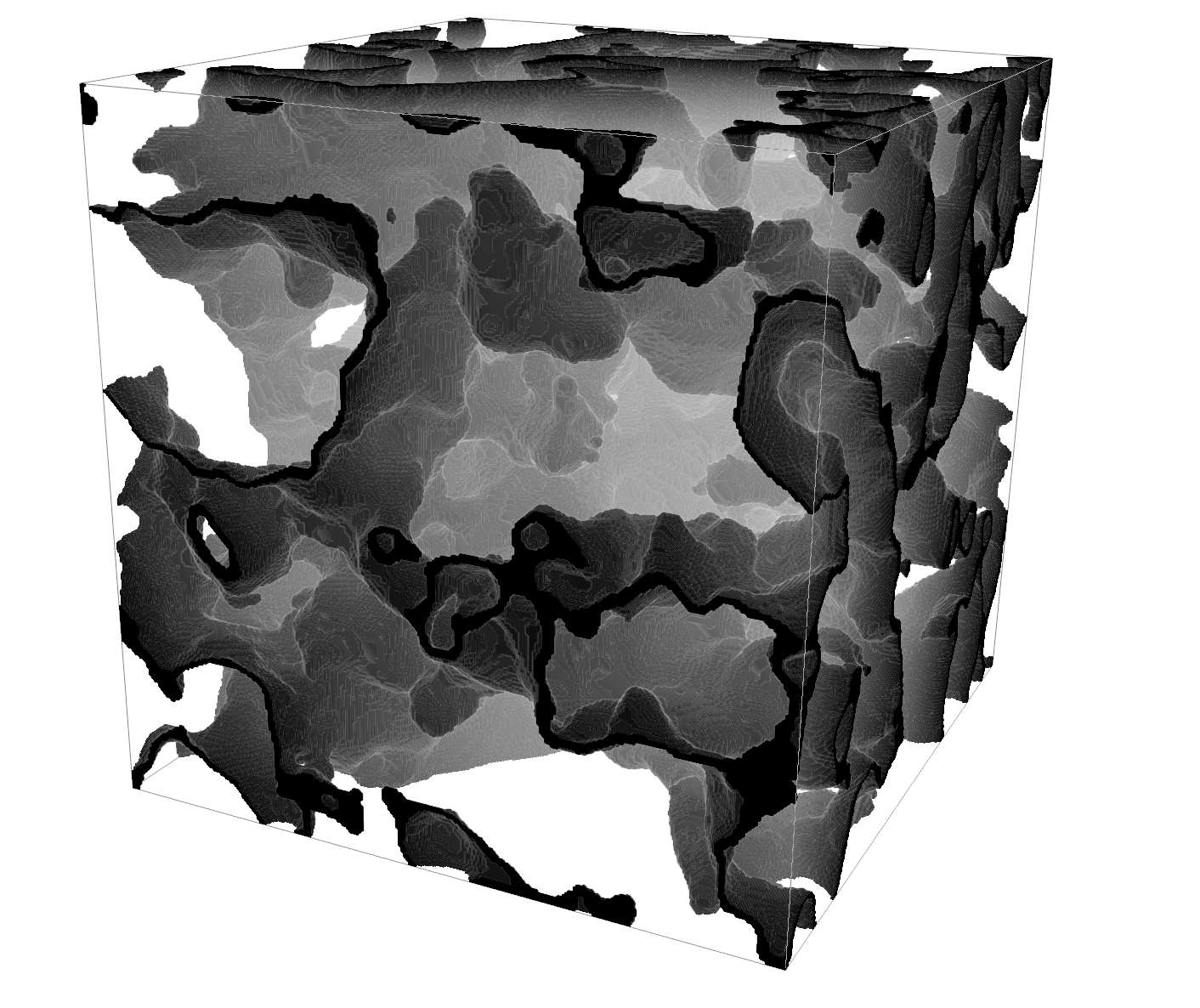}
		\caption{\scriptsize $\vfm = 0.1$, $\nu = 2$, $\corrlen=0.1$, \\ folded Gaussian intensity}\label{sbfig:ex3}
	\end{subfigure}
	
	\caption{Examples of Gaussian level-set model (\Cref{ex:Gaussian}):
		 (\subref{sbfig:ex1}) isotropic two-phase media; (\subref{sbfig:ex2}) anisotropic metric; (\subref{sbfig:ex3}) considering folded Gaussian distribution makes possible the distinction of connectivity of the phases.
	}		
	\label{fig:Gaussian}
\end{figure} 
\begin{figure}[!p]
	\centering	
	\newcommand{\size}{0.3\textwidth}
	\begin{subfigure}{\size}
		\centering
		\includegraphics[width=\textwidth]{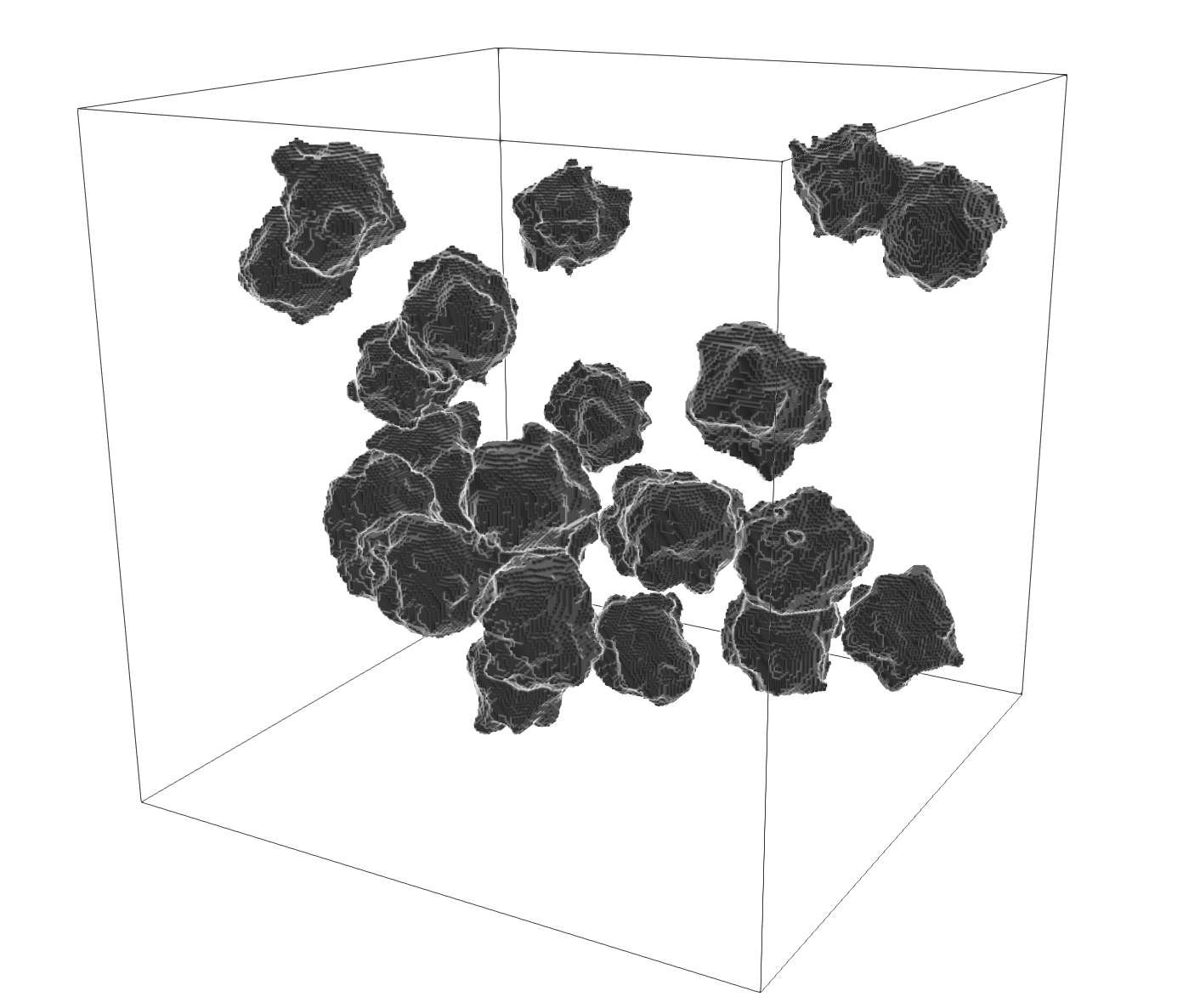}
		\caption{\scriptsize isotropic with $\tau=0.08$}\label{sbfig:spheres}
	\end{subfigure}
	\hfill
	\begin{subfigure}{\size}
		\centering
		\includegraphics[width=\textwidth]{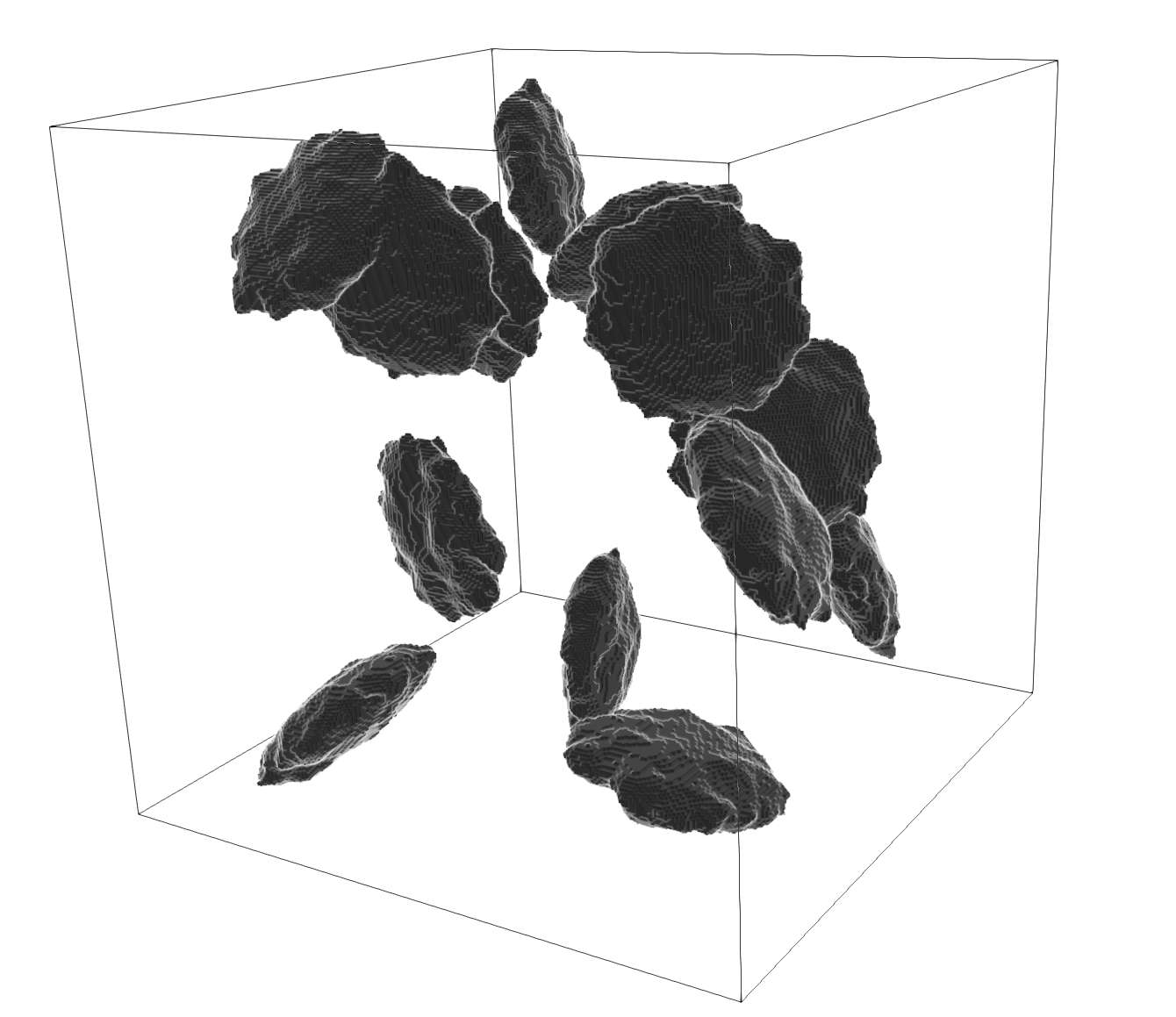}
		\caption{\scriptsize lengthscales $[3, 1, 3]$ with $\tau=0.05$}\label{sbfig:disks}
	\end{subfigure}	
	\hfill
	\begin{subfigure}{\size}
		\centering
		\includegraphics[width=\textwidth]{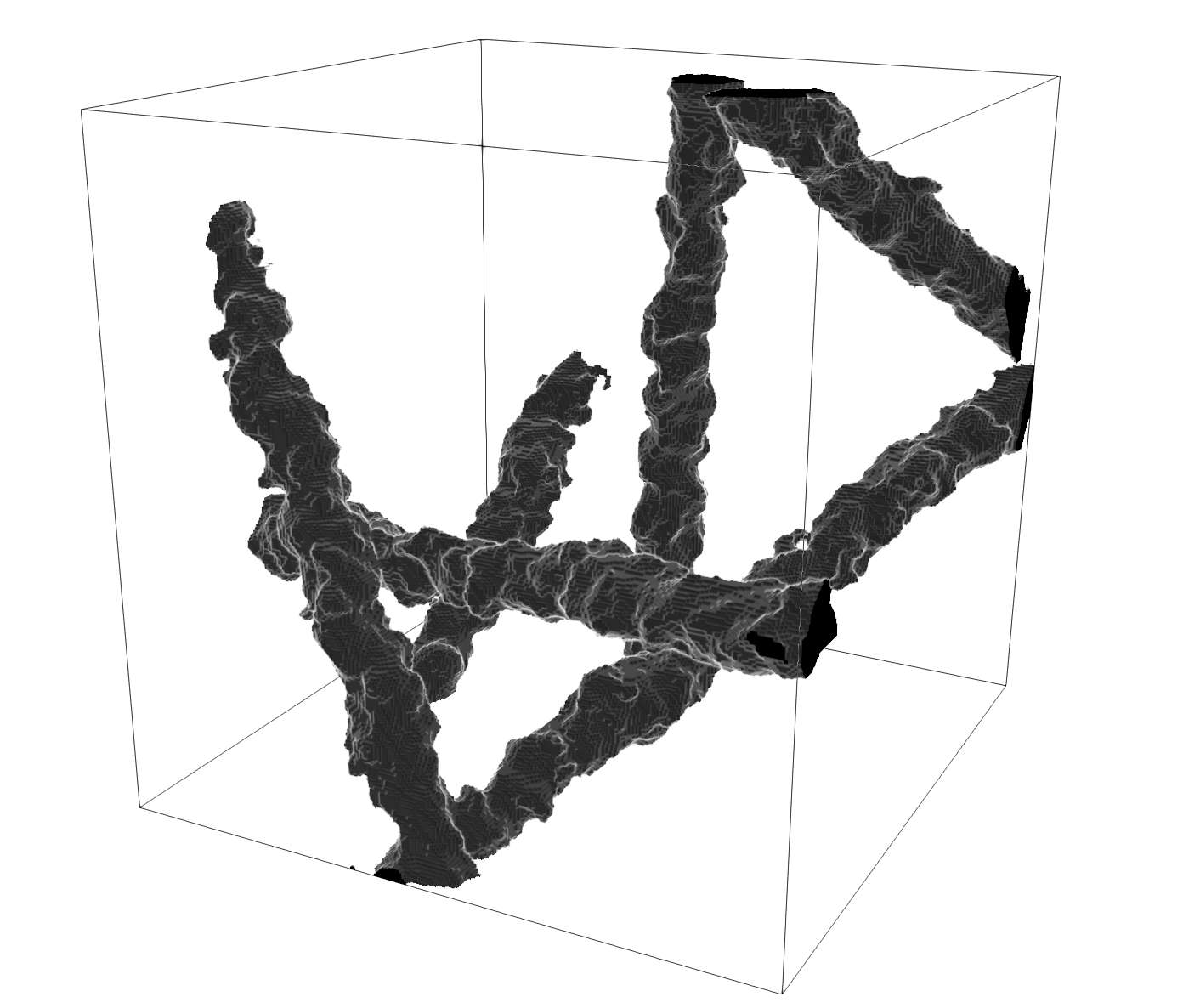}
		\caption{\scriptsize lengthscales $[200, 1, 1]$ with $\tau=0.05$}\label{sbfig:beams}
	\end{subfigure}	
	
	\caption{ Examples of the hybrid model with a collection of random points as support structure (see~\Cref{ex:Particles}):		
				(\subref{sbfig:spheres}) isotropic metric (spherical particles);		
				(\subref{sbfig:disks}) anisotropic metric (disks);
				(\subref{sbfig:beams}) anisotropic metric (beams).
				The surface imperfections are of level $\alpha=0.01$ with covariance of regularity $\nu=2$ and correlation length~$\corrlen=0.03$.
	}
	\label{fig:particles}
\end{figure}
\begin{figure}[!p]
\centering
\newcommand{\size}{0.45\textwidth}
\newcommand{\sfsize}{0.65\textwidth}
\begin{subfigure}{\size}
	\centering
	\includegraphics[width=\sfsize]{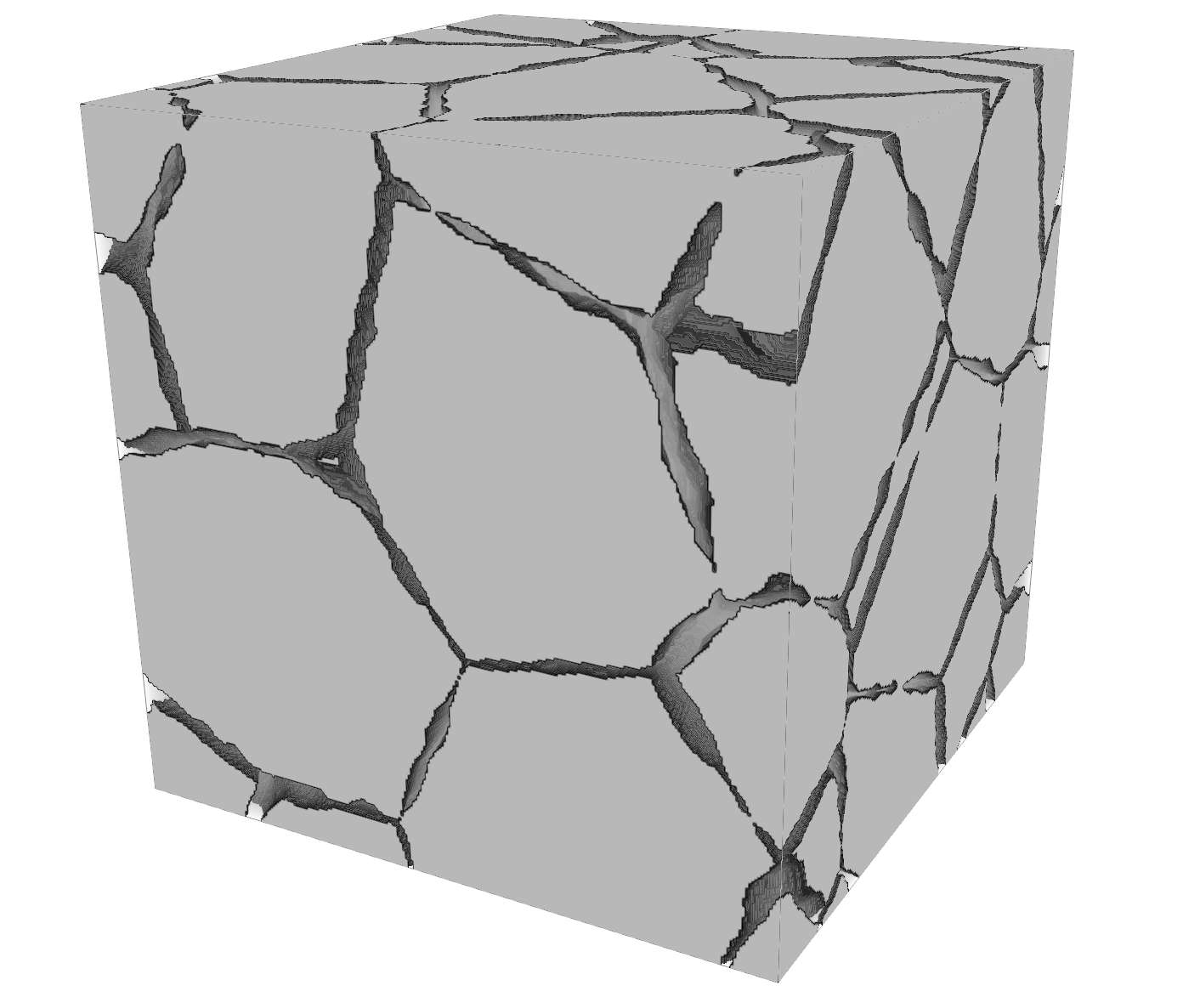}
	\caption{Synthetic cracks network (\Cref{ex:Cracks})}
	\label{sbfig:cracks}
\end{subfigure}
\hspace{5ex}
\begin{subfigure}{\size}
	\centering
	\includegraphics[width=\sfsize]{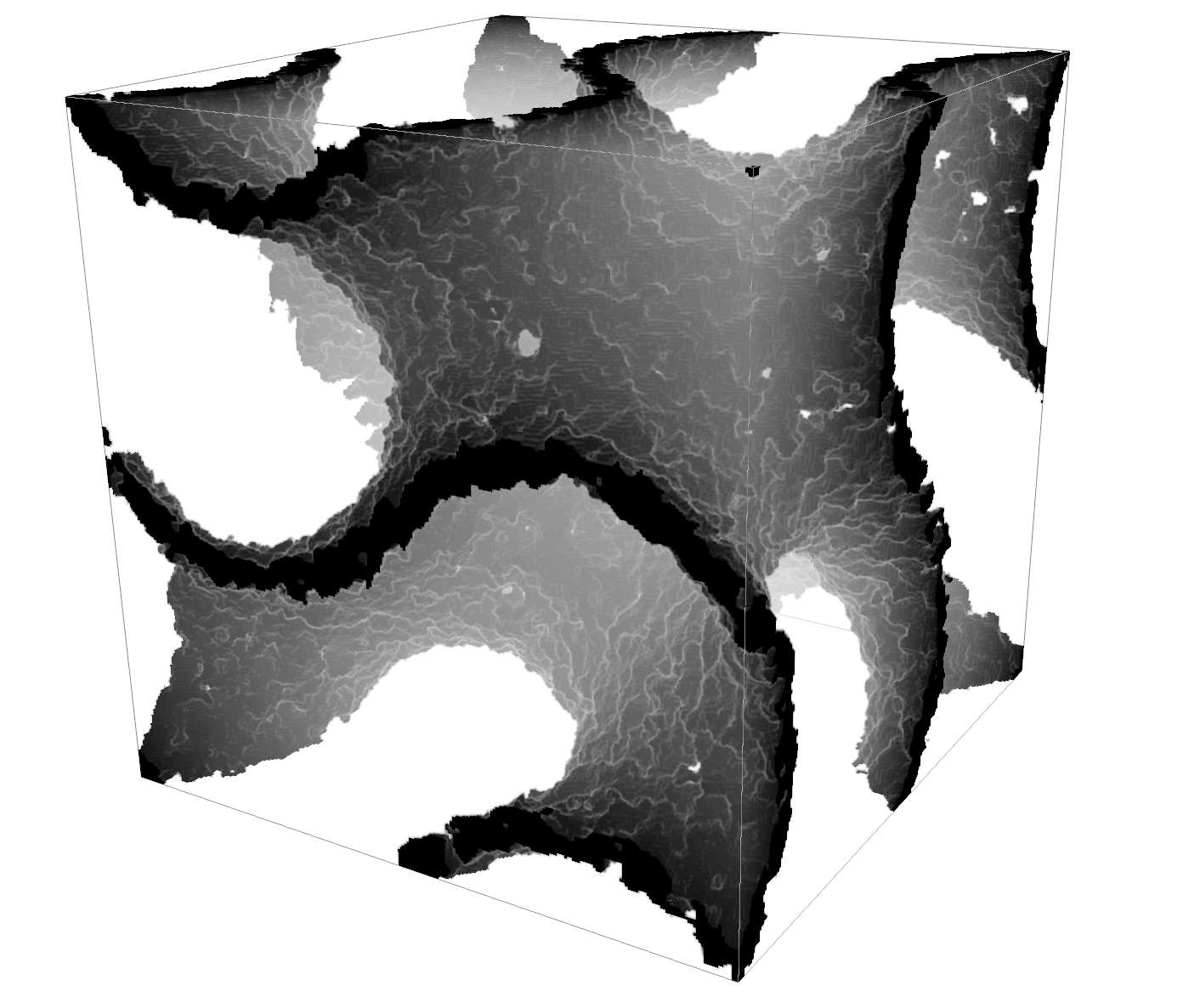}
	\caption{Imperfect gyroid structure (\Cref{ex:Gyroid})}
	\label{sbfig:gyroid}
\end{subfigure} 
\caption{(\subref{sbfig:cracks}) Example of a network of synthetic cracks (\Cref{ex:Cracks})
	with the mean opening $2\tau=0.02$ and imperfections of level $\alpha=0.005$ with covariance of regularity $\nu=2$ and  correlation length $\corrlen=0.03$.	
	(\subref{sbfig:gyroid}) Example of a synthetic gyroid structure (\Cref{ex:Gyroid}) subject to surface imperfections of level $\alpha=0.03$ with covariance of regularity $\nu=2$ and correlation length~$\corrlen=0.015$, $\tau=0.05$.
}
\end{figure}

\begin{example}[Gaussian level-set]\label{ex:Gaussian}
	
	Let $\alpha=0.5$, and let the structure term~$\Int_1$ in~\eqref{eq:HybridIntensity} be constant in~$\dom$, i.e., $\Int_1(\x)\equiv \tau\in\R$ for $\x\in\dom$.
	Then, the phase field~$\Phase$ is reduced to the threshold of a Gaussian random field with the mean~$\tau$ and covariance~$\cov(\x,\y)$, which corresponds to the common Gaussian level-set model; see, e.g.,~\cite{teubner1991level,lin2005properties,koutsourelakis2006simulation,ogorodnikov2018stochastic,zerhouni2021quantifying}.
	The resulting random field is statistically homogeneous and exhibits no particular topological shape.
	
	In~\Cref{fig:Gaussian}, the sample (\ref{sbfig:ex1}) presents a typical example of the Gaussian level-set model, while the samples (\ref{sbfig:ex2}) and (\ref{sbfig:ex3}) illustrate its modified variants.
	In particular, the sample (\ref{sbfig:ex2}) is obtained by introducing the anisotropic metric in~\eqref{eq:MaternCovariance}, i.e., considering the norm $\norm{\ten{A}_k(\omega)(\x-\y)}$ with the metric tensor~$\ten{A}_k$ defined by a definite positive $3\times3$ matrix.
	Its eigenvalues control the lengthscales in the directions defined by the corresponding eigenvectors, which leads to an anisotropic phase field.
	The sample (\ref{sbfig:ex3}) is obtained by replacing $u_2$ with $\abs{u_2}$  (folded Gaussian distribution), which allows the connectivity properties of the two phases to be distinguished.
	
	It is known that in the case of the Gaussian model, explicit closed-form formulas are available for statistical descriptors of the phase field~$\Phase$, such as the mean (expected volume fraction) $\bar{\Phase}$ and the two-point correlation function; see, e.g., 
	\cite{lantuejoul2001geostatistical,khristenko2019statistical}.
	In particular, the expected volume fraction $\bar{\Phase}$ is uniquely defined by the mean of the Gaussian intensity $\tau$.
	However, in the general case, explicit formulas for statistical moments are generally not available, and thus the statistics must be approximated numerically using, e.g., Monte Carlo methods.
	
	Though stationary random fields are easy to sample, they cannot reproduce the topological features of the statistically inhomogeneous media, such as randomly oriented particles or beams, cracks, imperfect lattice, etc.
	
\end{example}


\subsection{Topological support intensity}
\label{sec:particles}

A very simple surrogate for a porous media can be obtained as a collection of particles of ideal shape, e.g., ellipsoids~\cite{tarantino2019random,zerhouni2019numerically}.
Each such particle can be represented as level set of a cone in higher dimension~\cite{mantz2008utilizing}.
This inspires us to introduce the intensity field for an arbitrary geometrical structure in~\eqref{eq:HybridIntensity} as a generalized cone given by the formula
\begin{equation}\label{eq:StructureIntenscity}
	u_1(\x; \omega) = \tau - \dist(\x, \mathcal{G}(\omega)),
\end{equation}
where $\dist(\x, \mathcal{G})$ denotes the distance from the point $\x\in\Omega$ to a manifold~$\mathcal{G}$, which defines the geometrical structure; the parameter $\tau$ corresponds to the characteristic size of the structure and plays the role of the threshold in the level set model~\eqref{eq:levelcut}.
Thus, the phase field~\eqref{eq:levelcut} with intensity~\eqref{eq:StructureIntenscity} and $\alpha=0$, takes the value one in the points having a distance of at most $\tau$ from the manifold~$\mathcal{G}$.

\begin{example}[Collection of particles]\label{ex:Particles}
	An obvious example of structured material is a heterogeneous material with disjoint inclusions represented by collection of imperfect particles.
	Such structures can be constructed using the hybrid model~\eqref{eq:levelcut}-\eqref{eq:HybridIntensity} where the structure intensity~$\Int_1$ is given by the so-called \textit{maximum cones}~\cite{mantz2008utilizing} -- the maximum of the set of cones centered in randomly distributed points.
	Thus, it can be written in the form~\eqref{eq:StructureIntenscity} with the associated support structure given by a collection of randomly distributed points, particles centers, $\mathcal{G}(\omega) = \mathcal{G}_{\text{particles}}(\omega) = \{\vct{c}_k(\omega), \quad k=1,\ldots,N(\omega) \}$.
	Then, the distance to the manifold $\mathcal{G}_{\text{particles}}$ is given by the maximum of the distances to each of the centers:
	\begin{equation}\label{key}
	\dist(\x, \mathcal{G}_{\text{particles}}(\omega)) = \min\limits_{k\le N(\omega)}\norm{\ten{A}_k(\omega)(\x-\vec{c}_k(\omega))}_p.
	\end{equation}
	The particle shapes are defined by metric tensors~$\ten{A}_k(\omega)$ and the norm type $\norm{\cdot}_p$.
	In particular, the values $p=1$, $p=2$ or $p=\infty$ correspond to the diamond, ellipsoid or rectangular shapes, respectively.
	For each particle $k=1,\ldots,N$, the eigenvectors and eigenvalues of the metric tensor~$\tns{A}_k$ define the principal axes (particle orientation) and the corresponding lengthscales, respectively.
	
	\Cref{fig:particles} shows examples of imperfect inclusions with different eigenlengths.
	The particle orientation is random and obtained from three uniformly distributed Euler angles.
\end{example}


\begin{example}[Cracks]\label{ex:Cracks}
	A networks of cracks in the solids can also be represented in this framework.
	The structure support is the boundary of a Voronoi tessellation, $\mathcal{G}(\omega) = \mathcal{G}_{\text{Voronoi}}(\omega)$, associated with the set of random points $\mathcal{G}_{\text{particles}}(\omega) = \{\vct{c}_k(\omega), \quad k=1,\ldots,N(\omega) \}$ from the previous example.
	In particular, we define the distance from a point~$\x\in\Omega$ to the manifold $\mathcal{G}_{\text{Voronoi}}$ by	
	\begin{equation}\label{key}
	\dist(\x, \mathcal{G}_{\text{Voronoi}}(\omega)) = \frac{1}{2}\,\left|\frac{a^2(\x)-b^2(\x)}{c(\x)}\right|,
	\end{equation}
	where $a$ and $b$ are the distances from $\x$ to the two closest centers $k_1$ and $k_2$, i.e., two minimum distances of $\{\rho(\x, \vec{c}_k)=\norm{\x-\vec{c}_k}, \quad k=1,\ldots,N\}$;
	and $c$ is the distance between $\vec{c}_{k_1}$ and $\vec{c}_{k_2}$.	
	An example of a surrogate solid with cracks is given in~\Cref{sbfig:cracks}.
	Note that in this case, the parameter $\tau$ in~\eqref{eq:StructureIntenscity} corresponds to half of the mean crack opening size.
\end{example}


\begin{example}[Gyroid]\label{ex:Gyroid}
	As another example, we consider a \textit{gyroid} lattice~\cite{schoen1970infinite}, which is a popular architecture pattern in additive manufacturing.
	Although a gyroid is defined by the equations involving elliptical integrals~\cite{gandy2000exact}, its
	close approximation is given by the level surface of a simple trigonometric expression~\cite{wohlgemuth2001triply}.
	Thus, this level-set function naturally defines our distance from a point~$\x\in\Omega$ to the manifold $\mathcal{G}_{\text{gyroid}}$:	
	\begin{equation}\label{key}
	\dist(\x, \mathcal{G}_{\text{gyroid}})
	= [\sin(2\pi x)\cos(2\pi y)
	+\sin(2\pi y)\cos(2\pi z)
	+\sin(2\pi z)\cos(2\pi x)]^2,
	\end{equation}
	where $\x=(x,y,z)\in\Omega$.
	Note that without loss of generality, the square can be replaced with the absolute value.
	The parameter $\tau$ in~\eqref{eq:StructureIntenscity} controls the thickness of the structure.
	An example of a surrogate gyroid structure is given in~\Cref{sbfig:gyroid}.	
\end{example}


\begin{example}[Octet-truss lattice]\label{ex:lattice}
	\label{ex:octet_lattice}

\begin{figure}[t!]
	\centering
	\newcommand{\size}{0.48\textwidth}
	\newcommand{\sfsize}{0.22\textwidth}

	\begin{minipage}{0.45\textwidth}
		\centering
		\begin{subfigure}[b]{0.9\textwidth}
			\includegraphics[width=\textwidth,valign=c]{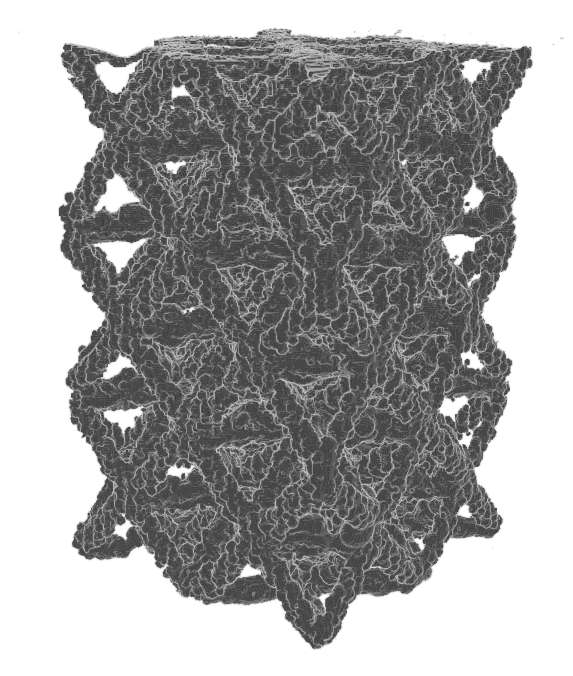}
			\vspace{1ex}
			\caption{Octet-truss lattice structure (CT scan)}
			\label{sbfig:lattice_CT}
		\end{subfigure}
	\end{minipage}	
	\begin{minipage}{0.45\textwidth}
		\centering
		\begin{subfigure}[t]{0.55\textwidth}
			\centering
			\includegraphics[width=\textwidth]{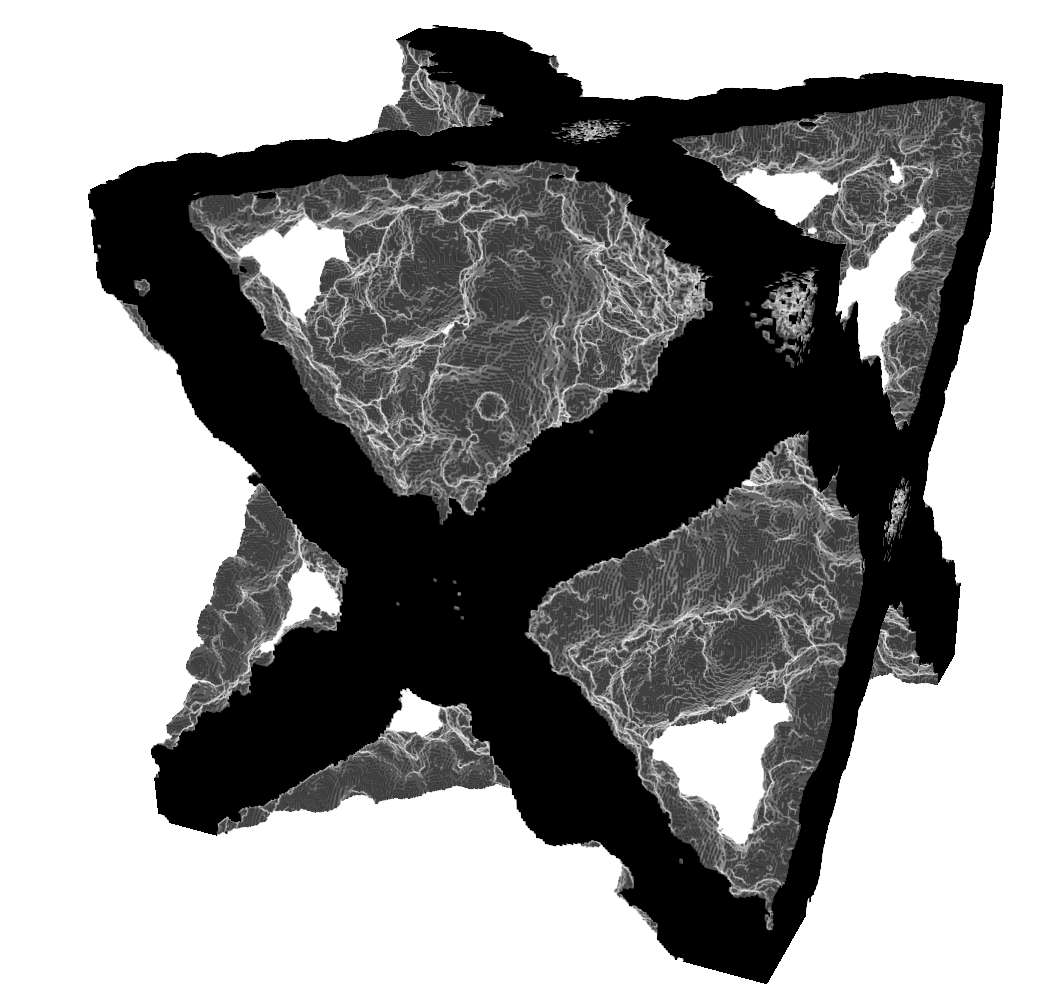}
			\caption{Unit cell (CT scan)}
			\label{sbfig:cell_original}
		\end{subfigure}
		
		\vspace{5ex}
		\begin{subfigure}[b]{0.45\textwidth}
			\centering
			\includegraphics[width=\textwidth,valign=c]{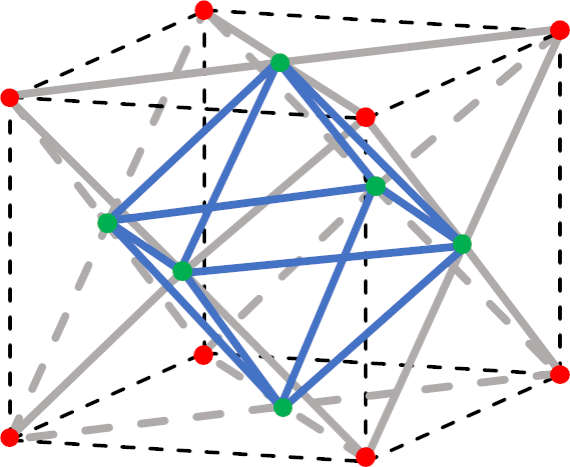}
			\caption{Cell connectivity}
			\label{sbfig:cell_scheme}
		\end{subfigure}
	\end{minipage}
	
	\caption{ Octet-truss lattice; see~\Cref{ex:lattice}. 
		(\subref{sbfig:lattice_CT}): Computed Tomography (CT) of a lattice structure,
		(\subref{sbfig:cell_original}):  CT of a unit cell,
		(\subref{sbfig:cell_scheme}): scheme of the truss connectivity in the unit cell.
	}
	\label{fig:octet_lattice}
\end{figure}

\begin{figure}[t!]
	\centering
	\newcommand{\size}{0.3\textwidth}
	\newcommand{\sfsize}{0.22\textwidth}
	
	\includegraphics[width=\size]{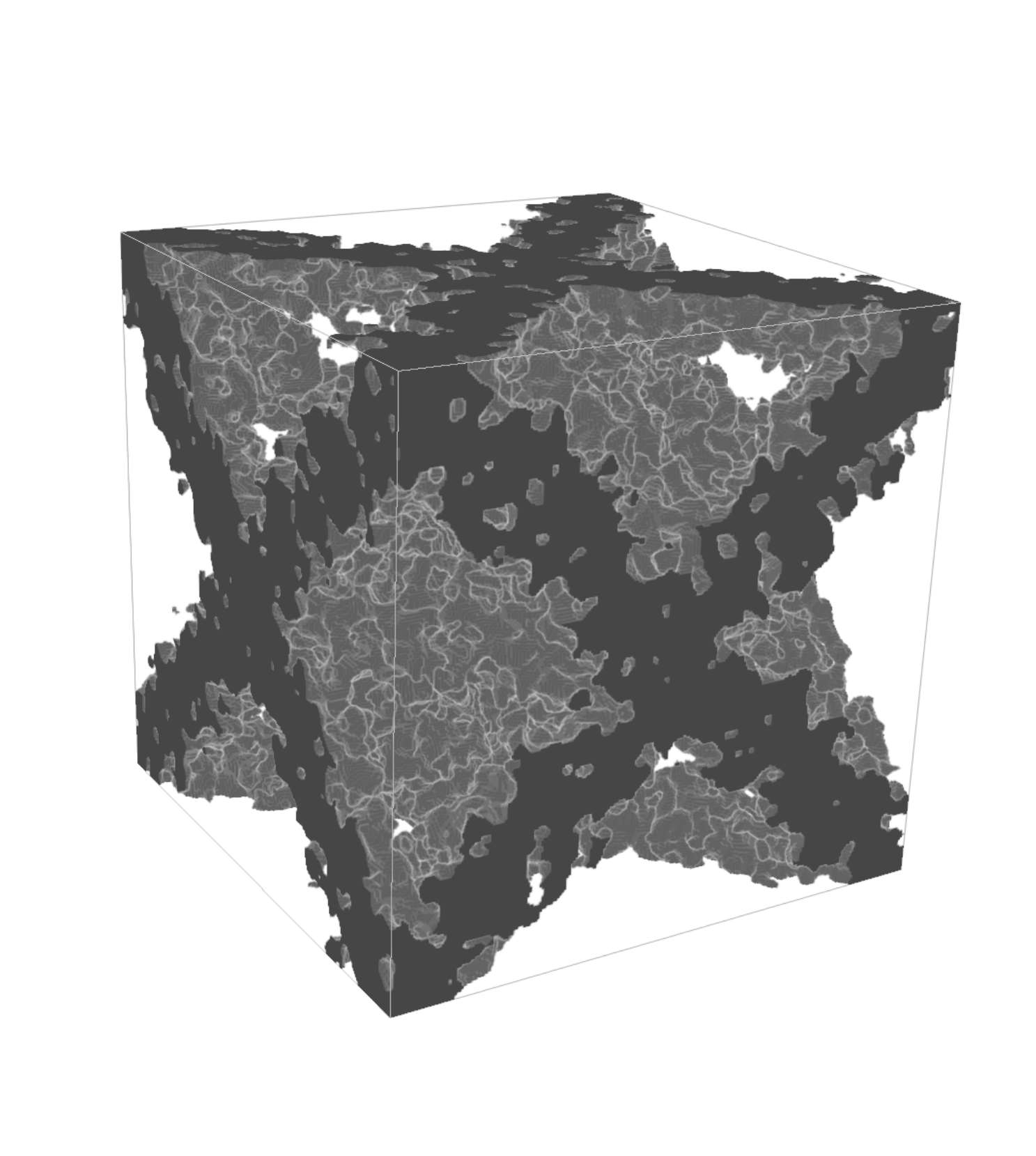}
	\hfill
	\includegraphics[width=\size]{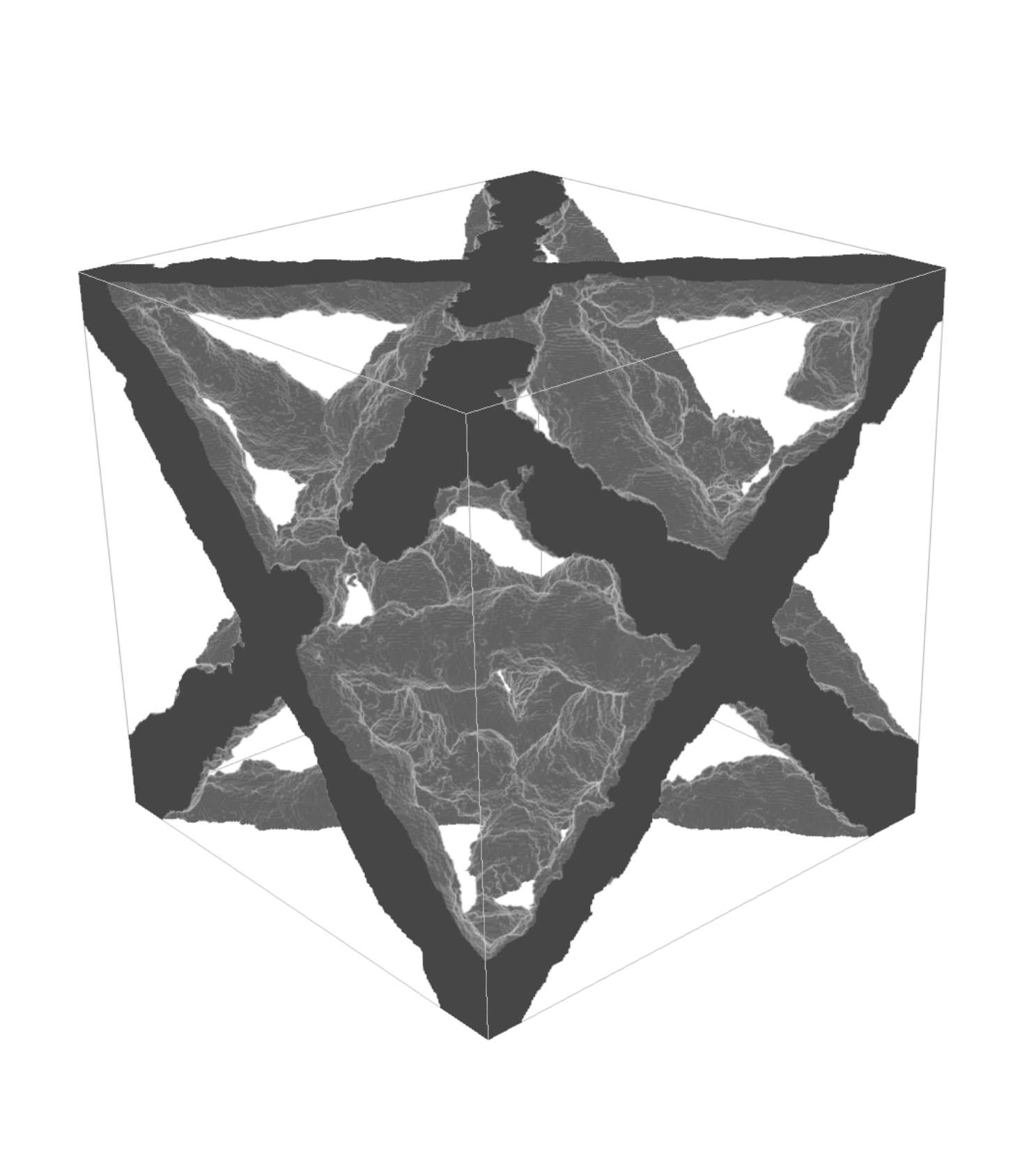}
	\hfill
	\includegraphics[width=\size]{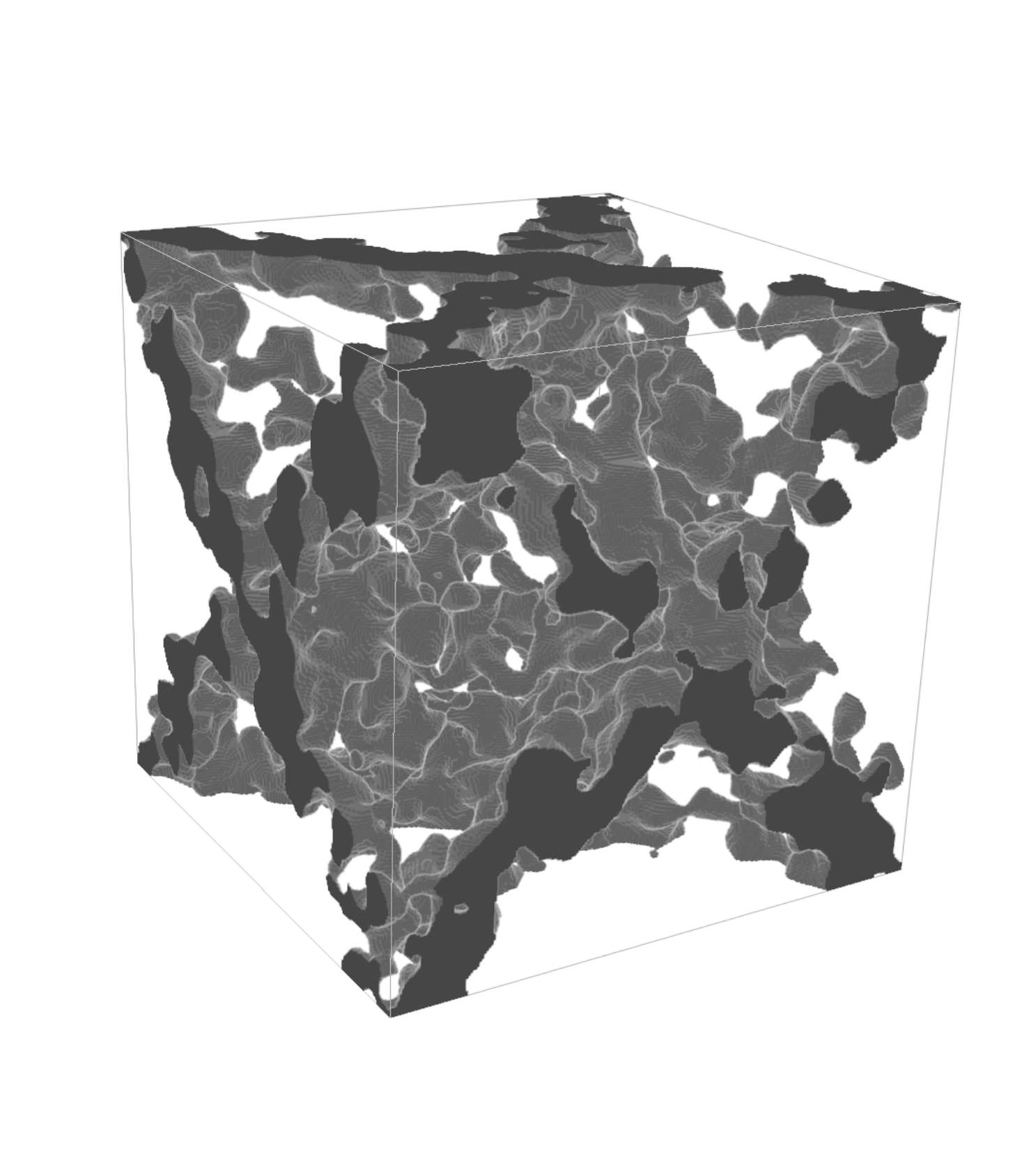}
	\caption{ Examples of the surrogate octet lattice cell with surface imperfections (\Cref{ex:lattice}) using different values for the imperfection level~$\alpha$ and the correlation length~$\corrlen$.
	}
	\label{fig:examples_octet_lattice}
\end{figure}

The octet-truss lattice structure (\Cref{fig:octet_lattice}) is another popular architecture pattern in additive manufacturing.
The connectivity of the struts in its unit cell is schematically depicted in~\Cref{sbfig:cell_scheme}.
Owing to the cell symmetry, it is sufficient to construct a quarter (subcell) of the cell.
The nodes of the subcell are given in local coordinates by
\begin{equation}\label{key}
	\{\vec{p}_k, \; k=1,\ldots,4\} = \{ (0,0,0), (0,1,1), (1,0,1), (1,1,0) \}.
\end{equation}
Then, the manifold $\mathcal{G}_{lattice}$ for the subcell is given by the set of the edges (struts):
\begin{equation}\label{key}
\{e_k=(\vec{e}_{k,1}, \vec{e}_{k,2}), \; k=1,\ldots,6\}
= \{ (\vec{p}_1, \vec{p}_2), (\vec{p}_1, \vec{p}_3), (\vec{p}_1, \vec{p}_4), (\vec{p}_2, \vec{p}_3), (\vec{p}_2, \vec{p}_4), (\vec{p}_3, \vec{p}_4)  \}.
\end{equation}
And, therefore, the distance to $\mathcal{G}_{lattice}$ is defined as
\begin{equation}\label{key}
	\dist(\x, \mathcal{G}_{\text{lattice}}) = \min\limits_{k\le 6}\dist(\x, e_k),
\end{equation}
where $\dist(\x, e_k)$ is the distance from $\x$ to the line $e_k$ which is given by the formula
\begin{equation}\label{key}
	\dist(\x, e_k) := \left[\frac{a^2 + b^2}{2} - \frac{c^2}{4} - \left(\frac{a^2-b^2}{2\,c}\right)^2\right]^{\frac{1}{2}},
\end{equation}
and
$a = \dist(\x, \vec{e}_{k,1})$, 
$b = \dist(\x, \vec{e}_{k,2})$,
$c = \dist(\vec{e}_{k,1}, \vec{e}_{k,2})$ are Euclidean distances between the points.

\Cref{fig:examples_octet_lattice} shows examples of the surrogate octet lattice cell with different values of the uncertainty level~$\alpha$ and the correlation length~$\corrlen$.
The average strut radius is defined by the parameter $\tau$ in~\eqref{eq:StructureIntenscity}.
Notice that with a high imperfection level $\alpha$, the struts may even lose their connectivity, which is an important case in risk-adverseness problems and in failure probability analysis.

\end{example}

	\section{Application example: Octet-truss lattice}\label{sec:Numerics}

\label{sec:Application}

In this section, we demonstrate the calibration of a surrogate material model.
To do this, we solve an optimization problem for finding the optimal design parameters of the model, minimizing the misfit between the synthetic model and the target material using a set of statistical/geometrical descriptors.
Then, we proceed with an application of the surrogate material in stochastic homogenization.
We use the calibrated model to generate an arbitrary number of synthetic samples within a Monte Carlo simulation for uncertainty quantification of the material effective properties.

\begin{figure}[h!]
	\newcommand{\size}{0.48\textwidth}	
	\centering
	
	\begin{subfigure}{\size}
		\centering
		\includegraphics[width=\textwidth]{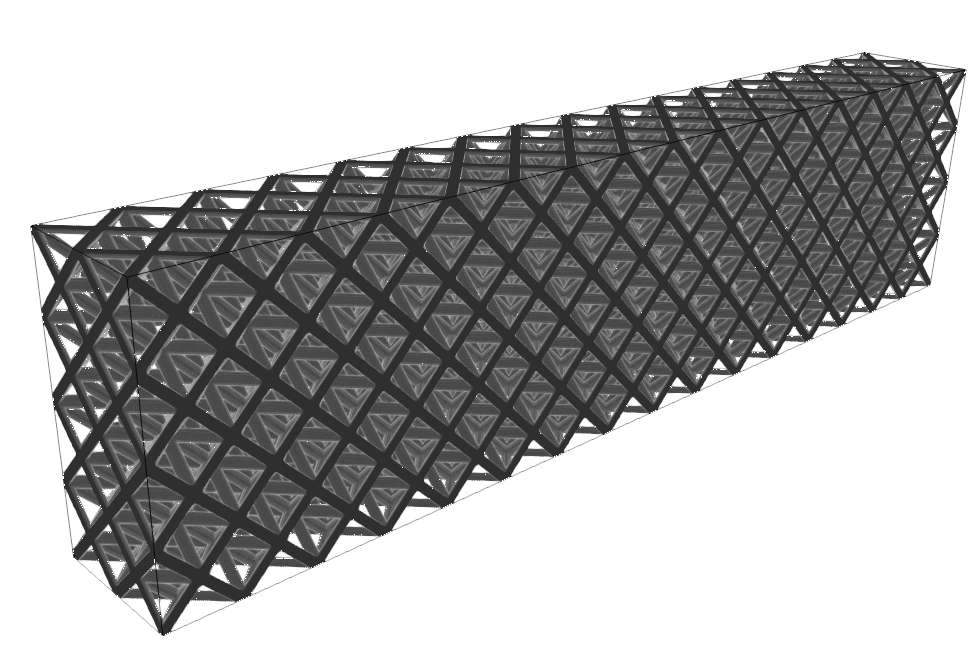}
		\caption{}\label{sbfig:perfect}
	\end{subfigure}
	\hfill
	\begin{subfigure}{\size}
		\centering
		\includegraphics[width=\textwidth]{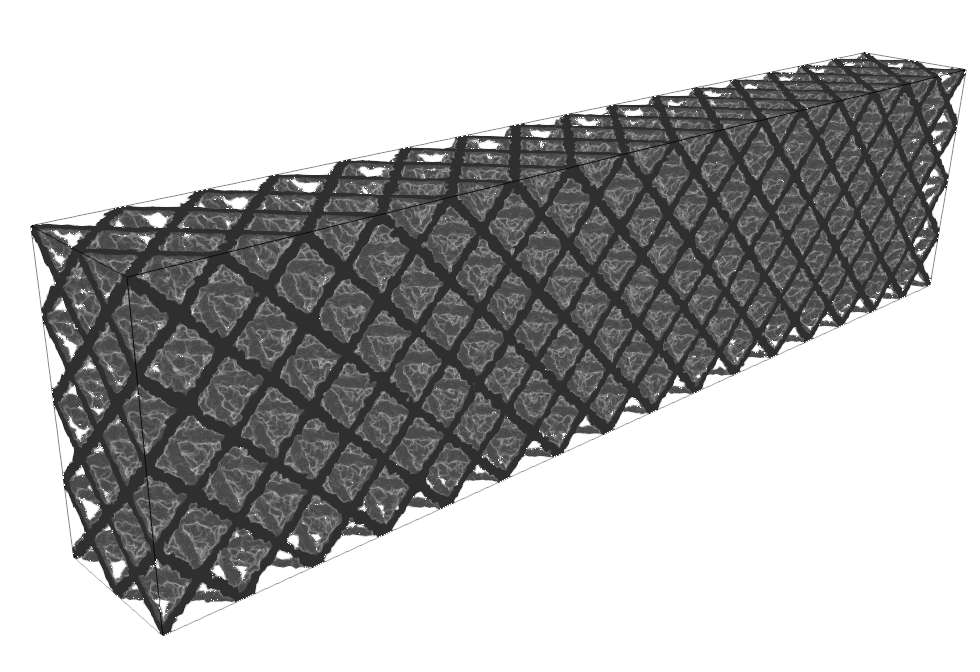}
		\caption{}\label{sbfig:imperfect}
	\end{subfigure}
	
	\caption{
		Example of the octet-truss lattice structure (generated using the surrogate model): (\subref{sbfig:perfect}) "as-designed", (\subref{sbfig:imperfect}) "as-manufactured".
	}
	\label{fig:full_lattice}
\end{figure}

In what follows, for our numerical experiments, we focus on a particular structure: octet-truss lattice cell (see \Cref{ex:octet_lattice}).
Owing to the manufacturing process, the resulting structure is perturbed by imperfections (see, e.g., \Cref{fig:full_lattice}) that strongly affects the properties of the manufactured material and produce uncertainties in the material behavior.
In particular, the difference in the effective properties of "as-designed" (defect-free) and "as-manufactured" (in the presence of imperfections) octet-truss lattices has been studied in~\cite{liu2017elastic,korshunova2021bending}.

\subsection{Calibration of the model parameters}
\label{sec:Calibration}

We consider the following stochastic optimization problem:
\begin{equation}\label{eq:minJ}
\min\limits_{\theta} \E_{\omega}\left[\loss(\theta, \omega)\right],
\end{equation}
where by $\loss$ we denote a distance measure between a surrogate sample~$\omega$ and the data.
In order to define such a measure, we use a set of \textit{geometrical descriptors} $\{\SD_i\}$.
The explicit choice of descriptors will be discussed below.
Then, we define $\loss$ as misfits between the corresponding descriptors:
\begin{equation}
\loss(\theta, \omega)
=
\frac{1}{2}\sum_{i}\frac{\norm{\SD_i(\Phase_{\theta}(\omega))-\SDdata_i}^2}{\norm{\SDdata_i}^2},
\end{equation}
where $\SDdata_i$ are the target descriptors obtained from the data, and the norm is the $\L{2}(\Domain)$-norm.

The design parameters include here the average strut radius~$\thickness>0$, the imperfection level~$\alpha\in[0,1]$ and the correlation length~$\corrlen>0$ of the surface imperfections.
In order to avoid constrained optimization, we define the vector of design parameters as follows:
\begin{equation}\label{eq:design_parameters}
\theta := \{ \log 2\thickness, \atanh(2\alpha-1), \log\corrlen\}.
\end{equation}

\begin{remark}
	Note that in~\eqref{eq:levelcut}, the Heaviside unit step function  is not differentiable.
	Thus, in the optimization process, we replace it with its smooth approximation:
	\begin{equation}\label{eq:levelcut_reg}
	\Phase(\x; \omega) = 
	\sigma(u(\x; \omega))
	\end{equation}
	with a smooth activation function approximating the unit step, e.g., $\sigma(x) = (1+\tanh(x/\epsilon))/2$, where $\epsilon$ is small.
	In particular, we use $\epsilon=h^2$, where $h$ stands for the voxel size.
\end{remark}


Let us now discuss the choice of the descriptors~$\{\SD_i\}$.
We use the material volume fraction and its specific surface area as the two first geometrical descriptors.
Specific surface area is the ratio of the surface area of the phase interface to the volume of the phase.
For the regularized version of the two-phase material~\eqref{eq:levelcut_reg}, we approximate the indicator function of the phase boundary with the absolute value of the gradient of the phase field.
We will refer to it as the interface field $\abs{\grad\Phase(\x)}$.
These descriptors are global and cannot capture the local structure of the sample.
To do this, we use the autocorrelation of the phase-field as another descriptor.
For a statistically homogeneous random field, the field autocorrelation approximates the two-point correlation function~\cite{torquato2013random}.
Here, though the perturbation field is considered to be homogeneous, the octet cell structure is not.
Moreover, the structure size dominates the perturbation correlation length scale.
That is, the autocorrelation can provide information on the structure size, but it barely detects the localized properties of the perturbation.
In order to get information on the perturbation, we need to consider the geometric features of the material surface.
To this end, we use the autocorrelation of the interface field as the fourth descriptor.
Thus, in our numerical experiments, we use the following geometrical descriptor of a sampled phase field~$\Phase$:
\begin{equation}\label{key}
	\begin{aligned}
	\SD_1(\Phase) &= \int\limits_{\Domain}\Phase(\x)\d\x,
	\\
	\SD_2(\Phase) &= \int\limits_{\Domain}\abs{\grad\Phase(\x)}\d\x, 
	\\	
	\SD_3(\Phase; \r) &= \int\limits_{\Domain}\left[\Phase(\x+\r)-\SD_1(\Phase)\right]\cdot\left[\Phase(\x)-\SD_1(\Phase)\right]\d\x, 
	\\	
	\SD_4(\Phase; \r) &= \int\limits_{\Domain}\left[\abs{\grad\Phase(\x+\r)}- \SD_2(\Phase)\right]\cdot\left[\abs{\grad\Phase(\x)}- \SD_2(\Phase)\right]\d\x, 
	\end{aligned}	
\end{equation}
with the spatial lag~$\r\in\Domain$.
Given an image of the real sample $X=\TargMat(\omega)$, the target descriptors are defined as $\SDdata_i:=\SD_i(X)$.
In case of a batch of the target material samples $\vec{X} = \{X_j = \TargMat(\omega_j),\quad j=1,\ldots,N_{data}\}$, the target descriptors can be defined as the average
$
\SDdata_i := \frac{1}{\abs{\vec{X}}}\sum_{X\in\vec{X}} \SD_i(X)
$.


\subsection{Calibration results}
\label{sec:calibration_results}

We calibrate the model design parameters by solving the stochastic optimization problem~\eqref{eq:minJ}.
In order to avoid oversampling, a progressive batching strategy is applied, when the appropriate number of samples (\textit{batch size} $\abs{S_k}$) is estimated at each iteration $k$ and is adaptively updated satisfying specific conditions; see~\cite{byrd2012sample,bollapragada2019exact,roosta2019sub,xie2020constrained}.
In our implementation, we follow the progressive batching {LBFGS} algorithm proposed in~\cite{bollapragada2018progressive}.
Technical details of the algorithm can be found in~\ref{sec:BFGS}.

The surrogate model generator is implemented using PyTorch~\cite{pytorch}, allowing to benefit from algorithmic differentiation for learning the model parameters.
In particular, we use PyTorch-LBFGS package~\cite{PyTorch-LBFGS} for implementation of the progressive batching minimization.

The target material is the octet-truss lattice structure manufactured using an SLM printer~\cite{solutions2017formup} from 316L stainless steel; see~\Cref{sbfig:lattice_CT}.
The relative density of the structure is 0.3.
The images of the octet lattice have been obtained on the as-printed specimens at the In Situ Innovative Set-ups for X-ray micro-tomography on the ISIS4D platform~\cite{limodin2013isis4d}.
Reconstruction of the tomographic data is performed with a filtered back-projection algorithm~\cite{kak2001principles} using {X-Act} software.
Similar X-ray tomography reconstructions have been used for exploring plasticity and fatigue phenomena on various materials, see, e.g.,~\cite{hosdez2019plastic,hosdez2020fatigue,shi2021analysis}.

\begin{figure}[!th]
	\centering\noindent
	\newcommand{\size}{0.31\textwidth}
	
	\begin{subfigure}{\size}
		\centering
		\includegraphics[width=\textwidth]{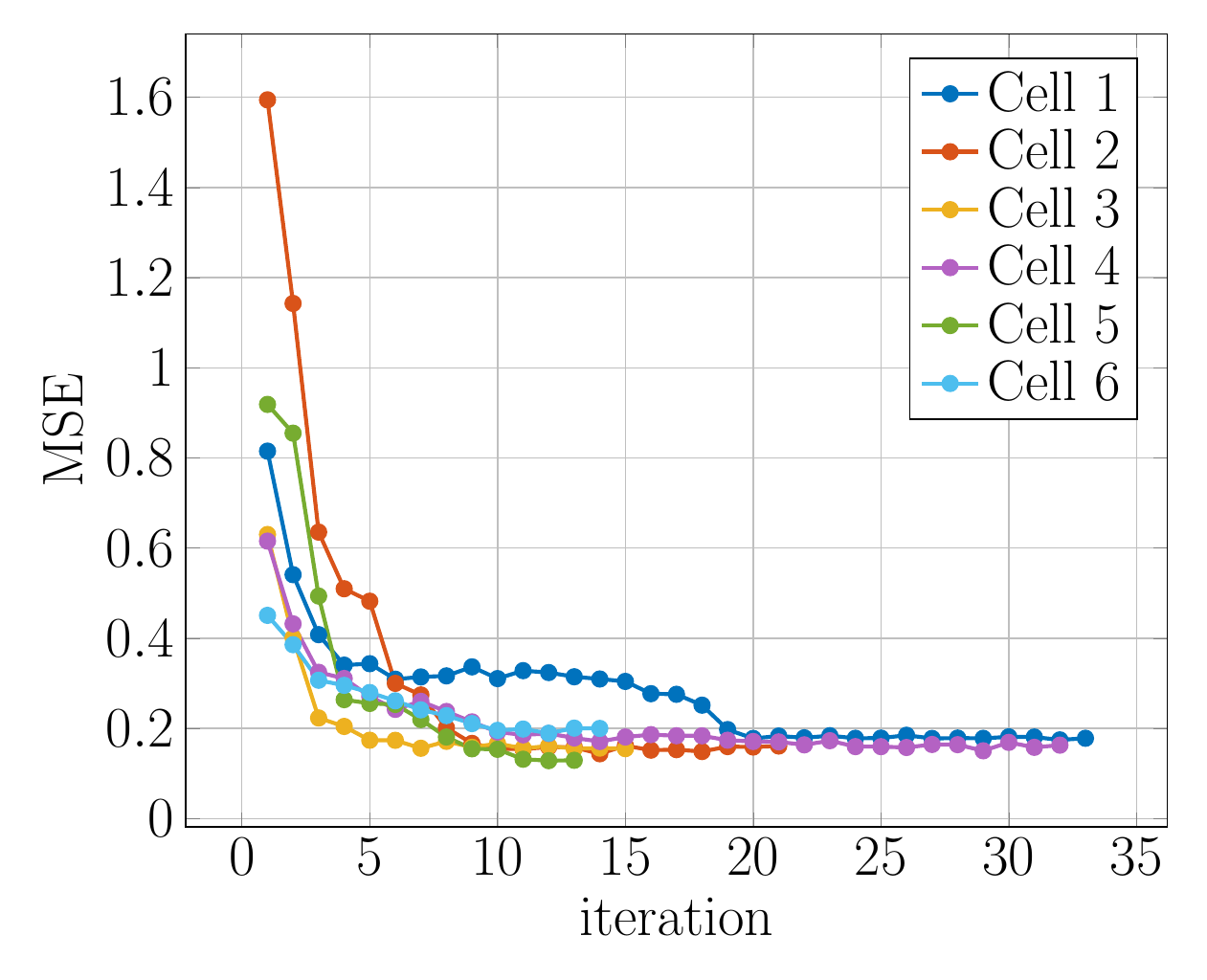}
		\caption{Loss}\label{sbfig:cells:loss}
	\end{subfigure}
	\hfill
	\begin{subfigure}{\size}
		\centering
		\includegraphics[width=\textwidth]{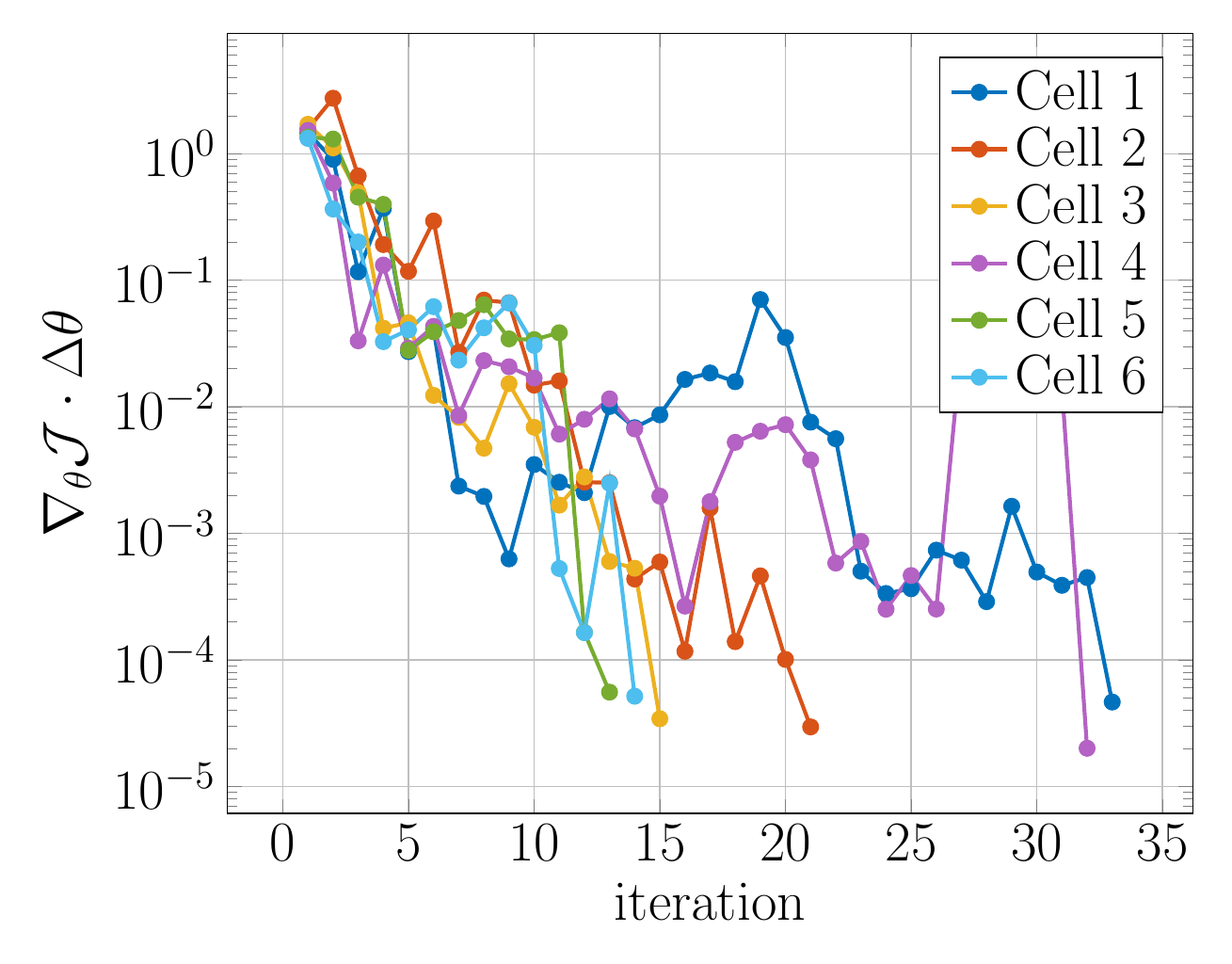}
		\caption{Increment}\label{sbfig:cells:crit}
	\end{subfigure}
	\hfill
	\begin{subfigure}{\size}
		\centering
		\includegraphics[width=\textwidth]{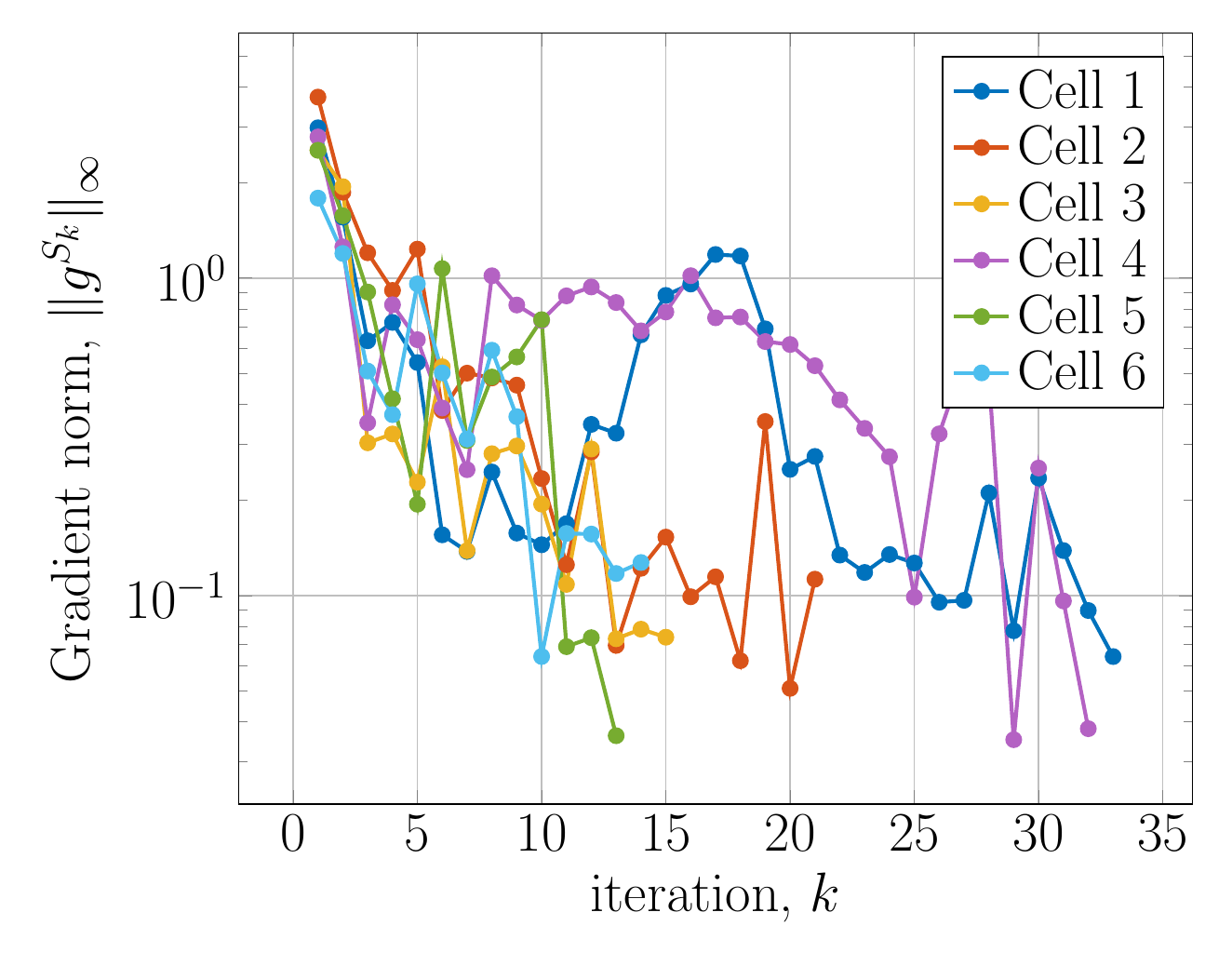}
		\caption{Gradient norm}\label{sbfig:cells:grad}
	\end{subfigure}
	
	\caption{
		Convergence of
		the loss function (\subref{sbfig:cells:loss}),
		the loss increment (\subref{sbfig:cells:crit})
		and the gradient norm (\subref{sbfig:cells:grad})
		for each individual cell fitting.
	}
	\label{fig:cells:group1}
\end{figure}

\begin{figure}[!th]
	\centering\noindent
	\newcommand{\size}{0.31\textwidth}
	
	\begin{subfigure}{\size}
		\centering
		\includegraphics[width=\textwidth]{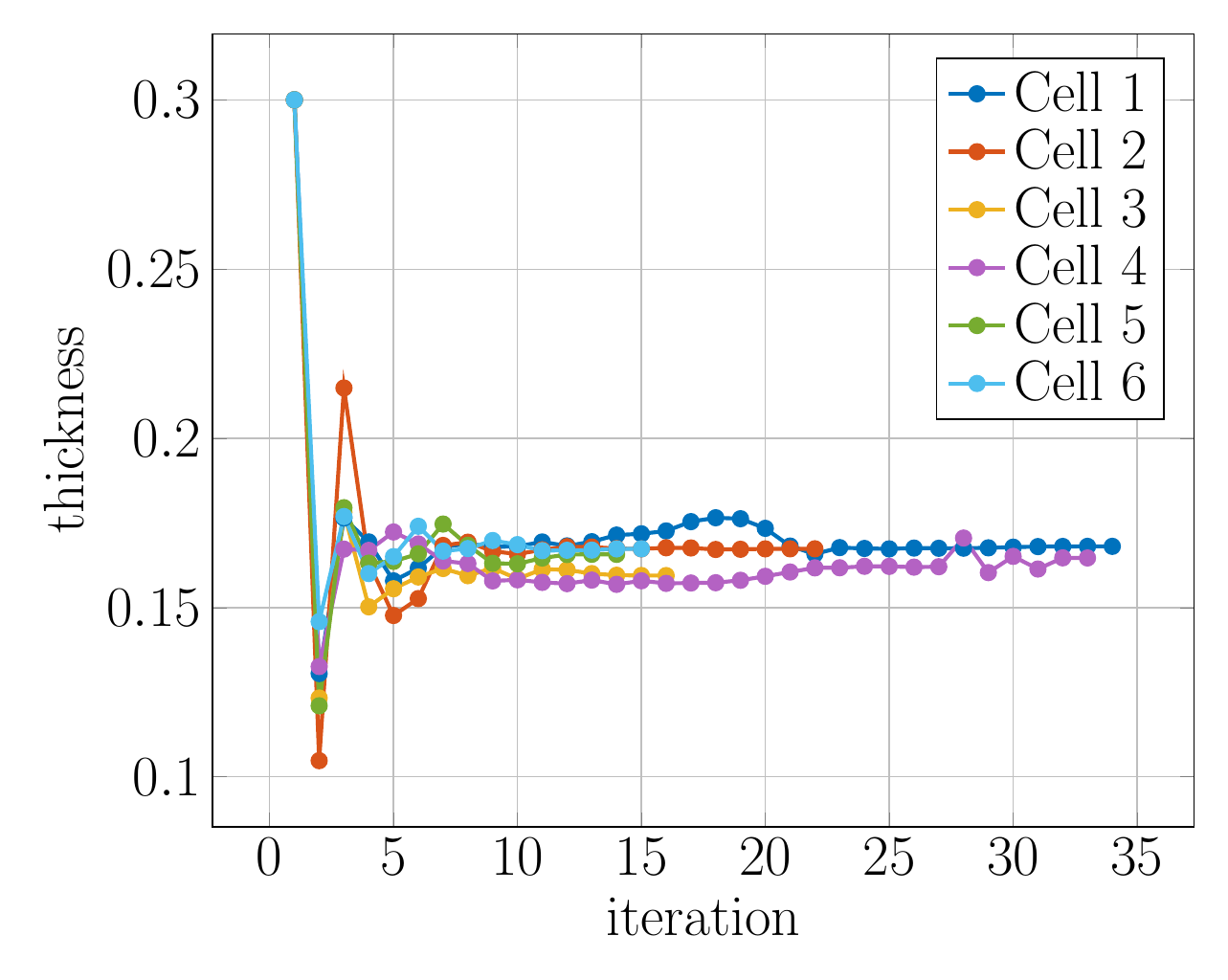}
		\caption{Strut thickness}\label{sbfig:cells:thickness}
	\end{subfigure}
	\hfill
	\begin{subfigure}{\size}
		\centering
		\includegraphics[width=\textwidth]{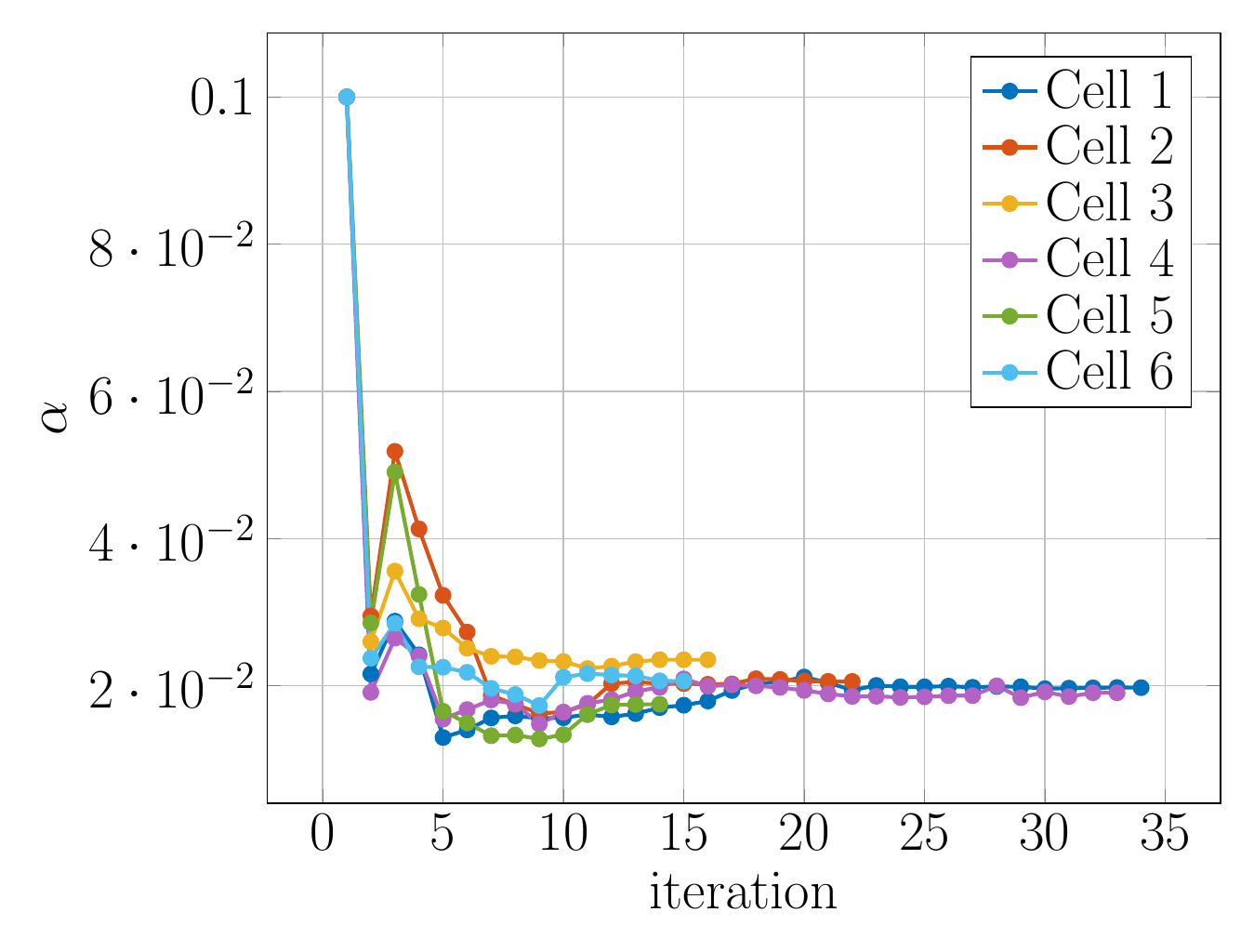}
		\caption{Imperfection level}\label{sbfig:cells:alpha}
	\end{subfigure}
	\hfill
	\begin{subfigure}{\size}
	\centering
	\includegraphics[width=\textwidth]{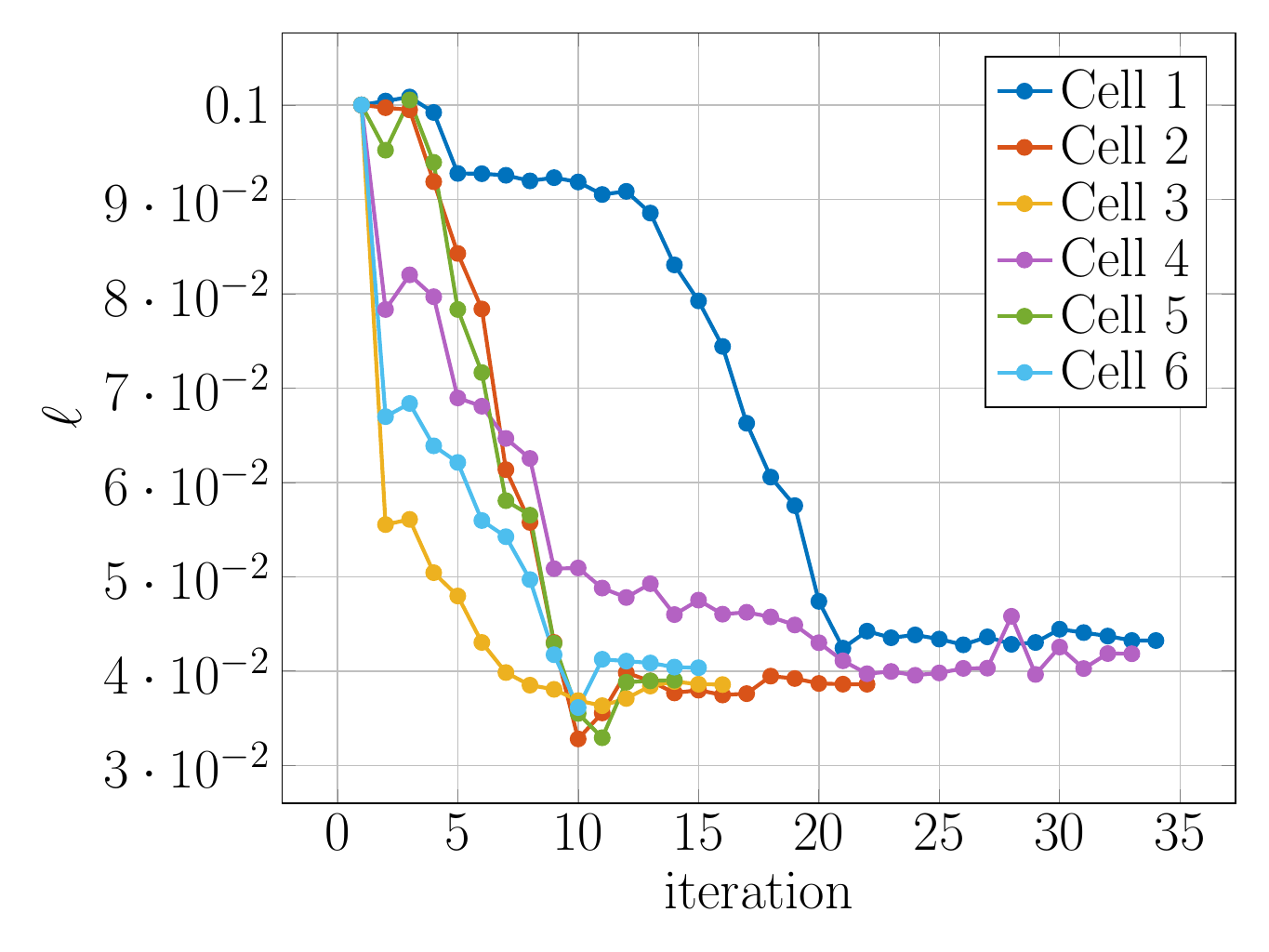}
	\caption{Correlation length}\label{sbfig:cells:corrlen}
	\end{subfigure}

	\caption{
		Convergence of the model parameters for each individual cell fitting:
		(\subref{sbfig:cells:thickness}) strut thickness $2\tau$;
		(\subref{sbfig:cells:alpha}) imperfection level $\alpha$;
		(\subref{sbfig:cells:corrlen}) correlation length $\corrlen$.
	}
	\label{fig:cells:group2}
\end{figure}
\begin{figure}[!h]
	\newcommand{\size}{0.31\textwidth}	
	\begin{subfigure}{\size}
		\centering
		\includegraphics[width=\textwidth]{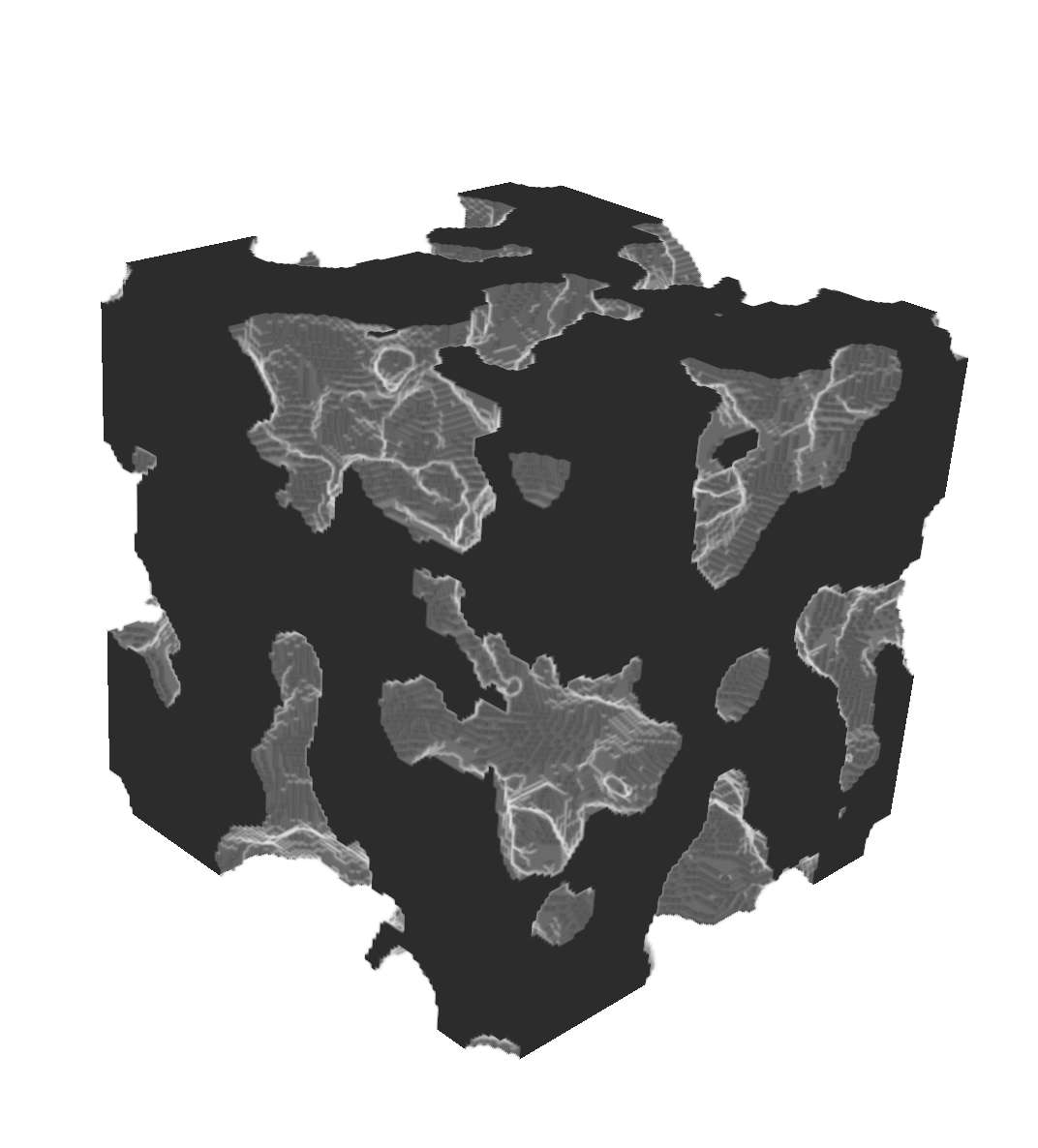}
		\caption{Initial guess (sample),\\
		$2\tau=0.3$, $\alpha=0.1$, $\corrlen=0.1$}\label{sbfig:initial}
	\end{subfigure}
	\hfill
	\begin{subfigure}{\size}
		\centering
		\includegraphics[width=\textwidth]{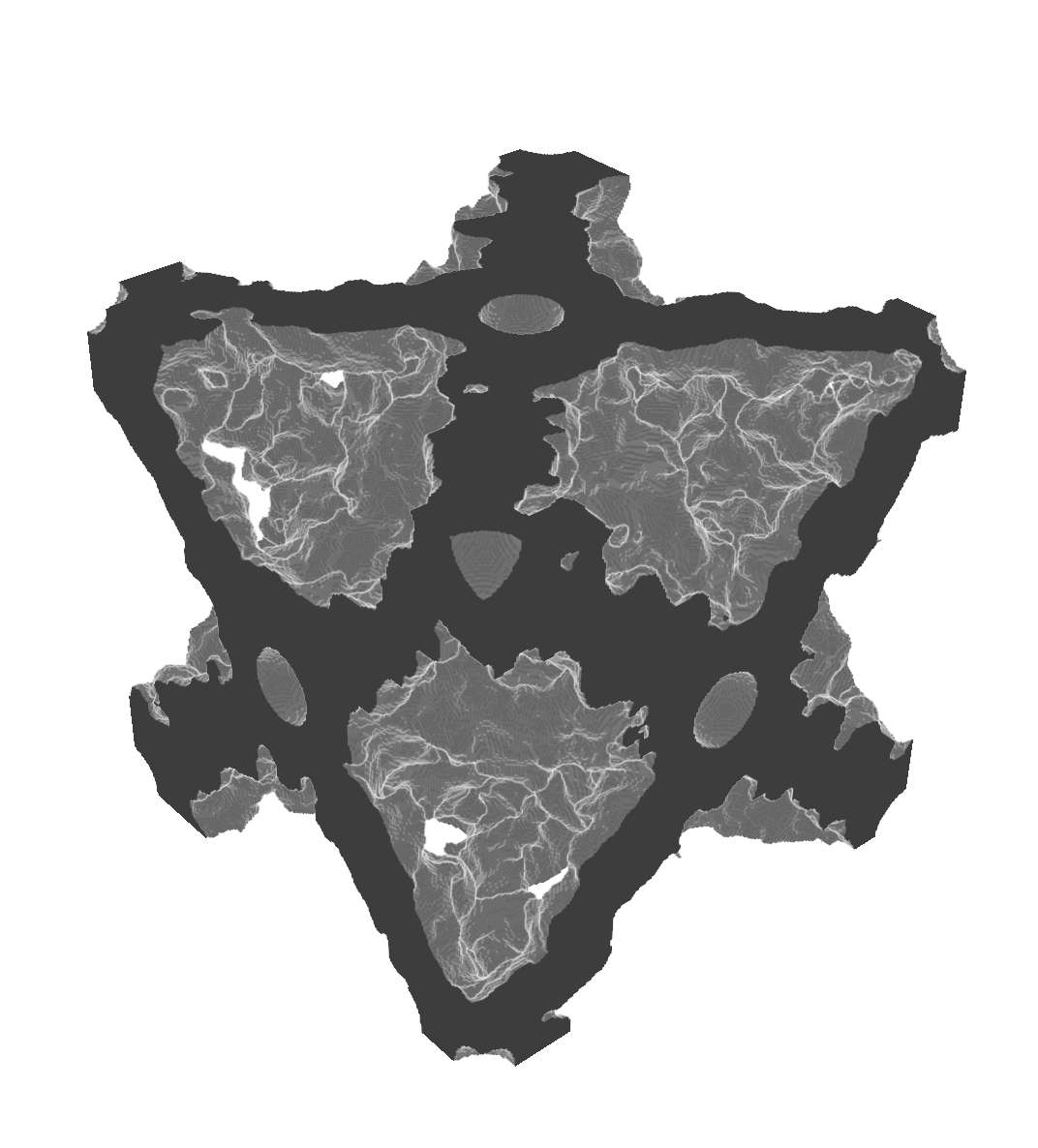}
		\caption{Calibration result (sample), \\
		$2\tau=0.168$, $\alpha=0.0198$, $\corrlen=0.0432$}\label{sbfig:result}
	\end{subfigure}
	\hfill
	\begin{subfigure}{\size}
		\centering
		\includegraphics[width=\textwidth]{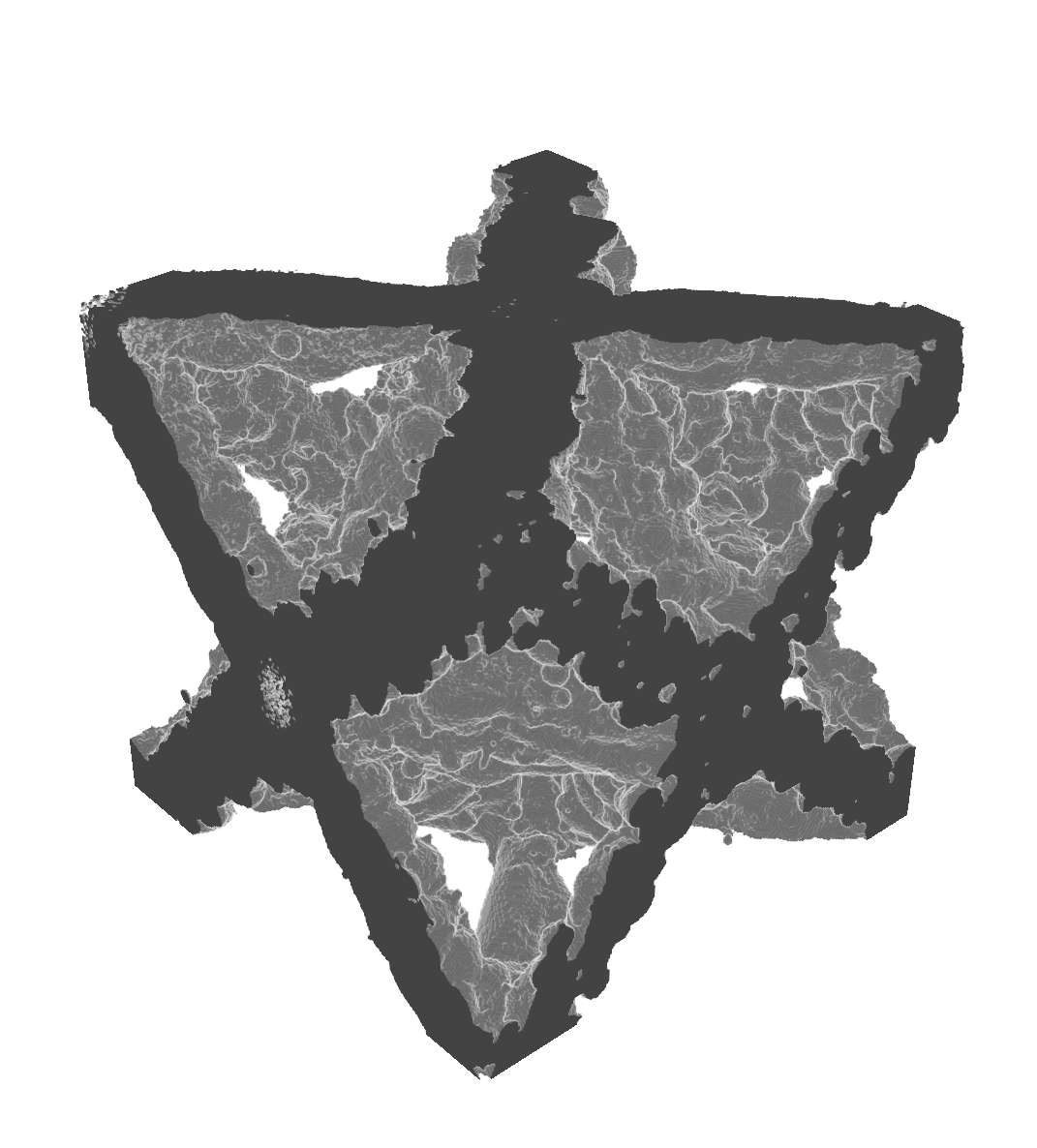}
		\caption{Target (CT data -- Cell~1)}\label{sbfig:target}
	\end{subfigure}

	\caption{
		Calibration of the surrogate model for the first cell: sample examples for
		the initial guess model (\subref{sbfig:initial}), the resulting model (\subref{sbfig:result}) and the target cell (\subref{sbfig:target}).
	}
	\label{fig:calibration_result}
\end{figure}

\begin{figure}[!h]
	\centering\noindent
	\includegraphics[width=0.35\textwidth]{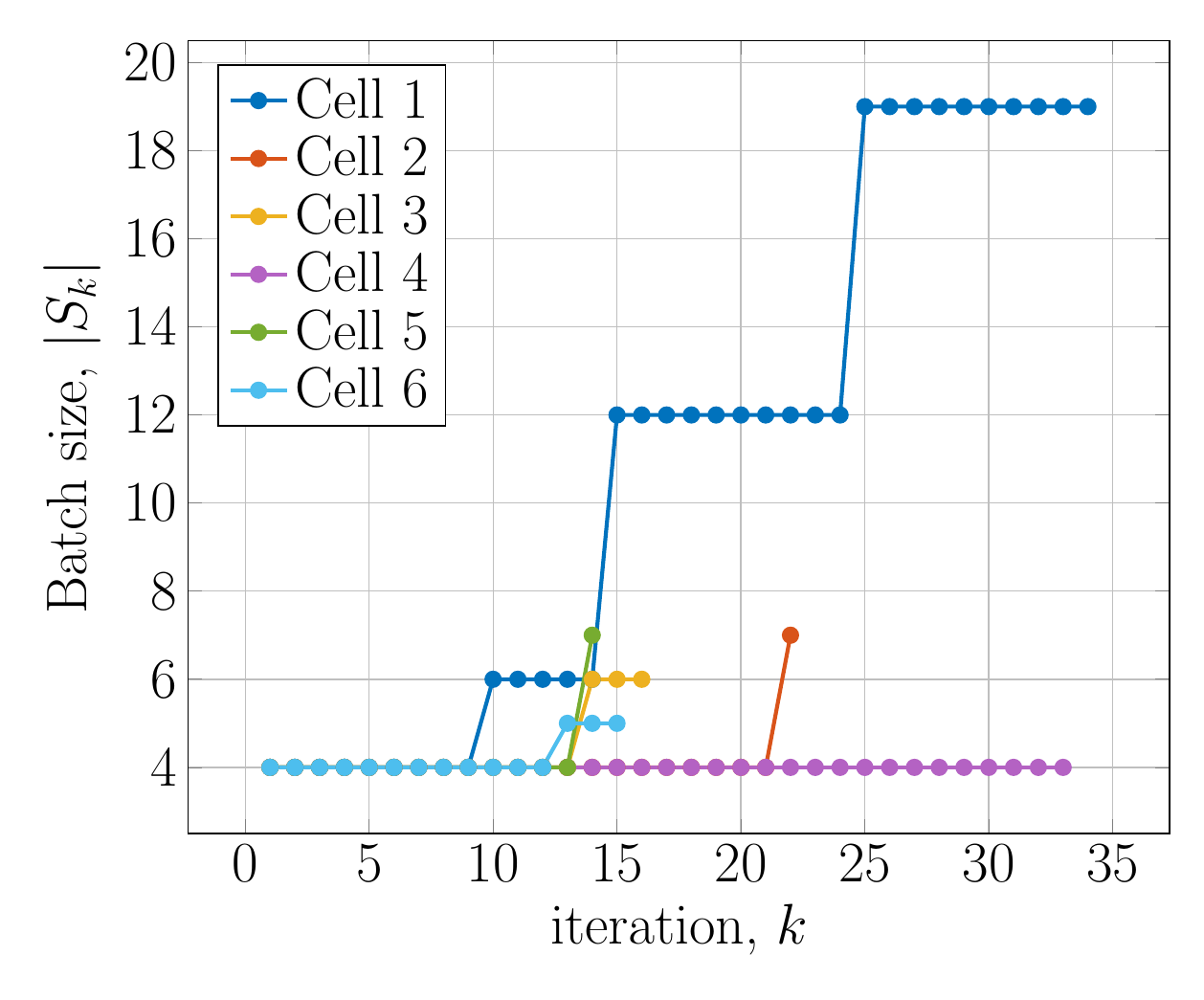}
	\caption{
		Progress of the batch size for each individual cell fitting.
	}
	\label{fig:cells:batchsize}
\end{figure}

We use six different lattice cells obtained from the CT image and fit the surrogate model individually on each of these targets, comparing at the end the results of these different fittings.
We fix $\nu=2$ assuming moderate smoothness of the imperfect interface.
Therefore, our three parameters to calibrate are the strut thickness $2\tau$, the imperfection level $\alpha$ and the correlation length~$\corrlen$.
The initial guess for the model parameters: $2\thickness_0=0.3$, $\alpha_0=0.1$, $\corrlen_0=0.1$.
The stopping criterion is reached when the loss increment is less than the fixed tolerance, precisely, $\nabla_{\theta}\bar{\loss}^{S_k}\cdot\Delta\theta<10^{-4}$.
For the model calibration as well for all our further numerical experiments, the size of the surrogate model samples is fixed at $2^7\times2^7\times2^7$ voxels.
\Cref{fig:cells:group1} shows the convergence of the loss function, its increment and the max-norm of its gradient.
And \Cref{fig:cells:group2} shows the evolution of the model parameters.
Comparison of a calibrated model sample with an initial guess sample as well as with the target (the first cell) can be found in~\Cref{fig:calibration_result}.
The progress of the batch size $\abs{S_k}$ is depicted in~\Cref{fig:cells:batchsize} for each individual cell fitting.
For all our samples the stopping criterion is reached within $35$ iterations, but in some cases $15$ iterations were sufficient.
Although only one data sample was considered for each case, we can observe that the deviation of the resulting parameters is small; see~\eqref{eq:paramters_distribution} below.
This is due to the fact that the distribution of the imperfections over the material surface  is statistically homogeneous, and even one sample provides enough information on the statistical properties of the uncertainties.
Note that in the general case, the required number of data samples can be larger depending on the deviation of the resulting design parameters.
In case more data samples are available, Bayesian inference techniques can also be employed.

\subsection{Uncertainty quantification of elasto-plastic response}

Owing to the structure imperfection, the material properties, such as effective bulk and shear moduli, computed on one material sample are random variables.
Once the surrogate material model is calibrated with respect to the real material data, we can proceed with uncertainty quantification of the material properties and, in particular, estimation of their probability distribution.
To this end, using synthetic samples, we employ a crude Monte Carlo method to approximate the expected value and the standard deviation of the \textit{quantities of interest} (QoIs).

In our application, the quantity of interest is the effective tangent modulus of the lattice material.
Let the base material of the lattice be assumed elasto-plastic with isotropic linear hardening, Young's modulus $200$ GPa, Poisson ratio $0.3$, yield stress $0.4$ GPa and the hardening modulus $0.6$ GPa.		
To compute the elasto-plastic material response, we use the software CraFT~\cite{CraFT} -- a mesh-free solver using a Fast-Fourier Transform algorithm and based on the Suquet-Moulinec method~\cite{suquet1990simplified,moulinec1994fast,moulinec1998numerical,michel2001computational}, in particular, using the Monchiet-Bonnet scheme~\cite{monchiet2012polarization}.
The contrast of the Young's moduli of the phases is fixed at $10^{-4}$.

Let $\vec{n}$ be a loading direction, $\norm{\vec{n}}=1$.
We impose in $25$ uniform loading steps (loading time $t\in[0,1]$) the macroscopic stress $\ten{\sigma}$ in the direction $\vec{n}$ (uniaxial extension) resulting in $1\%$ macroscopic deformation in the direction~$\vec{n}$.
That is, $\vec{n}\tp\strain\vec{n}= 0.01$ at $t=1$, where $\ten{\eps}$ denotes the resulting macro strain.
We consider $10$ different directions of~$\vec{n}$ from the regular grid ($4$ nodes per axis) in the triangle defined by the points $(1, 0, 0)$, $(1, 1, 0)$ and $(1, 1, 1)$.
All the other directions can be obtained by symmetry of the octet-truss lattice cell.

For each loading direction $\vec{n}$, we compute the elasto-plastic material responses of $20$ material samples, generated using the surrogate material model calibrated in~\Cref{sec:calibration_results}.	
In particular, we assume the design parameters~\eqref{eq:design_parameters}, $\theta = \{ \log 2\thickness, \atanh(2\alpha-1), \log\corrlen\}$,
to be random i.i.d. normally distributed with the mean and the standard deviation estimated using the calibration results for different cells from~\Cref{sec:calibration_results}, precisely, 
\begin{equation}\label{eq:paramters_distribution}
\theta 
=
\begin{bmatrix}
\log 2\thickness\\
\atanh(2\alpha-1)\\
\log\corrlen
\end{bmatrix}	
\sim \Gaussian\left(
\begin{bmatrix}
-1.8 \\
-1.94 \\
-3.21 \\
\end{bmatrix},	
\begin{bmatrix}
0.018^2 & &  \\
& 0.046^2 &  \\
& &  0.043^2   \\
\end{bmatrix}
\right).
\end{equation}
Thus, for each surrogate sample~$\omega$, we first sample the design parameters $\theta(\omega)$ and then the corresponding phase field realization $\Phase(\theta(\omega),\omega)$.

\begin{figure}[!h]
	\newcommand{\size}{0.45\textwidth}
	\centering\noindent
	\begin{subfigure}{\size}
		\centering
		\includegraphics[width=\textwidth]{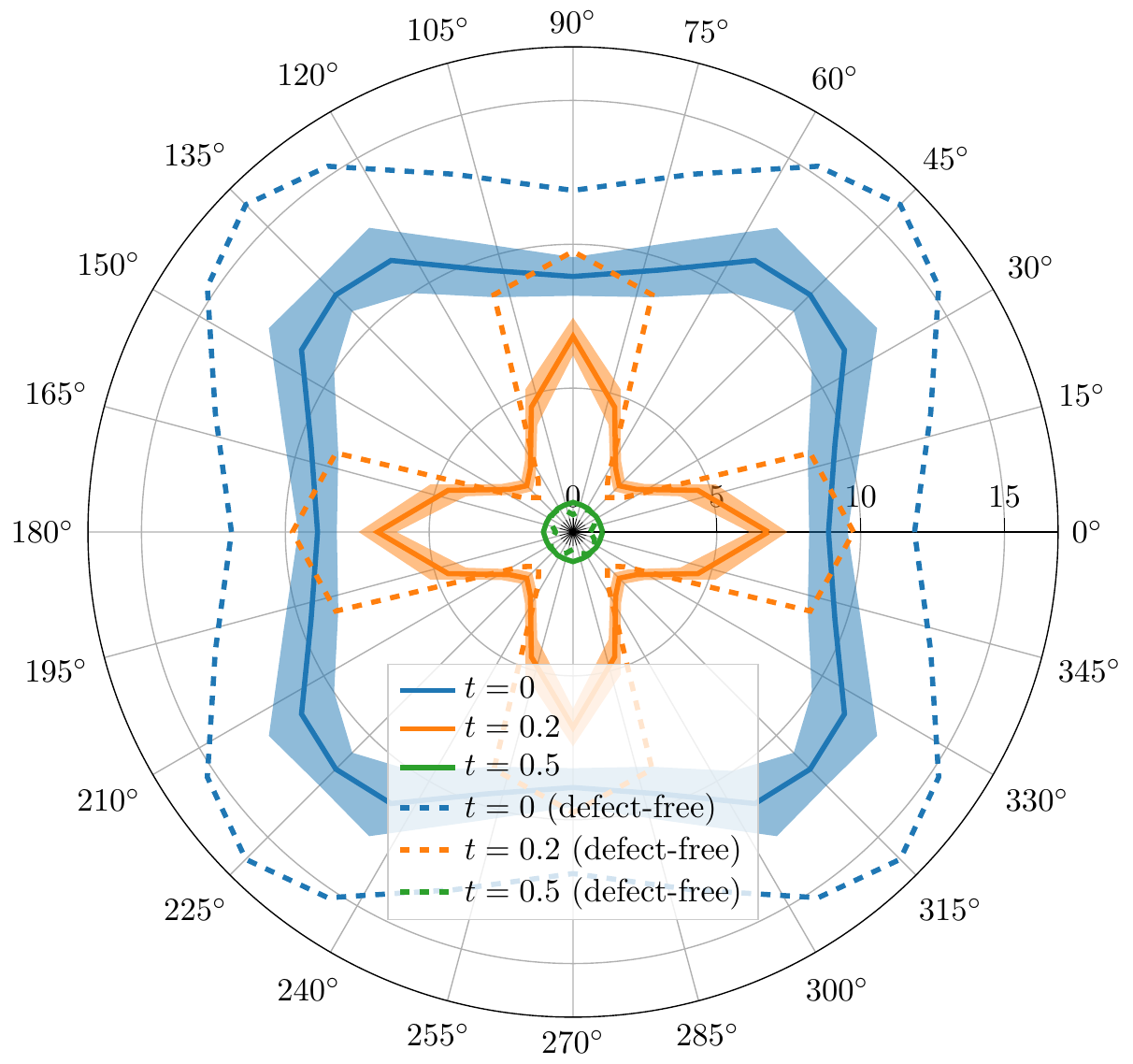}
		\caption{Plane A}
		\label{sfig:RosePlaneA}
	\end{subfigure}
	\hfill
	\begin{subfigure}{\size}
		\centering
		\includegraphics[width=\textwidth]{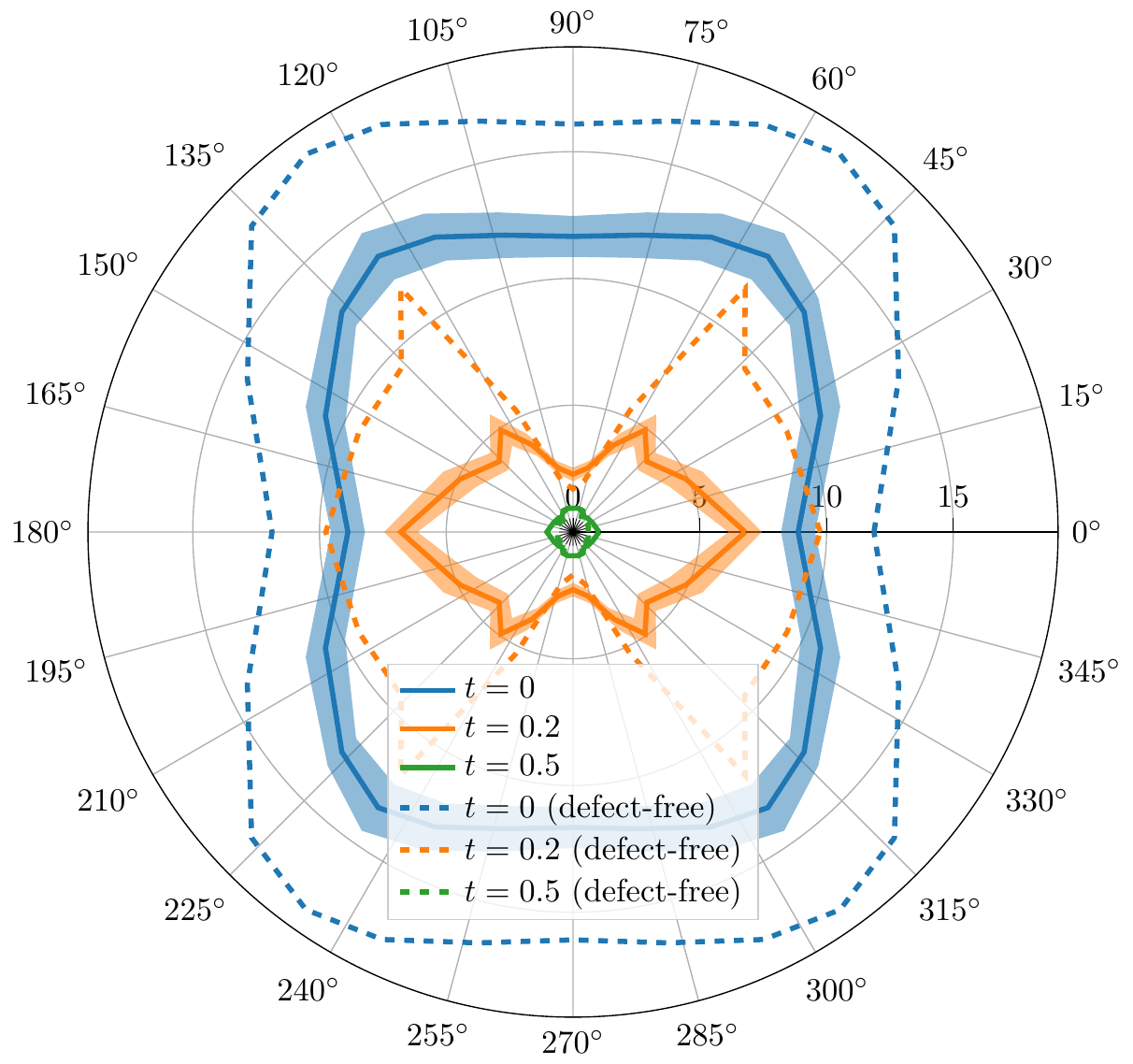}
		\caption{Plane B}
		\label{sfig:RosePlaneB}
	\end{subfigure}

	\vspace{3ex}
	\begin{subfigure}{\size}
		\centering
		\includegraphics[width=0.9\textwidth]{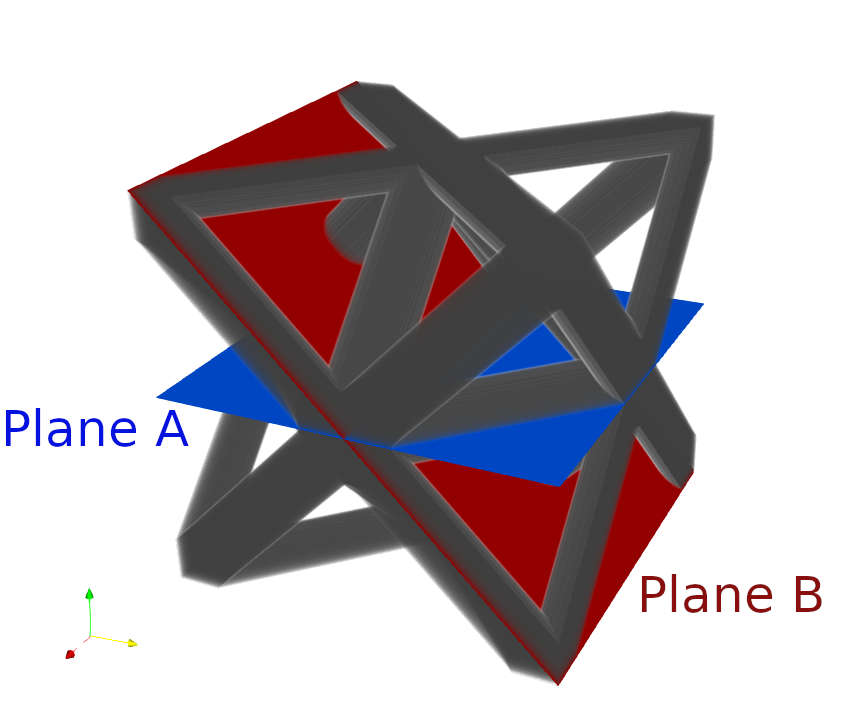}
		\caption{Scheme of the cross-sections
		}
		\label{sfig:Planes}
	\end{subfigure}
	\hfill
	\begin{subfigure}{\size}
		\centering
		\includegraphics[width=0.9\textwidth]{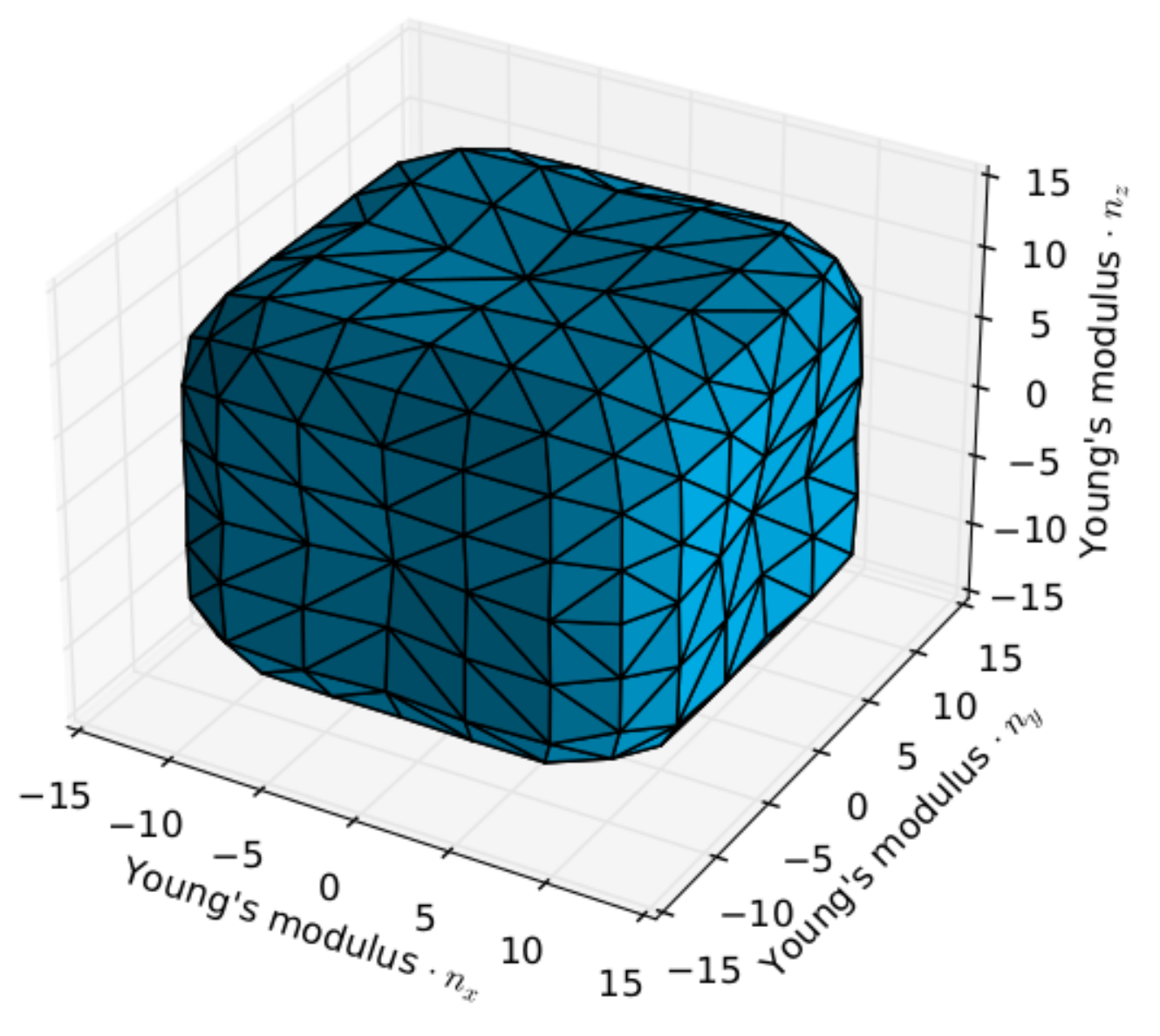}
		\caption{Young's modulus of the defect-free structure}
		\label{sfig:3DYoung}
	\end{subfigure}

	\caption{
		Directional variation of the tangent modulus at loading time $t=0, 0.2, 0.5$: (\subref{sfig:RosePlaneA}): in the plane A, (\subref{sfig:RosePlaneB}): in the plane B.
		Solid line -- expected value, transparent area -- standard deviation, dashed line -- defect-free case.			
		(\subref{sfig:Planes}): position of the planes A and B. Owing to the symmetry, all other cross-sections are equivalent to these two.
		(\subref{sfig:3DYoung}): 3D plot of directional variation of the Young's modulus for the defect-free structure.		
		We computed $10$ different 3D directions ($4$ for each of cross-sections A and B) in the triangle defined by the points $(1, 0, 0)$, $(1, 1, 0)$ and $(1, 1, 1)$.
		All the other directions are obtained by symmetry of the octet-truss lattice cell.
	}
	\label{fig:craft:YoungRose}
\end{figure}

\begin{figure}[!h]
	\centering\noindent
	\includegraphics[width=0.4\textwidth]{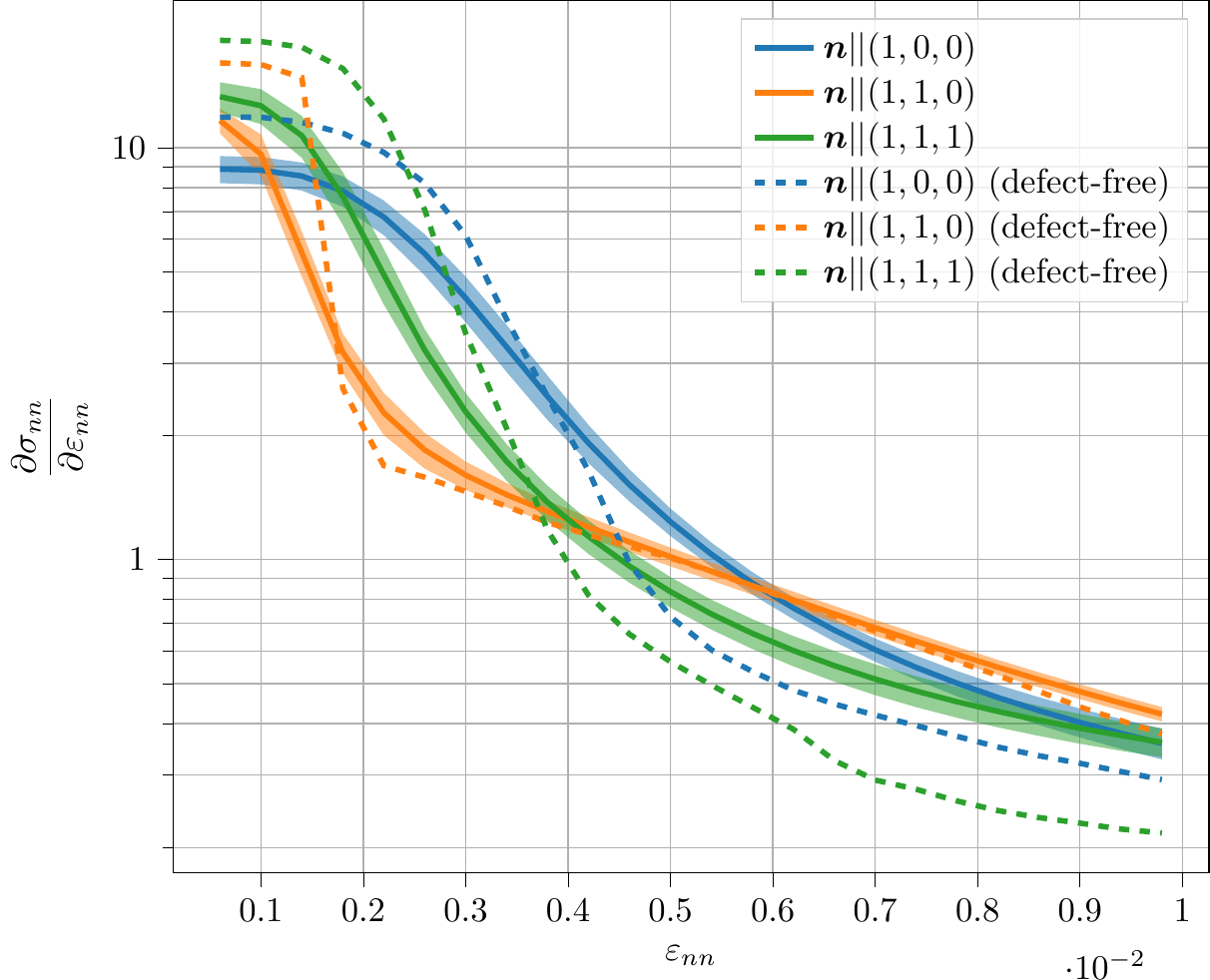}
	\caption{
		Tangent modulus in the directions co-linear to $(1, 0, 0)$, $(1, 1, 0)$ and $(1, 1, 1)$ as function of the strain~$\eps_{nn}$.
		Solid line -- expected value, transparent area -- standard deviation.
	}
	\label{fig:craft:TangentModulus}
\end{figure}

For each surrogate material sample, from the values $\eps_{nn}:=\vec{n}\tp\strain\vec{n}$ and $\sigma_{nn}:=\vec{n}\tp\stress\vec{n}$, the corresponding tangent moduli $\frac{\partial\sigma_{nn}}{\partial\eps_{nn}}$ are computed via finite differences.
From the obtained dataset, we estimate the expected value and the standard deviation of the tangent modulus.	
Exploiting the symmetry properties of the structure, we illustrate in \Cref{fig:craft:YoungRose}  the directional variation of the tangent modulus $\frac{\partial\sigma_{nn}}{\partial\eps_{nn}}$ for the surrogate model octet-lattice cell at different loading time $t$.
There, the polar plots (\subref{sfig:RosePlaneA})-(\subref{sfig:RosePlaneB}) correspond to the cell cross-sections (the planes A and B) shown in~\Cref{sfig:Planes}.
Owing to the symmetry, all other cross-sections are equivalent to these two.
The expected values of the tangent modulus are depicted with solid lines, while the transparent areas around represent the associated standard deviation.
We also plot with dashed lines the modulus values corresponding to the as-designed  (defect-free) structure, i.e., with the imperfection level $\alpha=0$.
A 3D plot of directional variation of the Young's modulus for the defect-free structure can be found in~\Cref{sfig:3DYoung}.
In~\Cref{fig:craft:TangentModulus}, we plot the tangent modulus $\frac{\partial\sigma_{nn}}{\partial\eps_{nn}}$ (expected value and deviation) in the directions co-linear to $(1, 0, 0)$, $(1, 1, 0)$ and $(1, 1, 1)$ as function of the applied deformation~$\eps_{nn}$.
The scale of $y$-axis is logarithmic.
We observe that the deviation magnitude changes with the strain proportionally to the expectation.
The volume fraction of the defect-free structure ($\alpha=0$) is $0.273$; in the presence of imperfections, the expected value for the volume fraction is $0.282$ with the  standard deviation $0.009$.
Note that the tangent modulus near $t=0$ corresponds to the Young's modulus and that the obtained values are in a good agreement with the numerical and experimental results, e.g., in~\cite{deshpande2001effective,qi2019mechanical}.
Besides, we observe that the plastic effects strongly depend on the loading direction; in particular, the most stiff direction varies with deformation.
Note also that the structure with imperfections is always less stiff than the as-designed one; moreover, the effective yield stress of the as-designed cell is higher than in the presence of imperfections.

\subsection{Analysis of imperfections effect}

\begin{figure}[!h]
	\newcommand{\size}{0.45\textwidth}
	\centering\noindent
	\begin{subfigure}{\size}
		\centering
		\includegraphics[width=\textwidth]{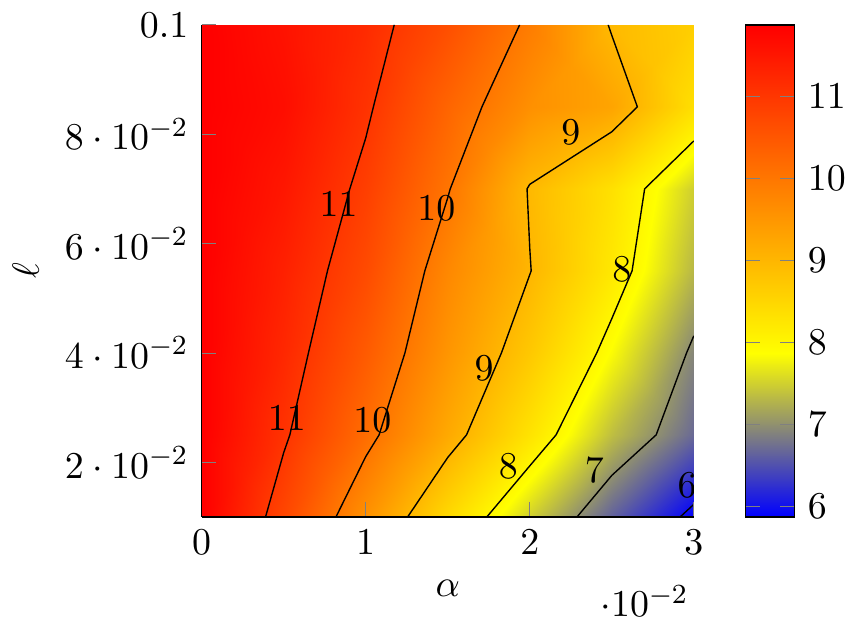}
		\caption{Directional minimum of Young's modulus}
		\label{sfig:contour:min}
	\end{subfigure}
	\hfill		
	\begin{subfigure}{\size}
		\centering
		\includegraphics[width=\textwidth]{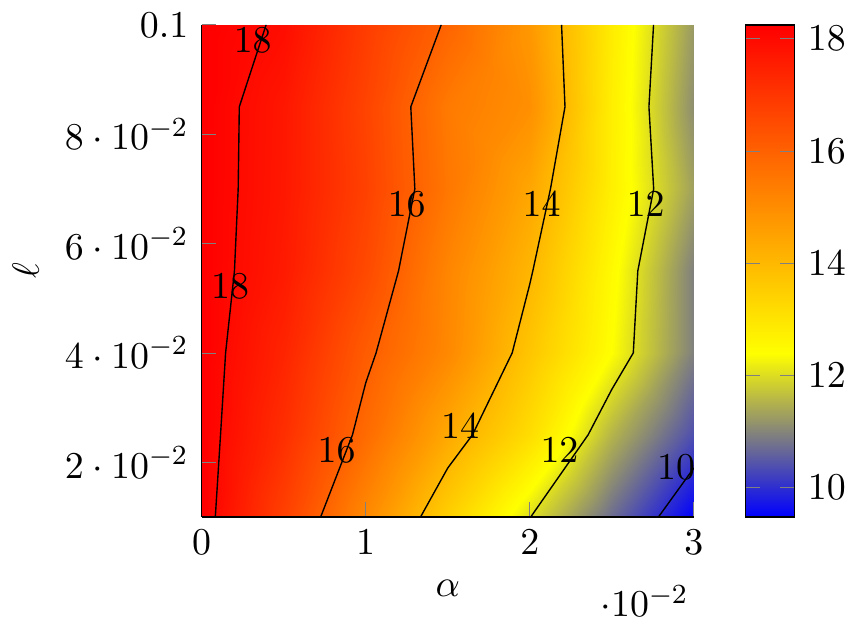}
		\caption{Directional maximum of Young's modulus}
		\label{sfig:contour:max}
	\end{subfigure}		
	\caption{
		Contour plots of Young's modulus in $(\alpha, \corrlen)$ plane illustrating the effects of imperfections of the octet-truss lattice cell; (\subref{sfig:contour:min}): directional minimum, (\subref{sfig:contour:max}): directional maximum.
	}
	\label{fig:craft:DefectsEffect}
\end{figure}

In this section, in order to assess the effect of the defects, we study the sensitivity of the Young's modulus to the model parameters describing the imperfections -- the level~$\alpha$ and the correlation length~$\corrlen$.
To this end, for variable values of the parameters $\alpha$ and $\corrlen$, we compute the Young's modulus in two directions:  $(1, 0, 0)$ -- a minimum stiffness direction, and $(1,1,1)$ -- a maximum stiffness direction.
The strut thickness is fixed to $2\tau=0.165$.
We consider the model parameters values in the domain $[0, 0.03]\times[0.01, 0.1]$ of the $(\alpha,\corrlen)$ plane ($7\times7$ mesh grid).
For each pair $(\alpha,\corrlen)$, we estimate the expected value of Young's modulus computed using $20$ surrogate material samples.
The resulting moduli are depicted in the contour plots in~\Cref{fig:craft:DefectsEffect} (using linear interpolation), where the abscissa corresponds to~$\alpha$, and the ordinate -- to~$\corrlen$.
The left plot corresponds to the directional minimum of the Young's modulus, while the right one corresponds to the maximum.
We observe that the Young's modulus decreases with increasing imperfections level.
Besides, for high enough $\alpha$, the correlation length starts to affect the Young's modulus which decreases with~$\corrlen$.

	\section{Generalizations}\label{sec:Extensions}

In this section, in addition, we briefly comment on a couple of generalizations of the proposed surrogate model: application to risk-aversion and an extension to multi-phase materials.

\subsection{Risk-aversion}
The optimization algorithm can be used not only for calibration of a surrogate model, but also for material design, i.e., for construction of a surrogate material with desired mechanical properties.
In this case, the expectations of the relevant quantities of interest can be used as the statistical descriptors.
Moreover, in the context of material design there naturally arises the risk aversion problem, when it is more important to minimize the probability of extreme values (tail of the distribution) than the expectation.	
In many industrial applications, a common choice for such risk measures is the so-called \textit{conditional value-at-risk} $\CVaR$ \cite{rockafellar2015engineering}:	
\begin{equation}\label{eq:CVaR1}
\CVaR_{\beta}(\xi) 
:= \E\left[\xi \;\big|\; \xi>\VaR_{\beta}(\xi)\right],
\end{equation}
where $\xi$ is a random variable, and the \textit{value-at-risk} $\VaR_{\beta}(\xi)$ is the $\beta$-quantile of $\xi$, $0<\beta<1$.
For example, let the random variable $\xi$ denote a stress measure of a structure (e.g., the maximum of von Mises stress).
When high values correspond to plastification or failure, lower values of $\xi$ are preferable.
The quantile~$\VaR_{\beta}(\xi)$ represents the most optimistic state which can be achieved in the worst $(1-\beta)\cdot 100\%$ of possible events, while $\CVaR_{\beta}(\xi)$ represents the expected value of~$\xi$ in these events.	
It is known (see \cite{rockafellar2000optimization}) that $\CVaR$ can be written via a scalar minimization problem:
\begin{equation}\label{eq:CVaR3}
\CVaR_{\beta}(\xi) = \inf\limits_{q\in\R}\E\left[q + \frac{1}{1-\beta}(\xi-q)_+\right],
\end{equation}
where $(\,\cdot\,)_+:=\max(0,\cdot)$.
Therefore, minimization of $\CVaR$ can be written in the form~\eqref{eq:minJ} with an additional parameter~$q$, which converges to the $\beta$-quantile.

\subsection{Multi-phase fields}
\begin{figure}[!ht]
	\centering
	\begin{subfigure}[b]{0.27\textwidth}
		\centering
		\includegraphics[width=\textwidth]{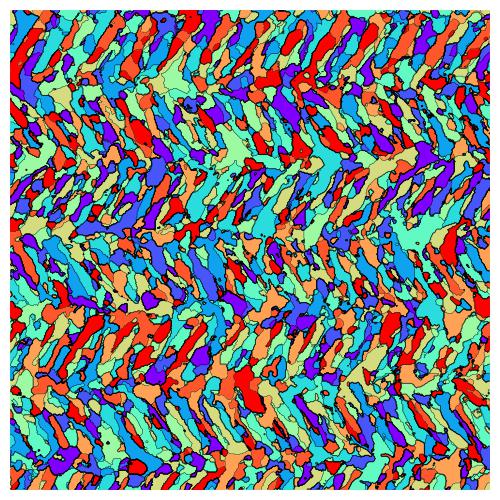}
		\caption{Synthetic grain texture}\label{sbfig:surrogate_grains}
	\end{subfigure}
	\hspace{8ex}
	\begin{subfigure}[b]{0.6\textwidth}
		\centering
		\includegraphics[width=\textwidth]{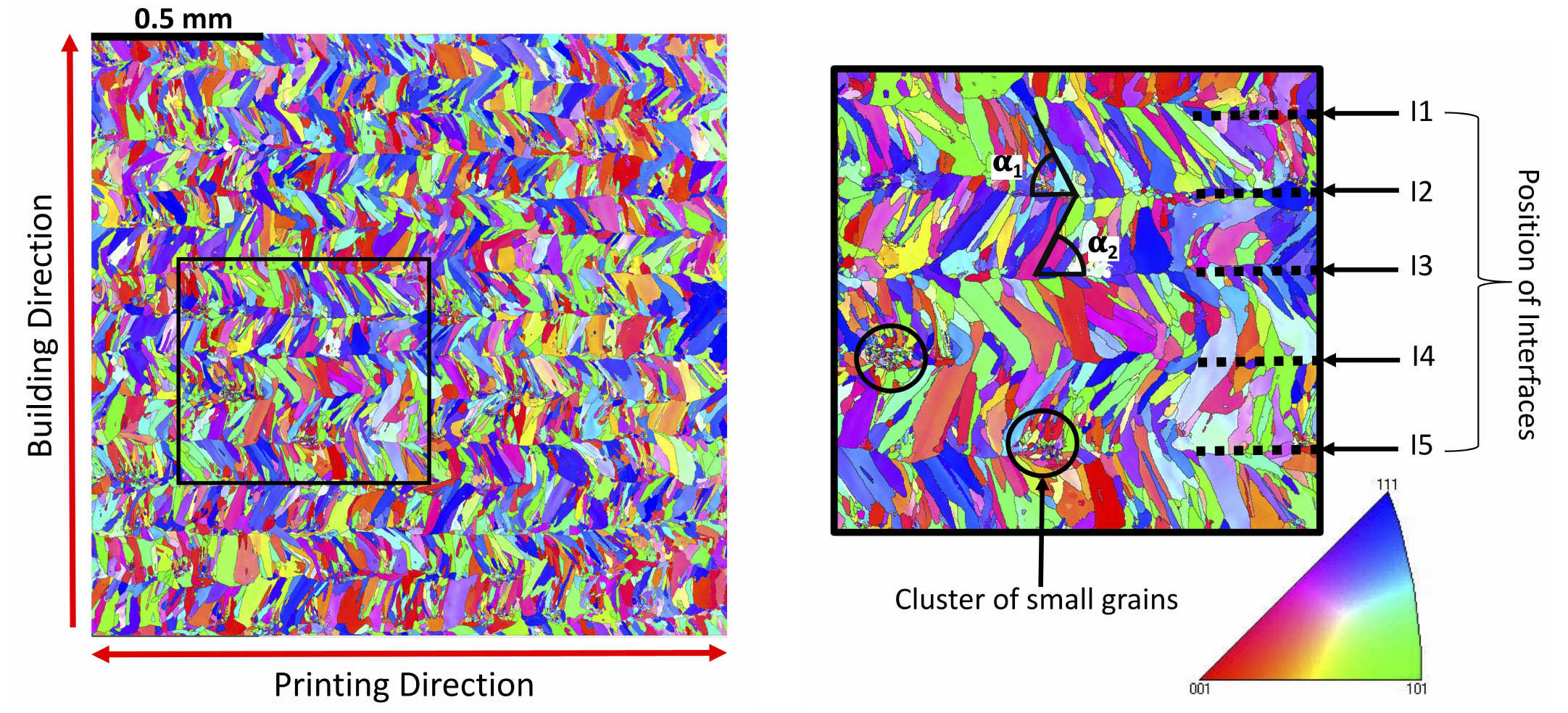}
		\caption{Real grain texture}\label{sbfig:real_grains}
	\end{subfigure}
	\caption{
		(\subref{sbfig:surrogate_grains}) Example of a synthetic multi-phase sample imitating polycrystalline grain texture in a manufactured material (not calibrated design parameters, $10$ phases).
		(\subref{sbfig:real_grains}) Electron Backscatter Diffraction of a specimen from bidirectionally-printed single-track thickness 316L stainless steel wall built by directed energy deposition (taken from~\cite{balit2020digital}), and a zoom showing the interfaces between layers, cluster of small grains at interfaces and the morphological grain angles for both direction of printing.
	}
	\label{fig:grains}
\end{figure}

As further extension of the discussed surrogate model, one can consider a multi-phase field:
\begin{equation}\label{eq:multiphase}
\Phase(\x) = \arg\max\ten{W}\vec{\Int}(\x),
\end{equation}
where $\vec{\Int}=\{\Int_1,\ldots,\Int_n\}$ is a vector random field, and $\ten{W}$ is a matrix of weights, such that $\ten{W}\vec{\Int}$ is a vector of intensities corresponding to $n$ different phases.
The \textit{argmax} activation returns the index of the most intense phase at the point.
Its smooth approximation is based on the \textit{softmax} function.
We remark that for $n=2$, $\vec{\Int}=\{\Int_1,\Int_2\}$ and $\ten{W}$ being diagonal with the entries $\alpha-1$ and $\alpha$, we recover the hybrid model \eqref{eq:levelcut}-\eqref{eq:HybridIntensity}.
A study of this extended model is a possible direction for future work.
Here, we only provide a 2D example in~\Cref{fig:grains} in order to outline the potential of this model in application to simulation of the morphologic and crystallographic textures of the polycrystalline grains in the additively manufactured materials~\cite{balit2020digital,balit2020high}.

	\section{Conclusion}\label{sec:Conclusion}

We proposed a surrogate material model which combines the topological shape and random imperfections.
Given in a unified form, the model is able to reproduce a variety of imperfect structures: porous media, fibers, cracks, lattice structures.
We used an imperfect octet-truss lattice cell for an illustration of the surrogate model application.
From a small amount of the real samples, we calibrated the model design parameters minimizing the misfit between the corresponding geometrical descriptors.
The implementation can benefit from algorithmic differentiation and progressive batching techniques.

Using the calibrated surrogate model, we are able to generate synthetic material samples and to use them in Monte Carlo type methods for uncertainty quantification of the material response.
We illustrated the applicability of the surrogate model in estimating the expectation and the deviation of the effective tangent modulus, as well as in investigating the sensitivity of the effective modulus to the imperfections.
Note that more sophisticated estimators of the probability distributions of QoIs can also be considered.
For example, by computing a number of statistical moments of QoIs, one can construct an approximation of the associated probability distribution function using a moment-matching method (such as the maximum entropy principle).

Note also that we used only geometrical descriptors obtained from the target material samples to calibrate the model and to use it consequently to sample the effective material properties.
In the general case, the effective moduli themselves can be also used as descriptors.
That is, in addition to the geometrical descriptors, one can use \textit{mechanical descriptors}, such as effective material moduli, maximum stress, etc..
In this case, while the mechanical descriptors of the surrogate samples have to be computed numerically, the corresponding target descriptor $\SD_i^*$ can be provided by experimental measurements.

Since the as-designed and as-manufactured materials have quite different effective properties, it is extremely important to take imperfections into account in the material design process.
Thus, the surrogate material model proposed in this work is shown to be a promising approach for numerical analysis of the impact of imperfections in the manufactured materials.


	\section*{Acknowledgments}
	UK and BW were supported by the German Research Foundation by grants WO671/11-1 and the European Union's Horizon 2020 research and innovation programme under grant agreement No 800898.
	AC and PLT were partly supported by the Andr\'e Citro\"en Chair.	
	We want to express our thanks to Jerome Hosdez and Nikolai Khailov for providing the image samples, and to Brendan Keith and Florian Beiser for helpful discussions on the optimization algorithm.


	\appendix

\section{Sub-sampled quasi-Newton minimization}
\label{sec:BFGS}

In this appendix, we present the details of the sub-sampled quasi-Newton minimization algorithm used for calibration of the surrogate model.
The method is proposed in~\cite{bollapragada2018progressive}.
We also refer the interested reader to~\cite{bollapragada2018adaptive,bollapragada2019exact,roosta2019sub,xie2020constrained}.

Let us consider a minimization problem in the form
\begin{equation}
\min\limits_{\theta} \bar{\loss}(\theta) = \E_{\omega}\left[\loss(\theta; \omega)\right].
\end{equation}
We define each particular sample with a random seed $\omega\in\mathcal{S} = \N\cap[0,10^3]$ of the pseudo random generator used to sample the uncertainties.
The sub-sampled quasi-Newton method iteratively computes a sequence $\theta_k$ of approximations of the optimal vector of design parameters, using the update formula
\begin{equation}\label{key}
\theta_{k+1} = \theta_{k} - \alpha_k H_k \bar{g}_k^{S_k},
\end{equation}
where $\bar{g}_k^{S_k}$ is the batch (sub-sampled) gradient of $\bar{\loss}$:
\begin{equation}\label{key}
\bar{g}_k^{S_k} := \frac{1}{\abs{S_k}}\sum_{\omega\in S_k}g_k(\omega),
\qquad
g_k(\omega) := \nabla_{\theta} \loss(\theta_k; \omega),
\end{equation}
with a set (batch) of material samples (random seeds) $S_k$ of the size~$\abs{S_k}$.
Above, $H_k$ is a positive definite quasi-Newton operator -- an approximation of the inverse Hessian matrix.
Using the limited memory BFGS method~\cite{liu1989limited}, it is defined by the recursive formula
\begin{equation}\label{eq:BFGS}
H_{k+1} = \left(\Id - \frac{\Delta\theta_{k}\, \Delta g_{k}\tp}{\Delta\theta_{k}\tp \Delta g_{k}}\right) H_k \left(\Id - \frac{\Delta\theta_{k}\, \Delta g_{k}\tp}{\Delta\theta_{k}\tp \Delta g_{k}}\right) + \frac{\Delta\theta_{k}\, \Delta\theta_{k}\tp}{\Delta\theta_{k}\tp \Delta g_{k}},
\end{equation}
with the curvature pair $\Delta g_{k} := \bar{g}_{k+1}^{S_k} - \bar{g}_{k}^{S_k}$ and $\Delta\theta_{k} := \theta_{k+1} - \theta_{k}$.
A common approach to compute the operator $H_k$ is given by the L-BFGS two-loop recursion algorithm~\cite{wright1999numerical}.

The key idea of a progressive batching approach is that at each iteration, the sample size is selected such that it satisfies the inner-product quasi-Newton test~\cite{bollapragada2018adaptive}:
\begin{equation}\label{eq:InnerProductTest}
\abs{S_k}\ge b_k := \frac{\Var_{\omega\in S_k}[(g_k(\omega))\tp H_k^2 \bar{g}_k^{S_k}]}{\kappa^2 \norm{H_k \bar{g}_k^{S_k}}^4},
\end{equation}
with some scalar parameter~$\kappa$ (we use $\kappa=1$).
Here, $\Var$ denotes the sampled variance of some random variable $f_k$, given by
\begin{equation}\label{key}
\Var_{\omega\in S_k}[f_k(\omega)]
:=
\frac{1}{\abs{S_k}-1}\sum_{\omega\in S_k}\norm{f_k(\omega)-\bar{f}_k^{S_k}}^2.
\end{equation}
Whenever condition~\eqref{eq:InnerProductTest} is not satisfied, the sample size~$\abs{S_k}$ is increased.

To find the steplength~$\alpha_k$, we perform a backtracking line search, satisfying the Armijo condition (see also~\cite{bollapragada2018progressive}):
\begin{equation}\label{eq:ArmijoTest}
 \bar{\loss}^{S_k}(\theta_{k} - \alpha_k H_k \bar{g}_k^{S_k})
\le
\bar{\loss}^{S_k}(\theta_{k}) - c_1\alpha_k\, [\bar{g}_k^{S_k}]\tp H_k \bar{g}_k^{S_k},
\end{equation}
where $\bar{\loss}^{S_k} := \frac{1}{\abs{S_k}}\sum_{\omega\in S_k}\loss(\theta_k; \omega)$ and the constant $c_1=10^{-4}$.
The initial steplength is $\alpha_k = \left(1 + \frac{\Var_{S_k}[\bar{g}_k]}{\abs{S_k}\cdot \norm{\bar{g}_k^{S_k}}
}\right)^{-1}$.
The full procedure is schematically outlined in~\Cref{alg:LBFGS}.

\begin{algorithm2e}[!t]
	\SetAlgoLined
	\SetKwInOut{Input}{Input}
	\KwResult{Approximated optimal design parmaters~$\theta_{opt}$}
	\Input{initial guess $\theta_0$, initial sample size $\abs{S_0}$}
	\Repeat{convergence}
	{
		Sample batch $S_k\subset\mathcal{S}$ of the size~$\abs{S_{k}}$\;
		Compute the bound $b_k$ by~\eqref{eq:InnerProductTest} \;
		\If{$\abs{S_k} < b_k$}
		{
			Sample $\Delta S_k\subset\mathcal{S}\setminus S_k$ with $\abs{\Delta S_k}=\ceil{b_k}-\abs{S_k}$  \;
			Set $S_k=S_k\cup \Delta S_k$ \;
		}
		Compute $\bar{g}_k^{S_k}$ \;
		Compute $p_k = - H_k \bar{g}_k^{S_k}$ using two-loop recursion\;
		Compute initial steplength $\alpha_k$ \;
		\While{the Armijo condition~\eqref{eq:ArmijoTest} is not satisfied}
		{
			Set $\alpha_k=\alpha_k/2$ \;
		}
		Update $\theta_{k+1} = \theta_{k} + \alpha_k p_k$ \;
		Update the curvature pairs $(\Delta g_{j}, \Delta\theta_{j})$ \;
		Set $\abs{S_{k+1}} = \abs{S_k}$ \;
		Increment $k$ \;
	}
	\caption{Sub-sampled L-BFGS}
	\label{alg:LBFGS}
\end{algorithm2e}

	\bibliographystyle{elsarticle-harv}

	\bibliography{references.bib}

\begin{thebibliography}{81}
\expandafter\ifx\csname natexlab\endcsname\relax\def\natexlab#1{#1}\fi
\providecommand{\url}[1]{\texttt{#1}}
\providecommand{\href}[2]{#2}
\providecommand{\path}[1]{#1}
\providecommand{\DOIprefix}{doi:}
\providecommand{\ArXivprefix}{arXiv:}
\providecommand{\URLprefix}{URL: }
\providecommand{\Pubmedprefix}{pmid:}
\providecommand{\doi}[1]{\href{http://dx.doi.org/#1}{\path{#1}}}
\providecommand{\Pubmed}[1]{\href{pmid:#1}{\path{#1}}}
\providecommand{\bibinfo}[2]{#2}
\ifx\xfnm\relax \def\xfnm[#1]{\unskip,\space#1}\fi
\bibitem[{Abrahamsen et~al.(2018)Abrahamsen, Kvernelv and
  Barker}]{abrahamsen2018simulation}
\bibinfo{author}{Abrahamsen, P.}, \bibinfo{author}{Kvernelv, V.},
  \bibinfo{author}{Barker, D.}, \bibinfo{year}{2018}.
\newblock \bibinfo{title}{{Simulation of Gaussian random fields using the Fast
  Fourier Transform (FFT)}}, in: \bibinfo{booktitle}{ECMOR XVI-16th European
  Conference on the Mathematics of Oil Recovery},
  \bibinfo{organization}{European Association of Geoscientists \& Engineers}.
  pp. \bibinfo{pages}{1--14}.
\bibitem[{Abramowitz and Stegun(1965)}]{abramowitz1965handbook}
\bibinfo{author}{Abramowitz, M.}, \bibinfo{author}{Stegun, I.A.},
  \bibinfo{year}{1965}.
\newblock \bibinfo{title}{Handbook of mathematical functions: with formulas,
  graphs, and mathematical tables}. volume~\bibinfo{volume}{55}.
\newblock \bibinfo{publisher}{Courier Corporation}.
\bibitem[{{AddUp Global Additive Solutions}({FormUp 350
  3D-printer})}]{solutions2017formup}
\bibinfo{author}{{AddUp Global Additive Solutions}}, \bibinfo{year}{{FormUp 350
  3D-printer}}.
\newblock \bibinfo{howpublished}{\url{https://addupsolutions.com}}.
\bibitem[{Allaire et~al.(2014)Allaire, Dapogny, Delgado and
  Michailidis}]{allaire2014multi}
\bibinfo{author}{Allaire, G.}, \bibinfo{author}{Dapogny, C.},
  \bibinfo{author}{Delgado, G.}, \bibinfo{author}{Michailidis, G.},
  \bibinfo{year}{2014}.
\newblock \bibinfo{title}{Multi-phase structural optimization via a level set
  method}.
\newblock \bibinfo{journal}{ESAIM: control, optimisation and calculus of
  variations} \bibinfo{volume}{20}, \bibinfo{pages}{576--611}.
\bibitem[{Allaire et~al.(2002)Allaire, Jouve and Toader}]{allaire2002level}
\bibinfo{author}{Allaire, G.}, \bibinfo{author}{Jouve, F.},
  \bibinfo{author}{Toader, A.M.}, \bibinfo{year}{2002}.
\newblock \bibinfo{title}{A level-set method for shape optimization}.
\newblock \bibinfo{journal}{Comptes Rendus Mathematique} \bibinfo{volume}{334},
  \bibinfo{pages}{1125--1130}.
\bibitem[{Allaire et~al.(2004)Allaire, Jouve and
  Toader}]{allaire2004structural}
\bibinfo{author}{Allaire, G.}, \bibinfo{author}{Jouve, F.},
  \bibinfo{author}{Toader, A.M.}, \bibinfo{year}{2004}.
\newblock \bibinfo{title}{Structural optimization using sensitivity analysis
  and a level-set method}.
\newblock \bibinfo{journal}{Journal of computational physics}
  \bibinfo{volume}{194}, \bibinfo{pages}{363--393}.
\bibitem[{Balit et~al.(2020a)Balit, Charkaluk and
  Constantinescu}]{balit2020digital}
\bibinfo{author}{Balit, Y.}, \bibinfo{author}{Charkaluk, E.},
  \bibinfo{author}{Constantinescu, A.}, \bibinfo{year}{2020}a.
\newblock \bibinfo{title}{Digital image correlation for microstructural
  analysis of deformation pattern in additively manufactured 316l thin walls}.
\newblock \bibinfo{journal}{Additive Manufacturing} \bibinfo{volume}{31},
  \bibinfo{pages}{100862}.
\bibitem[{Balit et~al.(2020b)Balit, Gu{\'e}venoux, Tanguy, Upadhyay, Charkaluk
  and Constantinescu}]{balit2020high}
\bibinfo{author}{Balit, Y.}, \bibinfo{author}{Gu{\'e}venoux, C.},
  \bibinfo{author}{Tanguy, A.}, \bibinfo{author}{Upadhyay, M.V.},
  \bibinfo{author}{Charkaluk, E.}, \bibinfo{author}{Constantinescu, A.},
  \bibinfo{year}{2020}b.
\newblock \bibinfo{title}{High resolution digital image correlation for
  microstructural strain analysis of a stainless steel repaired by directed
  energy deposition}.
\newblock \bibinfo{journal}{Materials Letters} \bibinfo{volume}{270},
  \bibinfo{pages}{127632}.
\bibitem[{Balit et~al.(2021)Balit, Margerit, Charkaluk and
  Constantinescu}]{balit2021crushing}
\bibinfo{author}{Balit, Y.}, \bibinfo{author}{Margerit, P.},
  \bibinfo{author}{Charkaluk, E.}, \bibinfo{author}{Constantinescu, A.},
  \bibinfo{year}{2021}.
\newblock \bibinfo{title}{Crushing of additively manufactured thin-walled
  metallic lattices: Two-scale strain localization analysis}.
\newblock \bibinfo{journal}{Mechanics of Materials} , \bibinfo{pages}{103915}.
\bibitem[{Bateman(1954)}]{bateman1954tables}
\bibinfo{author}{Bateman, H.}, \bibinfo{year}{1954}.
\newblock \bibinfo{title}{Tables of integral transforms}.
\newblock \bibinfo{publisher}{McGraw-Hill, N.Y.}
\bibitem[{Beiser et~al.(2020)Beiser, Keith, Urbainczyk and
  Wohlmuth}]{beiser2020adaptive}
\bibinfo{author}{Beiser, F.}, \bibinfo{author}{Keith, B.},
  \bibinfo{author}{Urbainczyk, S.}, \bibinfo{author}{Wohlmuth, B.},
  \bibinfo{year}{2020}.
\newblock \bibinfo{title}{Adaptive sampling strategies for risk-averse
  stochastic optimization with constraints}.
\newblock \bibinfo{journal}{arXiv preprint arXiv:2012.03844} .
\bibitem[{Bessa et~al.(2017)Bessa, Bostanabad, Liu, Hu, Apley, Brinson, Chen
  and Liu}]{bessa2017framework}
\bibinfo{author}{Bessa, M.}, \bibinfo{author}{Bostanabad, R.},
  \bibinfo{author}{Liu, Z.}, \bibinfo{author}{Hu, A.}, \bibinfo{author}{Apley,
  D.W.}, \bibinfo{author}{Brinson, C.}, \bibinfo{author}{Chen, W.},
  \bibinfo{author}{Liu, W.K.}, \bibinfo{year}{2017}.
\newblock \bibinfo{title}{A framework for data-driven analysis of materials
  under uncertainty: Countering the curse of dimensionality}.
\newblock \bibinfo{journal}{Computer Methods in Applied Mechanics and
  Engineering} \bibinfo{volume}{320}, \bibinfo{pages}{633--667}.
\bibitem[{Boittin et~al.()Boittin, Lab{\'e}, Moulinec, Silva and
  Suquet}]{CraFT}
\bibinfo{author}{Boittin, G.}, \bibinfo{author}{Lab{\'e}, A.},
  \bibinfo{author}{Moulinec, H.}, \bibinfo{author}{Silva, F.},
  \bibinfo{author}{Suquet, P.}, .
\newblock \bibinfo{title}{{CraFT ("Composite response and Fourier
  Transforms")}}.
\newblock \bibinfo{howpublished}{\url{https://lma-software-craft.cnrs.fr}}.
\bibitem[{Bollapragada et~al.(2018a)Bollapragada, Byrd and
  Nocedal}]{bollapragada2018adaptive}
\bibinfo{author}{Bollapragada, R.}, \bibinfo{author}{Byrd, R.},
  \bibinfo{author}{Nocedal, J.}, \bibinfo{year}{2018}a.
\newblock \bibinfo{title}{Adaptive sampling strategies for stochastic
  optimization}.
\newblock \bibinfo{journal}{SIAM Journal on Optimization} \bibinfo{volume}{28},
  \bibinfo{pages}{3312--3343}.
\bibitem[{Bollapragada et~al.(2019)Bollapragada, Byrd and
  Nocedal}]{bollapragada2019exact}
\bibinfo{author}{Bollapragada, R.}, \bibinfo{author}{Byrd, R.H.},
  \bibinfo{author}{Nocedal, J.}, \bibinfo{year}{2019}.
\newblock \bibinfo{title}{Exact and inexact subsampled newton methods for
  optimization}.
\newblock \bibinfo{journal}{IMA Journal of Numerical Analysis}
  \bibinfo{volume}{39}, \bibinfo{pages}{545--578}.
\bibitem[{Bollapragada et~al.(2018b)Bollapragada, Nocedal, Mudigere, Shi and
  Tang}]{bollapragada2018progressive}
\bibinfo{author}{Bollapragada, R.}, \bibinfo{author}{Nocedal, J.},
  \bibinfo{author}{Mudigere, D.}, \bibinfo{author}{Shi, H.J.},
  \bibinfo{author}{Tang, P.T.P.}, \bibinfo{year}{2018}b.
\newblock \bibinfo{title}{A progressive batching l-bfgs method for machine
  learning}, in: \bibinfo{booktitle}{International Conference on Machine
  Learning}, \bibinfo{organization}{PMLR}. pp. \bibinfo{pages}{620--629}.
\bibitem[{Byrd et~al.(2012)Byrd, Chin, Nocedal and Wu}]{byrd2012sample}
\bibinfo{author}{Byrd, R.H.}, \bibinfo{author}{Chin, G.M.},
  \bibinfo{author}{Nocedal, J.}, \bibinfo{author}{Wu, Y.},
  \bibinfo{year}{2012}.
\newblock \bibinfo{title}{Sample size selection in optimization methods for
  machine learning}.
\newblock \bibinfo{journal}{Mathematical programming} \bibinfo{volume}{134},
  \bibinfo{pages}{127--155}.
\bibitem[{De~Oliveira(2000)}]{de2000bayesian}
\bibinfo{author}{De~Oliveira, V.}, \bibinfo{year}{2000}.
\newblock \bibinfo{title}{Bayesian prediction of clipped gaussian random
  fields}.
\newblock \bibinfo{journal}{Computational Statistics \& Data Analysis}
  \bibinfo{volume}{34}, \bibinfo{pages}{299--314}.
\bibitem[{DebRoy et~al.(2021)DebRoy, Mukherjee, Wei, Elmer and
  Milewski}]{debroy2021metallurgy}
\bibinfo{author}{DebRoy, T.}, \bibinfo{author}{Mukherjee, T.},
  \bibinfo{author}{Wei, H.}, \bibinfo{author}{Elmer, J.},
  \bibinfo{author}{Milewski, J.}, \bibinfo{year}{2021}.
\newblock \bibinfo{title}{Metallurgy, mechanistic models and machine learning
  in metal printing}.
\newblock \bibinfo{journal}{Nature Reviews Materials} \bibinfo{volume}{6},
  \bibinfo{pages}{48--68}.
\bibitem[{Deshpande et~al.(2001)Deshpande, Fleck and
  Ashby}]{deshpande2001effective}
\bibinfo{author}{Deshpande, V.S.}, \bibinfo{author}{Fleck, N.A.},
  \bibinfo{author}{Ashby, M.F.}, \bibinfo{year}{2001}.
\newblock \bibinfo{title}{Effective properties of the octet-truss lattice
  material}.
\newblock \bibinfo{journal}{Journal of the Mechanics and Physics of Solids}
  \bibinfo{volume}{49}, \bibinfo{pages}{1747--1769}.
\bibitem[{Gandy and Klinowski(2000)}]{gandy2000exact}
\bibinfo{author}{Gandy, P.J.}, \bibinfo{author}{Klinowski, J.},
  \bibinfo{year}{2000}.
\newblock \bibinfo{title}{Exact computation of the triply periodic g (gyroid')
  minimal surface}.
\newblock \bibinfo{journal}{Chemical Physics Letters} \bibinfo{volume}{321},
  \bibinfo{pages}{363--371}.
\bibitem[{Garland et~al.(2020)Garland, White, Jared, Heiden, Donahue and
  Boyce}]{garland2020deep}
\bibinfo{author}{Garland, A.P.}, \bibinfo{author}{White, B.C.},
  \bibinfo{author}{Jared, B.H.}, \bibinfo{author}{Heiden, M.},
  \bibinfo{author}{Donahue, E.}, \bibinfo{author}{Boyce, B.L.},
  \bibinfo{year}{2020}.
\newblock \bibinfo{title}{Deep convolutional neural networks as a rapid
  screening tool for complex additively manufactured structures}.
\newblock \bibinfo{journal}{Additive Manufacturing} \bibinfo{volume}{35},
  \bibinfo{pages}{101217}.
\bibitem[{Gavazzoni et~al.(2022)Gavazzoni, Foletti and
  Pasini}]{gavazzoni2022cyclic}
\bibinfo{author}{Gavazzoni, M.}, \bibinfo{author}{Foletti, S.},
  \bibinfo{author}{Pasini, D.}, \bibinfo{year}{2022}.
\newblock \bibinfo{title}{Cyclic response of 3d printed metamaterials with soft
  cellular architecture: The interplay between as-built defects, material and
  geometric non-linearity}.
\newblock \bibinfo{journal}{Journal of the Mechanics and Physics of Solids}
  \bibinfo{volume}{158}, \bibinfo{pages}{104688}.
\bibitem[{Gayon-Lombardo et~al.(2020)Gayon-Lombardo, Mosser, Brandon and
  Cooper}]{gayon2020pores}
\bibinfo{author}{Gayon-Lombardo, A.}, \bibinfo{author}{Mosser, L.},
  \bibinfo{author}{Brandon, N.P.}, \bibinfo{author}{Cooper, S.J.},
  \bibinfo{year}{2020}.
\newblock \bibinfo{title}{{Pores for thought: generative adversarial networks
  for stochastic reconstruction of 3D multi-phase electrode microstructures
  with periodic boundaries}}.
\newblock \bibinfo{journal}{npj Computational Materials} \bibinfo{volume}{6},
  \bibinfo{pages}{1--11}.
\bibitem[{Gneiting and Guttorp(2012)}]{gneiting2012studies}
\bibinfo{author}{Gneiting, T.}, \bibinfo{author}{Guttorp, P.},
  \bibinfo{year}{2012}.
\newblock \bibinfo{title}{{Studies in the history of probability and statistics
  XLIX On the Matern correlation family}}.
\newblock \bibinfo{journal}{Biometrika} \bibinfo{volume}{93},
  \bibinfo{pages}{989--995}.
\bibitem[{Hida et~al.(2013)Hida, Kuo, Potthoff and Streit}]{hida2013white}
\bibinfo{author}{Hida, T.}, \bibinfo{author}{Kuo, H.H.},
  \bibinfo{author}{Potthoff, J.}, \bibinfo{author}{Streit, L.},
  \bibinfo{year}{2013}.
\newblock \bibinfo{title}{White noise: an infinite dimensional calculus}.
  volume \bibinfo{volume}{253}.
\newblock \bibinfo{publisher}{Springer Science \& Business Media}.
\bibitem[{Hosdez et~al.(2019)Hosdez, Langlois, Witz, Limodin, Najjar,
  Charkaluk, Osmond, Forre and Szmytka}]{hosdez2019plastic}
\bibinfo{author}{Hosdez, J.}, \bibinfo{author}{Langlois, M.},
  \bibinfo{author}{Witz, J.}, \bibinfo{author}{Limodin, N.},
  \bibinfo{author}{Najjar, D.}, \bibinfo{author}{Charkaluk, E.},
  \bibinfo{author}{Osmond, P.}, \bibinfo{author}{Forre, A.},
  \bibinfo{author}{Szmytka, F.}, \bibinfo{year}{2019}.
\newblock \bibinfo{title}{Plastic zone evolution during fatigue crack growth:
  Digital image correlation coupled with finite elements method}.
\newblock \bibinfo{journal}{International Journal of Solids and Structures}
  \bibinfo{volume}{171}, \bibinfo{pages}{92--102}.
\bibitem[{Hosdez et~al.(2020)Hosdez, Limodin, Najjar, Witz, Charkaluk, Osmond,
  Forr{\'e} and Szmytka}]{hosdez2020fatigue}
\bibinfo{author}{Hosdez, J.}, \bibinfo{author}{Limodin, N.},
  \bibinfo{author}{Najjar, D.}, \bibinfo{author}{Witz, J.},
  \bibinfo{author}{Charkaluk, E.}, \bibinfo{author}{Osmond, P.},
  \bibinfo{author}{Forr{\'e}, A.}, \bibinfo{author}{Szmytka, F.},
  \bibinfo{year}{2020}.
\newblock \bibinfo{title}{Fatigue crack growth in compacted and spheroidal
  graphite cast irons}.
\newblock \bibinfo{journal}{International Journal of Fatigue}
  \bibinfo{volume}{131}, \bibinfo{pages}{105319}.
\bibitem[{Kak and Slaney(2001)}]{kak2001principles}
\bibinfo{author}{Kak, A.C.}, \bibinfo{author}{Slaney, M.},
  \bibinfo{year}{2001}.
\newblock \bibinfo{title}{Principles of computerized tomographic imaging}.
\newblock \bibinfo{publisher}{SIAM}.
\bibitem[{Keith et~al.(2021)Keith, Khristenko and
  Wohlmuth}]{keith2021fractional}
\bibinfo{author}{Keith, B.}, \bibinfo{author}{Khristenko, U.},
  \bibinfo{author}{Wohlmuth, B.}, \bibinfo{year}{2021}.
\newblock \bibinfo{title}{A fractional pde model for turbulent velocity fields
  near solid walls}.
\newblock \bibinfo{journal}{Journal of Fluid Mechanics} \bibinfo{volume}{916}.
\bibitem[{Khristenko et~al.(2019)Khristenko, Constantinescu, Le~Tallec, Oden
  and Wohlmuth}]{khristenko2019statistical}
\bibinfo{author}{Khristenko, U.}, \bibinfo{author}{Constantinescu, A.},
  \bibinfo{author}{Le~Tallec, P.}, \bibinfo{author}{Oden, J.T.},
  \bibinfo{author}{Wohlmuth, B.}, \bibinfo{year}{2019}.
\newblock \bibinfo{title}{A statistical framework for generating
  microstructures of two-phase random materials: application to fatigue
  analysis}.
\newblock \bibinfo{journal}{arXiv preprint arXiv:1907.02412} .
\bibitem[{Korshunova et~al.(2021a)Korshunova, Alaimo, Hosseini, Carraturo,
  Reali, Niiranen, Auricchio, Rank and Kollmannsberger}]{korshunova2021bending}
\bibinfo{author}{Korshunova, N.}, \bibinfo{author}{Alaimo, G.},
  \bibinfo{author}{Hosseini, S.}, \bibinfo{author}{Carraturo, M.},
  \bibinfo{author}{Reali, A.}, \bibinfo{author}{Niiranen, J.},
  \bibinfo{author}{Auricchio, F.}, \bibinfo{author}{Rank, E.},
  \bibinfo{author}{Kollmannsberger, S.}, \bibinfo{year}{2021}a.
\newblock \bibinfo{title}{Bending behavior of octet-truss lattice structures:
  Modelling options, numerical characterization and experimental validation}.
\newblock \bibinfo{journal}{Materials \& Design} \bibinfo{volume}{205},
  \bibinfo{pages}{109693}.
\bibitem[{Korshunova et~al.(2021b)Korshunova, Alaimo, Hosseini, Carraturo,
  Reali, Niiranen, Auricchio, Rank and Kollmannsberger}]{korshunova2021image}
\bibinfo{author}{Korshunova, N.}, \bibinfo{author}{Alaimo, G.},
  \bibinfo{author}{Hosseini, S.B.}, \bibinfo{author}{Carraturo, M.},
  \bibinfo{author}{Reali, A.}, \bibinfo{author}{Niiranen, J.},
  \bibinfo{author}{Auricchio, F.}, \bibinfo{author}{Rank, E.},
  \bibinfo{author}{Kollmannsberger, S.}, \bibinfo{year}{2021}b.
\newblock \bibinfo{title}{Image-based numerical characterization and
  experimental validation of tensile behavior of octet-truss lattice
  structures}.
\newblock \bibinfo{journal}{Additive Manufacturing} , \bibinfo{pages}{101949}.
\bibitem[{Korshunova et~al.(2021c)Korshunova, Papaioannou, Kollmannsberger,
  Straub and Rank}]{korshunova2021uncertainty}
\bibinfo{author}{Korshunova, N.}, \bibinfo{author}{Papaioannou, I.},
  \bibinfo{author}{Kollmannsberger, S.}, \bibinfo{author}{Straub, D.},
  \bibinfo{author}{Rank, E.}, \bibinfo{year}{2021}c.
\newblock \bibinfo{title}{Uncertainty quantification of microstructure
  variability and mechanical behaviour of additively manufactured lattice
  structures}.
\newblock \bibinfo{journal}{arXiv preprint arXiv:2103.09550} .
\bibitem[{Koutsourelakis and Deodatis(2006)}]{koutsourelakis2006simulation}
\bibinfo{author}{Koutsourelakis, P.S.}, \bibinfo{author}{Deodatis, G.},
  \bibinfo{year}{2006}.
\newblock \bibinfo{title}{Simulation of multidimensional binary random fields
  with application to modeling of two-phase random media}.
\newblock \bibinfo{journal}{Journal of engineering mechanics}
  \bibinfo{volume}{132}, \bibinfo{pages}{619--631}.
\bibitem[{Kuo(2018)}]{kuo2018white}
\bibinfo{author}{Kuo, H.H.}, \bibinfo{year}{2018}.
\newblock \bibinfo{title}{White noise distribution theory}.
\newblock \bibinfo{publisher}{CRC press}.
\bibitem[{Lantu{\'e}joul(2001)}]{lantuejoul2001geostatistical}
\bibinfo{author}{Lantu{\'e}joul, C.}, \bibinfo{year}{2001}.
\newblock \bibinfo{title}{Geostatistical simulation: models and algorithms}.
\newblock \bibinfo{number}{1139}, \bibinfo{publisher}{Springer Science \&
  Business Media}.
\bibitem[{Le~Ravalec et~al.(2000)Le~Ravalec, Noetinger and Hu}]{le2000fft}
\bibinfo{author}{Le~Ravalec, M.}, \bibinfo{author}{Noetinger, B.},
  \bibinfo{author}{Hu, L.Y.}, \bibinfo{year}{2000}.
\newblock \bibinfo{title}{{The FFT moving average (FFT-MA) generator: An
  efficient numerical method for generating and conditioning Gaussian
  simulations}}.
\newblock \bibinfo{journal}{Mathematical Geology} \bibinfo{volume}{32},
  \bibinfo{pages}{701--723}.
\bibitem[{Limodin et~al.(2013)Limodin, Rougelot and Hauss}]{limodin2013isis4d}
\bibinfo{author}{Limodin, N.}, \bibinfo{author}{Rougelot, T.},
  \bibinfo{author}{Hauss, G.}, \bibinfo{year}{2013}.
\newblock \bibinfo{title}{Isis4d-in situ innovative set-ups under x-ray
  microtomography}.
\newblock \bibinfo{howpublished}{\url{http://isis4d.univ-lille1.fr}}.
\bibitem[{Lin and Clayton(2005)}]{lin2005properties}
\bibinfo{author}{Lin, P.S.}, \bibinfo{author}{Clayton, M.K.},
  \bibinfo{year}{2005}.
\newblock \bibinfo{title}{Properties of binary data generated from a truncated
  gaussian random field}.
\newblock \bibinfo{journal}{Communications in Statistics—Theory and Methods}
  \bibinfo{volume}{34}, \bibinfo{pages}{537--544}.
\bibitem[{Lindgren et~al.(2021)Lindgren, Bolin and Rue}]{lindgren2021spde}
\bibinfo{author}{Lindgren, F.}, \bibinfo{author}{Bolin, D.},
  \bibinfo{author}{Rue, H.}, \bibinfo{year}{2021}.
\newblock \bibinfo{title}{The spde approach for gaussian and non-gaussian
  fields: 10 years and still running}.
\newblock \href{http://arxiv.org/abs/2111.01084}{{\tt arXiv:2111.01084}}.
\bibitem[{Liu and Nocedal(1989)}]{liu1989limited}
\bibinfo{author}{Liu, D.C.}, \bibinfo{author}{Nocedal, J.},
  \bibinfo{year}{1989}.
\newblock \bibinfo{title}{On the limited memory bfgs method for large scale
  optimization}.
\newblock \bibinfo{journal}{Mathematical programming} \bibinfo{volume}{45},
  \bibinfo{pages}{503--528}.
\bibitem[{Liu et~al.(2017)Liu, Kamm, Garcia-Moreno, Banhart and
  Pasini}]{liu2017elastic}
\bibinfo{author}{Liu, L.}, \bibinfo{author}{Kamm, P.},
  \bibinfo{author}{Garcia-Moreno, F.}, \bibinfo{author}{Banhart, J.},
  \bibinfo{author}{Pasini, D.}, \bibinfo{year}{2017}.
\newblock \bibinfo{title}{Elastic and failure response of imperfect
  three-dimensional metallic lattices: the role of geometric defects induced by
  selective laser melting}.
\newblock \bibinfo{journal}{Journal of the Mechanics and Physics of Solids}
  \bibinfo{volume}{107}, \bibinfo{pages}{160--184}.
\bibitem[{Mantz et~al.(2008)Mantz, Jacobs and Mecke}]{mantz2008utilizing}
\bibinfo{author}{Mantz, H.}, \bibinfo{author}{Jacobs, K.},
  \bibinfo{author}{Mecke, K.}, \bibinfo{year}{2008}.
\newblock \bibinfo{title}{{Utilizing Minkowski functionals for image analysis:
  a marching square algorithm}}.
\newblock \bibinfo{journal}{Journal of Statistical Mechanics: Theory and
  Experiment} \bibinfo{volume}{2008}, \bibinfo{pages}{P12015}.
\bibitem[{Mat{\'{e}}rn(1986)}]{matern1986spatial}
\bibinfo{author}{Mat{\'{e}}rn, B.}, \bibinfo{year}{1986}.
\newblock \bibinfo{title}{{Spatial Variation}}. volume~\bibinfo{volume}{36} of
  \textit{\bibinfo{series}{Lecture Notes in Statistics}}.
\newblock \bibinfo{publisher}{Springer New York}, \bibinfo{address}{New York,
  NY}.
\bibitem[{Michel et~al.(2001)Michel, Moulinec and
  Suquet}]{michel2001computational}
\bibinfo{author}{Michel, J.}, \bibinfo{author}{Moulinec, H.},
  \bibinfo{author}{Suquet, P.}, \bibinfo{year}{2001}.
\newblock \bibinfo{title}{A computational scheme for linear and non-linear
  composites with arbitrary phase contrast}.
\newblock \bibinfo{journal}{International Journal for Numerical Methods in
  Engineering} \bibinfo{volume}{52}, \bibinfo{pages}{139--160}.
\bibitem[{Minasny and McBratney(2005)}]{minasny2005matern}
\bibinfo{author}{Minasny, B.}, \bibinfo{author}{McBratney, A.B.},
  \bibinfo{year}{2005}.
\newblock \bibinfo{title}{{The Mat\'ern function as a general model for soil
  variograms}}.
\newblock \bibinfo{journal}{Geoderma} \bibinfo{volume}{128},
  \bibinfo{pages}{192--207}.
\bibitem[{Monchiet and Bonnet(2012)}]{monchiet2012polarization}
\bibinfo{author}{Monchiet, V.}, \bibinfo{author}{Bonnet, G.},
  \bibinfo{year}{2012}.
\newblock \bibinfo{title}{A polarization-based fft iterative scheme for
  computing the effective properties of elastic composites with arbitrary
  contrast}.
\newblock \bibinfo{journal}{International Journal for Numerical Methods in
  Engineering} \bibinfo{volume}{89}, \bibinfo{pages}{1419--1436}.
\bibitem[{Moulinec and Suquet(1994)}]{moulinec1994fast}
\bibinfo{author}{Moulinec, H.}, \bibinfo{author}{Suquet, P.},
  \bibinfo{year}{1994}.
\newblock \bibinfo{title}{A fast numerical method for computing the linear and
  nonlinear mechanical properties of composites}.
\newblock \bibinfo{journal}{Comptes rendus de l'Acad{\'e}mie des sciences.
  S{\'e}rie II. M{\'e}canique, physique, chimie, astronomie.} .
\bibitem[{Moulinec and Suquet(1998)}]{moulinec1998numerical}
\bibinfo{author}{Moulinec, H.}, \bibinfo{author}{Suquet, P.},
  \bibinfo{year}{1998}.
\newblock \bibinfo{title}{A numerical method for computing the overall response
  of nonlinear composites with complex microstructure}.
\newblock \bibinfo{journal}{Computer methods in applied mechanics and
  engineering} \bibinfo{volume}{157}, \bibinfo{pages}{69--94}.
\bibitem[{Moussa et~al.(2021)Moussa, Melancon, El~Elmi and
  Pasini}]{moussa2021topology}
\bibinfo{author}{Moussa, A.}, \bibinfo{author}{Melancon, D.},
  \bibinfo{author}{El~Elmi, A.}, \bibinfo{author}{Pasini, D.},
  \bibinfo{year}{2021}.
\newblock \bibinfo{title}{Topology optimization of imperfect lattice materials
  built with process-induced defects via powder bed fusion}.
\newblock \bibinfo{journal}{Additive Manufacturing} \bibinfo{volume}{37},
  \bibinfo{pages}{101608}.
\bibitem[{Nika and Constantinescu(2019)}]{nika2019design}
\bibinfo{author}{Nika, G.}, \bibinfo{author}{Constantinescu, A.},
  \bibinfo{year}{2019}.
\newblock \bibinfo{title}{Design of multi-layer materials using inverse
  homogenization and a level set method}.
\newblock \bibinfo{journal}{Computer Methods in Applied Mechanics and
  Engineering} \bibinfo{volume}{346}, \bibinfo{pages}{388--409}.
\bibitem[{Ogorodnikov et~al.(2018)Ogorodnikov, Kablukova and
  Prigarin}]{ogorodnikov2018stochastic}
\bibinfo{author}{Ogorodnikov, V.A.}, \bibinfo{author}{Kablukova, E.G.},
  \bibinfo{author}{Prigarin, S.M.}, \bibinfo{year}{2018}.
\newblock \bibinfo{title}{Stochastic models of atmospheric clouds structure}.
\newblock \bibinfo{journal}{Statistical Papers} \bibinfo{volume}{59},
  \bibinfo{pages}{1521--1532}.
\bibitem[{Oliver(1995)}]{oliver1995moving}
\bibinfo{author}{Oliver, D.S.}, \bibinfo{year}{1995}.
\newblock \bibinfo{title}{Moving averages for gaussian simulation in two and
  three dimensions}.
\newblock \bibinfo{journal}{Mathematical Geology} \bibinfo{volume}{27},
  \bibinfo{pages}{939--960}.
\bibitem[{Pasini and Guest(2019)}]{pasini2019imperfect}
\bibinfo{author}{Pasini, D.}, \bibinfo{author}{Guest, J.K.},
  \bibinfo{year}{2019}.
\newblock \bibinfo{title}{Imperfect architected materials: Mechanics and
  topology optimization}.
\newblock \bibinfo{journal}{MRS Bulletin} \bibinfo{volume}{44},
  \bibinfo{pages}{766--772}.
\bibitem[{Paszke et~al.(2019)Paszke, Gross, Massa, Lerer, Bradbury, Chanan,
  Killeen, Lin, Gimelshein, Antiga, Desmaison, Kopf, Yang, DeVito, Raison,
  Tejani, Chilamkurthy, Steiner, Fang, Bai and Chintala}]{pytorch}
\bibinfo{author}{Paszke, A.}, \bibinfo{author}{Gross, S.},
  \bibinfo{author}{Massa, F.}, \bibinfo{author}{Lerer, A.},
  \bibinfo{author}{Bradbury, J.}, \bibinfo{author}{Chanan, G.},
  \bibinfo{author}{Killeen, T.}, \bibinfo{author}{Lin, Z.},
  \bibinfo{author}{Gimelshein, N.}, \bibinfo{author}{Antiga, L.},
  \bibinfo{author}{Desmaison, A.}, \bibinfo{author}{Kopf, A.},
  \bibinfo{author}{Yang, E.}, \bibinfo{author}{DeVito, Z.},
  \bibinfo{author}{Raison, M.}, \bibinfo{author}{Tejani, A.},
  \bibinfo{author}{Chilamkurthy, S.}, \bibinfo{author}{Steiner, B.},
  \bibinfo{author}{Fang, L.}, \bibinfo{author}{Bai, J.},
  \bibinfo{author}{Chintala, S.}, \bibinfo{year}{2019}.
\newblock \bibinfo{title}{{PyTorch: An Imperative Style, High-Performance Deep
  Learning Library}}, in: \bibinfo{editor}{Wallach, H.},
  \bibinfo{editor}{Larochelle, H.}, \bibinfo{editor}{Beygelzimer, A.},
  \bibinfo{editor}{d\textquotesingle Alch\'{e}-Buc, F.}, \bibinfo{editor}{Fox,
  E.}, \bibinfo{editor}{Garnett, R.} (Eds.), \bibinfo{booktitle}{Advances in
  Neural Information Processing Systems 32}. \bibinfo{publisher}{Curran
  Associates, Inc.}, pp. \bibinfo{pages}{8024--8035}.
\bibitem[{Qi et~al.(2019)Qi, Yu, Liu, Huang, Xu, Xia, Qian and
  Wu}]{qi2019mechanical}
\bibinfo{author}{Qi, D.}, \bibinfo{author}{Yu, H.}, \bibinfo{author}{Liu, M.},
  \bibinfo{author}{Huang, H.}, \bibinfo{author}{Xu, S.}, \bibinfo{author}{Xia,
  Y.}, \bibinfo{author}{Qian, G.}, \bibinfo{author}{Wu, W.},
  \bibinfo{year}{2019}.
\newblock \bibinfo{title}{Mechanical behaviors of {SLM} additive manufactured
  octet-truss and truncated-octahedron lattice structures with uniform and
  taper beams}.
\newblock \bibinfo{journal}{International Journal of Mechanical Sciences}
  \bibinfo{volume}{163}, \bibinfo{pages}{105091}.
\bibitem[{Rockafellar and Royset(2015)}]{rockafellar2015engineering}
\bibinfo{author}{Rockafellar, R.T.}, \bibinfo{author}{Royset, J.O.},
  \bibinfo{year}{2015}.
\newblock \bibinfo{title}{Engineering decisions under risk averseness}.
\newblock \bibinfo{journal}{ASCE-ASME Journal of Risk and Uncertainty in
  Engineering Systems, Part A: Civil Engineering} \bibinfo{volume}{1},
  \bibinfo{pages}{04015003}.
\bibitem[{Rockafellar et~al.(2000)Rockafellar, Uryasev
  et~al.}]{rockafellar2000optimization}
\bibinfo{author}{Rockafellar, R.T.}, \bibinfo{author}{Uryasev, S.}, et~al.,
  \bibinfo{year}{2000}.
\newblock \bibinfo{title}{Optimization of conditional value-at-risk}.
\newblock \bibinfo{journal}{Journal of risk} \bibinfo{volume}{2},
  \bibinfo{pages}{21--42}.
\bibitem[{Roininen et~al.(2014)Roininen, Huttunen and
  Lasanen}]{roininen2014whittle}
\bibinfo{author}{Roininen, L.}, \bibinfo{author}{Huttunen, J.M.},
  \bibinfo{author}{Lasanen, S.}, \bibinfo{year}{2014}.
\newblock \bibinfo{title}{{Whittle-Mat{\'e}rn priors for Bayesian statistical
  inversion with applications in electrical impedance tomography}}.
\newblock \bibinfo{journal}{Inverse Probl. Imaging} \bibinfo{volume}{8},
  \bibinfo{pages}{561--586}.
\bibitem[{Roosta-Khorasani and Mahoney(2019)}]{roosta2019sub}
\bibinfo{author}{Roosta-Khorasani, F.}, \bibinfo{author}{Mahoney, M.W.},
  \bibinfo{year}{2019}.
\newblock \bibinfo{title}{Sub-sampled newton methods}.
\newblock \bibinfo{journal}{Mathematical Programming} \bibinfo{volume}{174},
  \bibinfo{pages}{293--326}.
\bibitem[{Schoen(1970)}]{schoen1970infinite}
\bibinfo{author}{Schoen, A.H.}, \bibinfo{year}{1970}.
\newblock \bibinfo{title}{Infinite periodic minimal surfaces without
  self-intersections}.
\newblock \bibinfo{publisher}{National Aeronautics and Space Administration}.
\bibitem[{Shi and Mudigere()}]{PyTorch-LBFGS}
\bibinfo{author}{Shi, H.J.M.}, \bibinfo{author}{Mudigere, D.}, .
\newblock \bibinfo{title}{{PyTorch-LBFGS}}.
\newblock
  \bibinfo{howpublished}{\url{https://github.com/hjmshi/PyTorch-LBFGS}}.
\bibitem[{Shi et~al.(2021)Shi, Hosdez, Rougelot, Xie, Shao and
  Talandier}]{shi2021analysis}
\bibinfo{author}{Shi, H.L.}, \bibinfo{author}{Hosdez, J.},
  \bibinfo{author}{Rougelot, T.}, \bibinfo{author}{Xie, S.Y.},
  \bibinfo{author}{Shao, J.F.}, \bibinfo{author}{Talandier, J.},
  \bibinfo{year}{2021}.
\newblock \bibinfo{title}{Analysis of local creep strain field and cracking
  process in claystone by x-ray micro-tomography and digital volume
  correlation}.
\newblock \bibinfo{journal}{Rock Mechanics and Rock Engineering}
  \bibinfo{volume}{54}, \bibinfo{pages}{1937--1952}.
\bibitem[{Snow et~al.(2020)Snow, Nassar and Reutzel}]{snow2020review}
\bibinfo{author}{Snow, Z.}, \bibinfo{author}{Nassar, A.},
  \bibinfo{author}{Reutzel, E.W.}, \bibinfo{year}{2020}.
\newblock \bibinfo{title}{Review of the formation and impact of flaws in powder
  bed fusion additive manufacturing}.
\newblock \bibinfo{journal}{Additive Manufacturing} , \bibinfo{pages}{101457}.
\bibitem[{Stein(2012)}]{stein2012interpolation}
\bibinfo{author}{Stein, M.L.}, \bibinfo{year}{2012}.
\newblock \bibinfo{title}{Interpolation of spatial data: some theory for
  kriging}.
\newblock \bibinfo{publisher}{Springer Science \& Business Media}.
\bibitem[{Suquet(1990)}]{suquet1990simplified}
\bibinfo{author}{Suquet, P.}, \bibinfo{year}{1990}.
\newblock \bibinfo{title}{A simplified method for the prediction of homogenized
  elastic properties of composites with a periodic structure}.
\newblock \bibinfo{journal}{Comptes Rendus De L Academie Des Sciences Serie II}
  \bibinfo{volume}{311}, \bibinfo{pages}{769--774}.
\bibitem[{Tancogne-Dejean and Mohr(2018)}]{tancogne2018elastically}
\bibinfo{author}{Tancogne-Dejean, T.}, \bibinfo{author}{Mohr, D.},
  \bibinfo{year}{2018}.
\newblock \bibinfo{title}{Elastically-isotropic truss lattice materials of
  reduced plastic anisotropy}.
\newblock \bibinfo{journal}{International Journal of Solids and Structures}
  \bibinfo{volume}{138}, \bibinfo{pages}{24--39}.
\bibitem[{Tarantino et~al.(2019)Tarantino, Zerhouni and
  Danas}]{tarantino2019random}
\bibinfo{author}{Tarantino, M.}, \bibinfo{author}{Zerhouni, O.},
  \bibinfo{author}{Danas, K.}, \bibinfo{year}{2019}.
\newblock \bibinfo{title}{Random 3d-printed isotropic composites with high
  volume fraction of pore-like polydisperse inclusions and near-optimal elastic
  stiffness}.
\newblock \bibinfo{journal}{Acta Materialia} \bibinfo{volume}{175},
  \bibinfo{pages}{331--340}.
\bibitem[{Teubner(1991)}]{teubner1991level}
\bibinfo{author}{Teubner, M.}, \bibinfo{year}{1991}.
\newblock \bibinfo{title}{{Level surfaces of Gaussian random fields and
  microemulsions}}.
\newblock \bibinfo{journal}{EPL (Europhysics Letters)} \bibinfo{volume}{14},
  \bibinfo{pages}{403}.
\bibitem[{Torquato(2013)}]{torquato2013random}
\bibinfo{author}{Torquato, S.}, \bibinfo{year}{2013}.
\newblock \bibinfo{title}{Random heterogeneous materials: microstructure and
  macroscopic properties}. volume~\bibinfo{volume}{16}.
\newblock \bibinfo{publisher}{Springer Science \& Business Media}.
\bibitem[{Wang and Wang(2004)}]{wang2004color}
\bibinfo{author}{Wang, M.Y.}, \bibinfo{author}{Wang, X.}, \bibinfo{year}{2004}.
\newblock \bibinfo{title}{“color” level sets: a multi-phase method for
  structural topology optimization with multiple materials}.
\newblock \bibinfo{journal}{Computer Methods in Applied Mechanics and
  Engineering} \bibinfo{volume}{193}, \bibinfo{pages}{469--496}.
\bibitem[{Watson(1995)}]{watson1995treatise}
\bibinfo{author}{Watson, G.N.}, \bibinfo{year}{1995}.
\newblock \bibinfo{title}{{A treatise on the theory of Bessel functions}}.
\newblock \bibinfo{publisher}{Cambridge University press, Cambridge}.
\bibitem[{Whittle(1954)}]{whittle1954stationary}
\bibinfo{author}{Whittle, P.}, \bibinfo{year}{1954}.
\newblock \bibinfo{title}{On stationary processes in the plane}.
\newblock \bibinfo{journal}{Biometrika} , \bibinfo{pages}{434--449}.
\bibitem[{Whittle(1963)}]{whittle1963stochastic}
\bibinfo{author}{Whittle, P.}, \bibinfo{year}{1963}.
\newblock \bibinfo{title}{Stochastic-processes in several dimensions}.
\newblock \bibinfo{journal}{Bulletin of the International Statistical
  Institute} \bibinfo{volume}{40}, \bibinfo{pages}{974--994}.
\bibitem[{Williams and Rasmussen(2006)}]{williams2006gaussian}
\bibinfo{author}{Williams, C.K.}, \bibinfo{author}{Rasmussen, C.E.},
  \bibinfo{year}{2006}.
\newblock \bibinfo{title}{Gaussian processes for machine learning}.
\newblock \bibinfo{journal}{the MIT Press} \bibinfo{volume}{2},
  \bibinfo{pages}{4}.
\bibitem[{Wohlgemuth et~al.(2001)Wohlgemuth, Yufa, Hoffman and
  Thomas}]{wohlgemuth2001triply}
\bibinfo{author}{Wohlgemuth, M.}, \bibinfo{author}{Yufa, N.},
  \bibinfo{author}{Hoffman, J.}, \bibinfo{author}{Thomas, E.L.},
  \bibinfo{year}{2001}.
\newblock \bibinfo{title}{Triply periodic bicontinuous cubic microdomain
  morphologies by symmetries}.
\newblock \bibinfo{journal}{Macromolecules} \bibinfo{volume}{34},
  \bibinfo{pages}{6083--6089}.
\bibitem[{Wright and Nocedal(1999)}]{wright1999numerical}
\bibinfo{author}{Wright, S.}, \bibinfo{author}{Nocedal, J.},
  \bibinfo{year}{1999}.
\newblock \bibinfo{title}{Numerical optimization}.
\newblock \bibinfo{journal}{Springer Science} \bibinfo{volume}{35},
  \bibinfo{pages}{7}.
\bibitem[{Xie et~al.(2020)Xie, Bollapragada, Byrd and
  Nocedal}]{xie2020constrained}
\bibinfo{author}{Xie, Y.}, \bibinfo{author}{Bollapragada, R.},
  \bibinfo{author}{Byrd, R.}, \bibinfo{author}{Nocedal, J.},
  \bibinfo{year}{2020}.
\newblock \bibinfo{title}{Constrained and composite optimization via adaptive
  sampling methods}.
\newblock \bibinfo{journal}{arXiv preprint arXiv:2012.15411} .
\bibitem[{Zerhouni et~al.(2021)Zerhouni, Brisard and
  Danas}]{zerhouni2021quantifying}
\bibinfo{author}{Zerhouni, O.}, \bibinfo{author}{Brisard, S.},
  \bibinfo{author}{Danas, K.}, \bibinfo{year}{2021}.
\newblock \bibinfo{title}{Quantifying the effect of two-point correlations on
  the effective elasticity of specific classes of random porous materials with
  and without connectivity}.
\newblock \bibinfo{journal}{International Journal of Engineering Science}
  \bibinfo{volume}{166}, \bibinfo{pages}{103520}.
\bibitem[{Zerhouni et~al.(2019)Zerhouni, Tarantino and
  Danas}]{zerhouni2019numerically}
\bibinfo{author}{Zerhouni, O.}, \bibinfo{author}{Tarantino, M.},
  \bibinfo{author}{Danas, K.}, \bibinfo{year}{2019}.
\newblock \bibinfo{title}{Numerically-aided 3d printed random isotropic porous
  materials approaching the hashin-shtrikman bounds}.
\newblock \bibinfo{journal}{Composites Part B: Engineering}
  \bibinfo{volume}{156}, \bibinfo{pages}{344--354}.

\end{thebibliography}

\end{document}